\documentclass[a4paper,11pt]{article}

\usepackage[a4paper, top=25mm, bottom =25mm, 
textwidth=165mm]{geometry}

\usepackage{amsthm,amsmath,amssymb,amsfonts}
\usepackage{enumitem}

\usepackage[dvipsnames]{xcolor}

\usepackage{epsfig}
\usepackage{url,hyperref}
\newcommand{\doilink}[1]{\href{https://doi.org/#1}{\nolinkurl{doi:#1}}}
\usepackage{tikz}
\usetikzlibrary{calc,patterns,arrows,positioning,cd,calc}
\usetikzlibrary{decorations.pathreplacing,decorations.markings,shapes,decorations.pathmorphing}
\tikzset{
	on each segment/.style={
		decorate,
		decoration={
			show path construction,
			moveto code={},
			lineto code={
				\path [#1]
				(\tikzinputsegmentfirst) -- (\tikzinputsegmentlast);
			},
			curveto code={
				\path [#1] (\tikzinputsegmentfirst)
				.. controls
				(\tikzinputsegmentsupporta) and (\tikzinputsegmentsupportb)
				..
				(\tikzinputsegmentlast);
			},
			closepath code={
				\path [#1]
				(\tikzinputsegmentfirst) -- (\tikzinputsegmentlast);
			},
		},
	},
	mid arrow/.style={postaction={decorate,decoration={
				markings,
				mark=at position .5 with {\arrow[#1]{stealth}}
	}}},
}

\tikzset{
	styleArrow/.style={postaction={decorate},decoration={markings,mark=at position 0.7 with {\arrow{Stealth}}}},
	blackCircle/.style={fill=black,thick,radius=0.1,inner sep=0},
	mid arrow/.style={postaction={decorate,decoration={
				markings,
				mark=at position .5 with {\arrow[#1]{stealth}}
	}}},
	whiteCircle/.style={fill=white,thick,radius=0.1,inner sep=0},
	loop style/.style={
		styleArrow, 
	}
}

\tikzset{styleNode/.style={
		draw, ellipse, inner sep=1
}}

\tikzset{styleNodeFr/.style={
		rectangle,draw,inner sep=2
	}
}

\tikzset{
	special arrow/.style={
		decoration={
			markings,
			mark=at position #1 with {\arrow{Stealth}}
		},
		postaction={decorate}
	}
}

\makeatletter

\def\Appendix{\appendix
	\def\@seccntbformat##1{Appendix~\csname the##1\endcsname.~~}}
\makeatother

\usepackage{caption}  % For more control over captions

\counterwithin{figure}{section} % Automatically resets figure counter with each section
\newtheorem{Proposition}{Proposition}[section]
\newtheorem{Lemma}[Proposition]{Lemma}
\newtheorem*{Lemma*}{Lemma}

\newtheorem{Theorem}[Proposition]{Theorem}

\newtheorem*{Theorem*}{Theorem}

\theoremstyle{definition}
\newtheorem{Definition}[Proposition]{Definition}

\newtheorem{Remark}[Proposition]{Remark}
\newtheorem{Example}[Proposition]{Example}

\makeatletter

\@addtoreset{equation}{section}
\makeatother

\def \wt{\operatorname{wt}}

\newcommand{\re}{{\mathrm{e}}}
\newcommand{\rd}{{\mathrm{d}}}

\title{Cluster integrable systems}
\date{}
\author{Mikhail Bershtein}

\begin{document}

\maketitle

\begin{abstract}
In these lecture notes, we give an introduction to cluster integrable systems.
The topics include relativistic Toda systems, moduli spaces of framed local systems, Goncharov--Kenyon integrable systems, and quantization.
\end{abstract}

\tableofcontents

\section{Introduction}

Cluster integrable systems form a relatively new, interesting, and important class of integrable systems.
One of their basic features is that they are multiplicative (or, in physical terms, relativistic).
Another essential feature is the natural construction of discrete flows and quantization.
They have a very important (partially conjectural) relation to the moduli spaces of vacua (for instance, Coulomb branches) of supersymmetric theories.

The notes are based on my course in \emph{Training School on Contemporary Trends in Integrable Systems}, at the University of Lisbon, July 2024 \footnote{See school website with links to video recordings [\href{https://sites.google.com/view/ctis24/titles-and-abstracts}{\texttt{https://sites.google.com/view/ctis24/titles-and-abstracts}}] }.
I also extensively used my course materials of the same title at Skoltech\footnote{See the video recordings [\href{https://www.youtube.com/playlist?list=PLLGkFbxve671hc2_lbADs_1hWTsN-Tdc6}{\texttt{https://www.youtube.com/playlist?list=PLLGkFbxve671hc2\_lbADs\_1hWTsN\-Tdc6}}]}.

For these notes, it is desirable to have some basic acquaintance with integrable systems and Lie groups.
Note that no prior acquaintance with cluster algebras and varieties is assumed.
On the contrary, studying integrable structures associated with clusters can serve as a good introduction to this field.

The first (smaller) half of the notes (Sections \ref{Sec:Poisson}--\ref{Sec:Rel Toda}) is devoted to the main example of a cluster integrable system: the open relativistic Toda system.
Along the way, we recall the basic concepts of Poisson geometry, Poisson--Lie groups, and the definition of a cluster variety.

The large Sections~\ref{Sec:FG} and \ref{Sec:GK} are devoted to two different developments of the material of the first half and are independent of each other.
Namely, in Section~\ref{Sec:FG} we discuss the cluster structure on moduli spaces of framed local systems following Fock and Goncharov.
In Section~\ref{Sec:GK} we introduce Goncharov--Kenyon integrable systems.
It is worth mentioning that this class of systems was introduced by Goncharov and Kenyon in \cite{Goncharov:2013} together with the term \emph{cluster integrable systems} that gives the title of these notes.
Section~\ref{Sec:quantum} is devoted to quantization.

The text of the notes reflects the spirit of the lectures.
The main goal is to introduce notions and ideas of this field.
Proofs are often replaced by ideas, sketches, illustrative examples, or just references to the literature.
Sometimes, definitions are also replaced by explanatory figures.
Formulas are used primarily to express ideas, some conventions (especially in signs) may differ between sections.

We give references during the text, but a few general ones need to be mentioned now.
As a general introduction to cluster algebras, we recommend the book \cite{Fomin:2016introduction}, \cite{Fomin:2017introduction}, \cite{Fomin:2020introduction}, \cite{Fomin:2021introduction}.
As a general reference on the theory of integrable systems, see, e.g., \cite{Babelon:2003introduction}.
These notes are mostly based on articles \cite{Fock:2006cluster}, \cite{Fock:2006moduli}, \cite{Goncharov:2013}, \cite{Fock:2016}, \cite{Schrader:2018b}.
See also the recent review \cite{Gekhtman:2024integrable} on cluster integrable systems, which complements the material in these lectures.

\paragraph{Acknowledgments} I am grateful to A.~Grigorev, A.~Gurenkova, S.~Kosyak, A.~Marshakov, M.~Prokushkin, D.~Rachenkov, A.~Shapiro, M.~Shapiro for their comments on the subject and preliminary parts of the notes.
I am especially grateful to I.~Sechin and M.~Semenyakin for the encouragement and many lengthy discussions; in some places, these notes closely follow their advice.
I am also grateful to referees for their comments. 

AI tools (GPT-5.6) were used to improve the exposition. 

The Training School was supported by CA21109 COST Action CaLISTA.
I am grateful to S.~Abenda, G.~Cotti, G.~Bonelli, D.~Guzzetti, and A.~Tanzini for organizing the school.
Preparing these notes, I was also supported by the European Research Council under the European Union's Horizon 2020 research and innovation programme under grant agreement No 948885.

%\newpage

\section{Poisson manifolds and classical integrable systems}
\label{Sec:Poisson}

\subsection{Poisson manifolds} The exposition in this section is minimal, see, e.g., books \cite{Vaisman:2012lectures}, \cite{Laurent:2012poisson} for more details and proofs.

\begin{Definition}
	A \emph{Poisson algebra} \(A\) is a commutative algebra with a bilinear operation 
	\begin{equation}
		\{\cdot, \cdot\} \colon A\otimes A \rightarrow A
	\end{equation}
	which satisfies 
	\begin{subequations}
		\begin{align}
			&\text{anti-symmetry}&\;\;  &\{f,g\}=-\{g,f\},&\;\; &\forall f,g \in A,
			\\
			&\text{Jacobi identity}&\;\;  &\{f,\{g, h\}\}+\{h,\{f, g\}\}+\{g,\{h, f\}\}=0,&\;\; &\forall f,g,h \in A,
			\\
			&\text{Leibniz identity}&\;\;  &\{f,g h\}=\{f,g\}h+g\{f, h\},&\;\; &\forall f,g,h \in A.
		\end{align}
	\end{subequations}
\end{Definition}

\begin{Definition}
	A \emph{Poisson manifold} is a smooth manifold with a Poisson algebra structure on the algebra of smooth functions \(C^\infty(M)\).
\end{Definition}

In the definition, only Poisson brackets of global functions are defined, however, in the real smooth setting, it extends to the Poisson brackets of the functions \(C^\infty(U)\) for small open subsets \(U\subset M\) (since we can use smooth extension by zero).

Let \(M\) be a Poisson manifold.
Let \(U\subset M\) be a chart with local coordinates \(x_1,\dots, x_n\).
Let \(\Pi_{i,j}=\{x_i,x_j\}\).
Then for any functions \(f,g\) we have
\begin{equation}
	\{f,g\}=\sum\nolimits_{i,j} \Pi_{i,j} \frac{\partial f}{\partial x_i}\frac{\partial g}{\partial x_j}.	
\end{equation}
In other words, the Poisson bracket is determined by the Poisson bivector
\(\Pi=\frac12 \sum \Pi_{i,j}\partial_{x_i}\wedge \partial_{x_j} \in \Gamma(M,\Lambda^2 T_M)\). 
This ensures skew-commutativity and Leibniz identity, while the Jacobi identity is equivalent to 
\begin{equation}\label{eq:Schouten-Nijenhuis}
	\sum\nolimits_r \Big(\Pi_{r,i} \partial_{x_r}\Pi_{j,k}+\Pi_{r,j} \partial_{x_r}\Pi_{k,i}+\Pi_{r,k} \partial_{x_r}\Pi_{i,j}\Big)=0,\quad \forall i,j,k.
\end{equation}
This can also be written as \([\Pi,\Pi]=0\), where \([\cdot,\cdot]\) is the Schouten--Nijenhuis bracket.

In the algebraic or analytical setting, the existence of \(\Pi\in \Gamma(M,\Lambda^2 T_M)\) which satisfies \([\Pi,\Pi]=0\) should be taken as a definition of the Poisson structure, since there may not be sufficiently many global functions.

\begin{Example}[Constant bracket] The most basic example of the Poisson manifold is \(\mathbb{R}^{2n}\) with the bracket given by
	\begin{equation}
		\{f,g\}=\sum\nolimits_i \Big(\frac{\partial f}{\partial p_i}\frac{\partial g}{\partial x_i}-\frac{\partial f}{\partial x_i}\frac{\partial g}{\partial p_i}\Big).
	\end{equation}	
	More generally, for any constant matrix \(\Pi_{ij}\), the equation~\eqref{eq:Schouten-Nijenhuis} is clearly satisfied, so the bivector \(\Pi\) defines a Poisson bracket.
\end{Example}

In particular, \(\Pi=0\) also defines a Poisson structure.
This example shows that the matrix \(\Pi_{i,j}\) can be degenerate.

\begin{Example} The generic Poisson bracket on affine space, which is linear in coordinates, has the form
	\begin{equation}
		\{x_i,x_j\}=\sum\nolimits_k c_{i,j}^k x_k
	\end{equation}	
	The constants \(c_{i,j}^k\) should be antisymmetric \(c_{i,j}^k=-c_{j,i}^k\) and the Jacobi identity for \(\Pi\) leads to the relation 
	\begin{equation}
		\sum\nolimits_r \Big(c_{r,i}^l c_{j,k}^r+c_{r,j}^l c_{k,i}^r+c_{r,k}^l c_{i,j}^r\Big)=0,\quad \forall i,j,k,l.
	\end{equation}
	This is equivalent to the fact that \(c_{i,j}^k\) are structure constants of some Lie algebra \(\mathfrak{g}\).
	The Poisson manifold is identified with the dual space \(\mathfrak{g}^*\).
	If we restrict ourselves to the algebraic functions, then the Poisson algebra is  \(S^\bullet(\mathfrak{g})=\mathbb{C}[\mathfrak{g}^*]\).
	This is called the \emph{Kirillov--Kostant--Souriau} Poisson bracket.
\end{Example}

We will discuss examples of quadratic Poisson brackets below; actually, they will serve as the main example for us.

\begin{Example}[Symplectic manifolds]
	Recall that a symplectic manifold is a smooth manifold \(M\) equipped with a non-degenerate closed 2-form \(\omega\).
	For any function \(f\), we can assign a vector field \(V_f\) such that \(i_U(df)=\omega (U, V_f)\) for any vector field \(U\) (equivalently, we just raise the indices for the 1-form \(df\) using the form \(\omega\)).
	Then the Poisson bracket of two functions is defined as
	\begin{equation}
		\{f,g\}=\omega(V_f,V_g).
	\end{equation}
	In local coordinates, the matrix \(\Pi_{i,j}\) of the Poisson bivector is inverse to the matrix of the symplectic form \(\omega=\frac12 \sum \omega^{i,j}dx_i dx_j\).
	Here we used the non-degeneracy of \(\omega^{i,j}\) and the closeness of \(\omega\) implies the vanishing of the Schouten--Nijenhuis bracket~\eqref{eq:Schouten-Nijenhuis}.
	
	On the contrary, if the matrix \(\Pi_{i,j}\) is non-degenerate at any point, then its inverse \(\Pi^{-1}_{i,j}\) defines a symplectic structure.
\end{Example}

\subsection{Symplectic leaves} A generic Poisson manifold can be viewed as a union of symplectic manifolds.

We can view Poisson bivector \(\Pi\) as a map \(\Pi \colon T^*_M \to T_M \).
Let \(T^\Pi\) denote the image of this map.
Equivalently, for any point \(x \in M\) let \(\Pi=\sum_{i=1}^r \Pi^i_{(1)}\otimes \Pi^i_{(2)} \) be a minimal decomposition into a sum of decomposable tensors, then
\begin{equation}
	T^\Pi_x=\langle \Pi_{(1)}^i \mid 1\le i \le r\rangle = \langle \Pi_{(2)}^i \mid 1\le i \le r \rangle.
\end{equation}
Note that the foliation \(T^\Pi\) in general has non-constant rank, namely, the rank of \(T^\Pi\) at point \(x\) is equal to the rank of the matrix \(\Pi\) at \(x\).

It is easy to see that the commutator of two vector fields tangent to \(T^\Pi\) is also tangent to \(T^\Pi\).
Hence, on the open subset where \(T^\Pi\) has a maximal rank, this foliation is integrable by the Frobenius theorem.
In general, the symplectic leaves are defined as submanifolds tangent to the foliation \(T^\Pi\).

Let us give the definition more accurately.
For any function \(f\) let \(V_f=\Pi (df \otimes 1) \in \Gamma(M, TM)\) be the corresponding \emph{Hamiltonian vector field}.
We will say that a curve \(\gamma\)  is a \emph{Hamiltonian path} from \(x\) to \(y\) if \(\gamma\) is defined in an open neighbourhood of \([0,1]\), \(\gamma(0)=x\), \(\gamma(1)=y\), and \(\gamma\) is an integral curve of a Hamiltonian vector field \(V_f\), where \(f\) is defined in an open neighbourhood of \(\gamma([0,1])\).
\begin{Definition}
	We will say that \(x \sim y\) if there is a piecewise Hamiltonian path which goes from \(x\) to \(y\).
	Then \(\sim \) is an equivalence relation, and equivalence classes of \(\sim\) are called \emph{symplectic leaves}.
\end{Definition}

Here we defined symplectic leaves as connected subsets.
Note that in general symplectic leaves are not submanifolds, see Example~\ref{Ex:symplecti leaves} below.
Instead each of them is an image of the immersion \(\iota \colon S  \to M\), where \(S\) is symplectic and \(\Pi_M|_{\operatorname{Im} \iota}= \iota_* \Pi_S\).
On the open subset where \(T^\Pi\) has a maximal rank, this follows from Frobenius theorem mentioned above.
In general, this follows from the Weinstein splitting theorem \cite{Weinstein:1983}.

\begin{Example}\label{Ex:symplecti leaves}
Consider \(M\) to be a real 3-dimensional torus \(M=\mathbb{R}^3/\mathbb{Z}^3\) and \(\Pi=\partial_x \wedge (\partial_y+\alpha \partial_z)\).
	Then for \(\alpha \not \in \mathbb{Q}\), each symplectic leaf is dense in \(M\).
\end{Example}

\begin{Definition}
	For a Poisson algebra \(A\), the \emph{Poisson center} is 
	\begin{equation}
		Z(A)=\{f\in A \mid \{f,g\}=0, \forall g \in A\}.
	\end{equation} 
	Any element of \(Z(A)\) is called a \emph{Casimir function}. 
\end{Definition}

Often generic symplectic leaves are defined as level sets of Casimir functions (generators of the Poisson center).

\begin{Example} 
	Let \(M=\mathbb{R}^3\) and \(\Pi=\partial_x \wedge \partial_y\).
	Then \(T_p^\Pi=\langle\partial_x,\partial_y \rangle\), and symplectic leaves are horizontal planes \(z=const\).
	The Poisson center is an algebra of functions on \(z\).
	
More generally, let \(M=\mathbb{R}^n\) and \(\Pi\) be constant.
	As above, let us consider \(\Pi\) as an operator \(\Pi\colon \mathbb{R}^n\to \mathbb{R}^n\).
	Then the symplectic leaves are planes parallel to \(\operatorname{Im}\Pi\), and the Poisson center as an algebra is generated by \(\operatorname{Ker}\Pi\).
\end{Example}	

\begin{Example}
	Consider the Poisson bracket in \(\mathbb{R}^3\) given by $\{x, y\} = 0$, $\{z, x\} = x$, $\{z, y\} = y$.
	Then any point \((0,0,z)\) is a zero-dimensional symplectic leaf. The other symplectic leaves are the connected components of the planes \(\{ax+by=0 \}\) with the \(z\)-axis removed.
	Locally, these leaves are separated by the Casimir function  \(x/y\).
	However, there is no global Casimir function, since it cannot be continuously extended to the \(z\)-axis.
\end{Example}

\begin{Example} 
Let \(\mathfrak{g}\) be a Lie algebra, consider \(\mathfrak{g}^*\) with \(\Pi_{KKS}\).
Let \(\alpha\in \mathfrak{g}^*\) be any point and \(x,y\in \mathfrak{g}\) considered as linear functions on \(\mathfrak{g}^*\).
	Then we have
	\begin{equation}\label{eq:x,y, alpha}
		\{x,y\}(\alpha)=\alpha([x,y])=-\operatorname{ad}^*_x(\alpha)(y).
	\end{equation}
	Here \(\operatorname{ad}^*\) denotes the coadjoint action of \(\mathfrak{g}\) on \(\mathfrak{g}^*\).
	The formula~\eqref{eq:x,y, alpha} means that \(\Pi(dx)(\alpha)	=-\operatorname{ad}^*_x(\alpha)\) where \(\Pi\) is considered as an operator \(T^*_M\rightarrow T_M \).
Hence \(T^\Pi\) at point \(\alpha\) is generated by the Lie algebra elements acting on \(\alpha\), i.e. by \(\operatorname{ad}^*_x(\alpha)\).
Therefore, the symplectic leaf is given by the coadjoint orbit \(Ad^*(G)\alpha\), where \(G\) is a (connected) Lie group corresponding to \(\mathfrak{g}\).
	
	In particular, this implies that any coadjoint orbit is even-dimensional.
	
Let a function \(f \in S^\bullet (\mathfrak{g}) =\mathbb{C}[\mathfrak{g}^*]\) be an element of the Poisson center.
This is equivalent to \(\{f,x_i\}=0\) \(\forall i\), where \(\langle x_i\rangle\) is a basis of \(\mathfrak{g}\).
Hence, the Poisson center of \(S^\bullet (\mathfrak{g})\) coincides with the subalgebra of invariants under the adjoint action \(S^\bullet (\mathfrak{g})^{\mathfrak{g}}\) (or, equivalently, \(S^\bullet (\mathfrak{g})^{G}\)).
	
	For example, let us consider \(\mathfrak{g}=\mathfrak{gl}_N\).
We can also identify the dual space \(\mathfrak{g}^*\) with the space of \(N\times N\) matrices using the \(\operatorname{Tr}\) form.
	Then we can view \(\alpha, x, y\) as matrices and relation~\eqref{eq:x,y, alpha} means that
	\begin{equation}
		\operatorname{Tr}(\alpha[x,y])=-\operatorname{Tr}([x,\alpha]y).
	\end{equation}

	The coadjoint orbit of \(\alpha\) is a conjugation class \(\{g \alpha g^{-1} \mid g \in GL_N\}\).
	It is well known that conjugation classes are parameterized by Jordan normal forms.
	On the open set where eigenvalues are distinct, the conjugation classes are distinguished by the symmetric functions of eigenvalues of the matrix \(\alpha\), or, equivalently, the coefficients of the characteristic polynomial
	\begin{equation}
		\det(\operatorname{Id}+\lambda\alpha)=\sum\nolimits_{j=0}^N \lambda^j \operatorname{Tr}\bigwedge\nolimits^j\alpha.
	\end{equation} 
	Therefore \(S^\bullet (\mathfrak{gl}_N)^{\mathfrak{gl}_N}\) is generated by \(N\) algebraically independent functions \(\operatorname{Tr}\bigwedge\nolimits^j\alpha\), \(1\le j \le N\).
	Hence, the dimension of a generic coadjoint orbit equals \(N^2-N\).
\end{Example}

\subsection{Classical Integrable systems}
For any function \(H\) on the Poisson manifold, let us define the corresponding \emph{Hamiltonian vector field} by \(V_H =\Pi (dH \otimes 1) \in \Gamma(M, T_M)\).
The trajectories along this vector field are given by the Hamiltonian equations
\begin{equation}\label{eq:Ham EOM}
	\frac{\rd }{\rd t}g=\{H,g\}.
\end{equation}
The Hamiltonian vector fields are tangent to \(T^\Pi\).
The corresponding trajectories preserve symplectic leaves.
The integrals of motion (functions preserved by the flow) are the functions that Poisson commute with \(H\).

\begin{Definition}
	An \emph{integrable system} is a symplectic manifold \(M\) of dimension \(2n\) and \(n\) functionally independent functions \(H_1,\dots,H_n\) such that \(\{H_i,H_j\}=0\), \(\forall i,j\).
\end{Definition}

In the algebraic setting, functional independence is replaced by algebraic independence.
The number \(n\) is the maximal size of the independent Poisson commuting set, and any function \(f\) that Poisson commutes with \(H_1,\dots H_n\) would be functionally (algebraically in the algebraic setting) dependent on them.
The manifold \(M\) is called the \emph{phase space} of the integrable system.
The functions \(H_1,\dots,H_n\) are called \emph{Hamiltonians} of the integrable system.

We will also discuss integrable systems in the Poisson manifold setting.
Let \(M\) be a Poisson manifold of dimension \(2n+k\) and assume that the Poisson center of \(C^\infty(M)\) is (locally) generated by \(k\) functionally independent functions.
Then a generic symplectic leaf has dimension \(2n\).
The integrable system is a system of \(n+k\) functionally independent and Poisson commuting functions.
Its restriction defines an integrable system on a generic symplectic leaf.
Informally, one can think that an integrable system contains \(k\) Casimir functions and \(n\) Hamiltonians.

Usually, integrable systems are defined locally, on some chart or open submanifold.
Then, the extension of the integrable system to the compactification is an interesting geometric question, which we ignore in these lectures.

\begin{Example}
	Consider a Poisson algebra with generators \(J_x,J_y,J_z\) and brackets 
	\begin{equation}
		\{J_x,J_y\}=J_z,\;\; \{J_z,J_x\}=J_y,\;\; \{J_y,J_z\}=J_x.
	\end{equation}
	These are the brackets of angular momentum components in three dimensions.
	Equivalently, this is the KKS Poisson bracket for \(\mathfrak{so}(3)\) (which is isomorphic to $\mathfrak{sl}_2$ over complex numbers).
	
	There is one Casimir function given by \(J^2=J_x^2+J_y^2+J_z^2\).
	An integrable system can be given by a pair of Poisson commuting and functionally independent functions \(J_z, J^2\).
\end{Example}

\begin{Example}[Gelfand--Tsetlin integrable system] 
	Consider the phase space \(M=\mathfrak{gl}_N^*\).
	We can identify this space with the space of \(N\times N\) matrices $L$.
	The matrix elements \(L_{a,b}\), \(1\le a,b \le N\)  are generators of the algebra of functions.
	The KKS bracket, in terms of these functions, has the form
	\begin{equation} 
		\{L_{a,b}, L_{c,d}\}= \delta_{b,c} L_{a,d}- \delta_{a,d} L_{c,b}.
	\end{equation}
	The Poisson center is generated by the functions \(\operatorname{Tr}\bigwedge\nolimits^j L\) (traces of exterior powers of the matrix~\(L\)), \(1\le j \le N\). 
	
	Let \(L^{(k)}\), \(1\le k \le N\) denote the matrix which is formed by the first \(k\) rows and columns of \(L\).
	Consider functions \(H_{j,k}=\operatorname{Tr}\bigwedge\nolimits^j L^{(k)}\), \(1 \le k \le N\), \(1 \le j \le k\).
It is easy to see that they Poisson commute with each other.
	Indeed, let us consider \(H_{j,k}\) and \(H_{j',k'}\).
	Without loss of generality, we can assume that \(k' \le k\).
	Note that \(H_{j,k}\) Poisson commutes with any function of the matrix elements of the submatrix~\(L^{(k)}\).
	Hence \(\{H_{j,k},H_{j',k'}\}=0\).
	
It can be proved (see, e.g., \cite{Kostant:2006gelfand}) that functions \(\{H_{j,k} \mid 1 \le k \le N, 1 \le j \le k\}\) are functionally independent.
	Overall, we have \(\frac{N(N+1)}{2}=N+\frac{N(N-1)}{2}\) functions, where \(N\) is the number of Casimirs and \(\frac{N(N-1)}{2}\) is half of the dimension of a generic symplectic leaf.
	Hence, we defined an integrable system.
	It is called the Gelfand--Tsetlin integrable system.
\end{Example}

\begin{Example}[Open Toda system] \label{Ex:Toda}
	Consider a space with coordinates \(p_1,\dots, p_N, q_1,\dots, q_N\) and canonical Poisson brackets 
	\begin{equation}
		\{p_i,q_j\}=\delta_{i,j},\quad \{p_i,p_j\}=\{q_i,q_j\}=0, \quad \forall i,j.
	\end{equation} 
	The Toda Hamiltonian is 
	\begin{equation}\label{eq:Toda Ham}
		H_2=\frac12 \sum\nolimits_{i=1}^N p_i^2+ \sum\nolimits_{i=1}^{N-1} \re^{q_i-q_{i+1}}.
	\end{equation}
	
	The formulas above depend on the exponentials of the \(q_i\).
	So the phase space of the system is \(T^*(\mathbb{R}^*)^N\).
	Here \((\re^{q_1},\dots,\re^{q_N})\) are coordinates on \((\mathbb{R}^*)^N\) and \((p_1,\dots,p_N)\) are coordinates in the fiber of the cotangent bundle.
	Moreover, usually, we complexify the phase space to \(T^*(\mathbb{C}^*)^N\).
	
	The equations of motion~\eqref{eq:Ham EOM} for the Hamiltonian~\eqref{eq:Toda Ham} have the form
	\begin{equation}
		\begin{cases}
		&\frac{\rd }{\rd t} q_i=p_i,\quad 1 \le i \le N; 
		\\
		&\frac{\rd }{\rd t} p_1=-\re^{q_1-q_2};\;\; \frac{\rd }{\rd t} p_N=\re^{q_{N-1}-q_N};
		\\
		&\frac{\rd }{\rd t} p_i=\re^{q_{i-1}-q_{i}}-\re^{q_i-q_{i+1}},\quad 1 < i < N.
		\end{cases}
	\end{equation}
	
	We want to construct an integrable system, i.e., embed \(H_2\) into a system of commuting Hamiltonians.
	One standard approach to this is based on the so-called Lax matrix.
	Namely, consider the following \(N\times N\) matrices \(L\) and \(M\)
	\begin{equation}\label{eq:Lax additive}
		L=\begin{pmatrix}
			p_1 & a_1 & 0 & \dots & 0 & 0
			\\
			a_1 & p_2 & a_2 & \dots & 0 & 0
			\\ 
			0 & a_2 & p_3 & \dots & \vdots  & \vdots 
			\\ 
			\vdots  & \ddots & \ddots & \ddots & \ddots & \vdots 
			\\ 
			0 & 0& \dots & 0 & a_{N-1} & p_N 
		\end{pmatrix}
		\quad 		
		M=\frac12\begin{pmatrix}
			0 & a_1 & 0 & \dots & 0 & 0
			\\
			-a_1 & 0 & a_2 & \dots & 0 & 0
			\\ 
			0 & -a_2 & 0 & \dots & \vdots  & \vdots 
			\\ 
			\vdots  & \ddots & \ddots & \ddots & \ddots & \vdots 
			\\ 
			0 & 0& \dots & 0 & -a_{N-1} & 0
		\end{pmatrix},
	\end{equation}
	where \(a_i=\exp(\frac12(q_i-q_{i+1}))\).
	It is straightforward to check that the equations of motion are equivalent to the Lax equations
	\begin{equation}
		\frac{\rd}{\rd t}L=[M,L].
	\end{equation}
	It follows from this equation that the spectrum of \(L\) is preserved, i.e., functions \(H_k=\frac1{k!}\operatorname{Tr}L^k\) are integrals of motion.
	Clearly \(H_1=\sum p_i\) is a momentum operator and \(H_2\) is a Toda Hamiltonian~\eqref{eq:Toda Ham}.
	
	Note, however, that we have not proved Poisson commutativity of \(H_k\), \(k=1,\dots, N\); we only noted that they are integrals, i.e., commute with \(H_2\).
	In order to compute Poisson brackets between \(H_k\) and \(H_m\), it is convenient first to find Poisson brackets between matrix elements of the matrix \(L\).
	They can be written in the form
	\begin{equation}\label{eq:PB r additive}
		\{L_1, L_2\} =[r,L_1+L_2].
	\end{equation}
	Let us explain the notations.
	The identity~\eqref{eq:PB r additive} is in \(\operatorname{End}(\mathbb{C}^N\otimes \mathbb{C}^N)\), i.e., effectively, an identity between matrices of size \(N^2\times N^2\).
	Here and below we denote \(L_1=L\otimes 1\) and \(L_2=1\otimes L\).
	And \(r\) is an \(N^2\times N^2\) matrix of the form
	\begin{equation}\label{eq:r matrix}
		r=\frac12 \sum\nolimits_{a<b} \left(E_{a,b}\otimes E_{b,a}-E_{b,a}\otimes E_{a,b} \right ).
	\end{equation}
	It is called a \emph{classical \(r\)-matrix}.
	Here \(E_{a,b}\) denotes a matrix unit.
	Using all definitions above, we can rewrite formula~\eqref{eq:PB r additive} as
	\begin{equation}
		\{L_{a_1,b_1}, L_{a_2,b_2}\}=\sum\nolimits_c \left( r_{a_1,a_2}^{c,b_2}L_{c,b_1}+r_{a_1,a_2}^{b_1,c}L_{c,b_2}- L_{a_1,c}r_{c,a_2}^{b_1,b_2}- L_{a_2,c}r_{a_1,c}^{b_1,b_2}\right).
	\end{equation}
	
	Therefore we get 
	\begin{multline}
		\{\operatorname{Tr}L^k,\operatorname{Tr}L^m\} = \operatorname{Tr}_{12}\{L_1^k,L_2^m\}
		= \operatorname{Tr}_{12} \sum\nolimits_{i=1}^k \sum\nolimits_{j=1}^m L_1^{i-1}L_2^{j-1}\{L_1,L_2\} L_1^{k-i}L_2^{m-j}
		\\ 
		= \operatorname{Tr}_{12} \left( mk\, [r,L_1+L_2] L_1^{k-1}L_2^{m-1} \right)
		=	\operatorname{Tr}_{12} \left( m [r,L_1^k]L_2^m+ k [r,L_2^m]L_1^k \right) =0,
	\end{multline}
	where \(\operatorname{Tr}_{12}\) denotes the trace of an operator acting on the tensor product \(\mathbb{C}^N\otimes \mathbb{C}^N\) of two vector spaces.
	One can also show that \(H_1,\dots, H_N\) are functionally independent.
	Indeed, we have \(k!
	H_k=\sum p_i^k+(\text{terms of lower degree in the }p_i)\).
	Hence \(\det(\partial H_i /\partial p_j)\) has leading term equal to the Vandermonde determinant \(\prod_{i<j}(p_i-p_j)\) and is therefore nonzero at a generic point.
	
	Hence, we proved the integrability of the (open) Toda system.
	See, e.g., \cite{Babelon:2003introduction} for more details about it.
\end{Example}

\begin{Remark}\label{rem:bialgebra Poisson dual}
	We used formula~\eqref{eq:PB r additive} above as a definition of the Poisson bracket on the space of tridiagonal matrices of the form~\eqref{eq:Lax additive}.
	On the other hand, one can use the same formula to define the Poisson bracket on the space of all \(N\times N\) matrices.
	This bracket is linear, so it is the KKS bracket for some Lie algebra.
	One can show that this Lie algebra is isomorphic to \(\{L^+,L^-\mid L^{\pm}\in \mathfrak{b}_{\pm}, \operatorname{pr}_+(L^+)+\operatorname{pr}_-(L^-)=0\}\), where \(\mathfrak{b}_+\) and \(\mathfrak{b}_-\) are Lie algebras of upper and lower triangular matrices and \(\operatorname{pr}_\pm \colon \mathfrak{b}_{\pm} \to \mathfrak{h} \) are projections onto the Lie algebra of diagonal matrices.
	This is the Lie algebra dual to \(\mathfrak{gl}_N\) with the standard \(r\)-matrix bialgebra structure.
\end{Remark}

\section{Poisson--Lie groups} \label{Sec:PL groups}
\subsection{Poisson--Lie groups} In the examples above, the phase space was additive at least in some directions (vector space in the case of the Gelfand--Tsetlin system and total space of a vector bundle in the case of the Toda system).
Now we move to the \emph{multiplicative setting}; informally, this is a step from Lie algebras to Lie groups.

In terms of classical integrable systems, we will get \emph{relativistic} systems (morally with a replacement like \(p^2 \mapsto \re^p+\re^{-p}\)).
After quantization, this would correspond to the step from the differential operators to the \emph{difference} operators.\footnote{Indeed, in the usual presentation \(\hat{p}=\hbar \partial_x\), Taylor's formula gives the shift operator \(\re^{\hat{p}} F(x)=F(x+\hbar)\).}

In order to do this, we need a reasonable Poisson structure on the group.
First, we recall the standard definition from Poisson geometry.
\begin{Definition}
	Let \((X,\Pi_X)\) and \((Y,\Pi_Y)\) be Poisson manifolds with Poisson bivectors \(\Pi_X\) and \(\Pi_Y\), respectively.
	The manifold \((X\times Y, \Pi_X+\Pi_Y)\) is called a \emph{product of Poisson manifolds}.
	
	A map \(\varphi \colon X\to Y\) is called a \emph{Poisson map} if  \(\varphi_* \Pi_X=\Pi_Y\). 
\end{Definition}

\begin{Definition}
	A \emph{Poisson--Lie} group is a Lie group \(G\) with a Poisson structure such that the multiplication \(m\colon G \times G \to G \) is a Poisson map.
\end{Definition}
More explicitly, this property means that for any two functions \(\phi, \psi \in C^\infty(G)\) we have 
\begin{equation}\label{eq:PL functions}
	\{\phi,\psi\}(gh)= \{\phi,\psi\}(gh)|_{g \text{ fixed}}+\{\phi,\psi\}(gh)|_{h \text{ fixed}}.
\end{equation}
In terms of the Poisson bivector \(\Pi\) the Poisson--Lie property means 
\begin{equation}\label{eq:Poison bivector on PL group}
	\Pi(gh)=(\rho_h\times \rho_h)_* \Pi(g)+(\lambda_g\times \lambda_g )_* \Pi(h),
\end{equation}
where \(\rho_g\colon G \to G\) is multiplication by \(g\) to the right and \(\lambda_g\colon G \to G\) is multiplication by \(g\) to the left.
In particular, it follows from this formula that \(\Pi(e)=0\), hence a Poisson--Lie group cannot be symplectic (if \(\dim G \neq 0\)).

Note that no conditions are imposed on the inversion map \(g \mapsto g^{-1}\).
It follows from the Poisson property of the multiplication that inversion is an anti-Poisson map, see, e.g., \cite[Sec. 2.1]{Etingof:2002Lectures}.

\begin{Example}
	Let \(\mathfrak{g}\) be a Lie algebra.
	Consider \(\mathfrak{g}^*\) with the addition operation and the KKS Poisson bracket.
	This is a Poisson--Lie group.
	Indeed, it is sufficient to check property~\eqref{eq:PL functions} for linear functions, say \(x_i,x_j\), since linear functions generate the algebra of all functions.
	We have
	\begin{equation}
		\{x_i,x_j\}=\sum c_{i,j}^k x_k =\sum c_{i,j}^k x^{(1)}_k+\sum c_{i,j}^k x^{(2)}_k
		=\{x^{(1)}_{i}+x^{(2)}_{i},x^{(1)}_j+x^{(2)}_j\}
		%=\{x^{(1)}_{i},x^{(1)}_j\}+\{x^{(2)}_{i},x^{(2)}_j\}
		,
	\end{equation}
	where \(x^{(1)}_{i}\) and \(x^{(2)}_{i}\) denote linear functions on the first and second factors of \(\mathfrak{g}^*\times \mathfrak{g}^*\), respectively.
	On the left side, we used the Poisson bracket on \(\mathfrak{g}^*\), while on the right side, we used the Poisson bracket on~\(\mathfrak{g}^*\times \mathfrak{g}^*\).
\end{Example}

Now let us consider the main example of a Poisson--Lie group.
Let \(G=GL_N\) (most constructions below also work for \(G=SL_N\) or \(G=PGL_N\)).
The Poisson bracket between matrix elements \(L_{a,b}\) of the matrix \(L \in G\) is given by the formula
\begin{equation}\label{eq:PB r mult}
	\{L_1, L_2\} =[r,L_1L_2].
\end{equation}
This bracket is called the \emph{Sklyanin bracket}.
Notations here are the same as in~\eqref{eq:PB r additive}, namely \(L_1=L\otimes 1\), \(L_2=1\otimes L\), and \(r\) is given by formula~\eqref{eq:r matrix}.
Explicitly formula~\eqref{eq:PB r mult} has the form
\begin{equation}
	\{L_{a_1,b_1}, L_{a_2,b_2}\}=r_{a_1,a_2}^{c_1,c_2}L_{c_1,b_1}L_{c_2,b_2}- L_{a_1,c_1}L_{a_2,c_2}r_{c_1,c_2}^{b_1,b_2}.
\end{equation}
In particular, we see that this Poisson structure is quadratic.

Anti-commutativity of the Sklyanin bracket follows from anti-symmetry of the \(r\)-matrix.
The Poisson--Lie property~\eqref{eq:PL functions} can be shown as
\begin{multline}
	\{L_1,L_2\}=[r,L_1L_2]=[r,L_1^{(1)}L_1^{(2)}L_2^{(1)}L_2^{(2)}]
	=[r,L_1^{(1)}L_2^{(1)}]L_1^{(2)}L_2^{(2)}+L_1^{(1)}L_2^{(1)}[r,L_1^{(2)}L_2^{(2)}]
	\\
	=\{L_1^{(1)},L_2^{(1)}\}L_1^{(2)}L_2^{(2)} + L_1^{(1)}L_2^{(1)}\{L_1^{(2)},L_2^{(2)}\} =\{L_1^{(1)}L_1^{(2)},L_2^{(1)}L_2^{(2)}\},
\end{multline}
where upper indices in \(L^{(1)}\) and \(L^{(2)}\) correspond to the first and second factors of the product \(G \times G\).

In order to show that formula~\eqref{eq:PB r mult} actually defines a Poisson--Lie group structure on \(G\), it remains to show the Jacobi identity.
It appears (from Drinfeld's theorems, see \cite[Th. 2.2 and Th. 3.1]{Etingof:2002Lectures}) that the Jacobi identity follows from the \emph{modified classical Yang--Baxter} relation satisfied by \(r\)
\begin{equation}\label{eq:modified classical YB}
	[r_{12},r_{13}]+[r_{12},r_{23}]+[r_{13},r_{23}]	=c \Omega.
\end{equation}
This is an identity in \(\mathfrak{g}^{\otimes 3}\).
By \(r_{ij}\) we denote the \(r\)-matrix acting on \(i\)th and \(j\)th factors, e.g., \(r_{12}=r\otimes 1\). \(\Omega=\sum E_{ab}\wedge E_{bc}\wedge E_{ca}\) denotes the unique (up to scalar) \(\mathfrak{g}\)-invariant element in \(\Lambda^3\mathfrak{g}\), and \(c\) stands for a scalar, which is not important.

See, e.g., \cite{Etingof:2002Lectures} for more details about Poisson--Lie groups.

\subsection{Double Bruhat cells}
In order to define integrable systems, it is reasonable to have some information about symplectic leaves on \(G\).
It appears easier to describe certain Poisson submanifolds first.

\begin{Definition}
	Let \(M\) be a Poisson manifold.
	The submanifold \(N\subset M\)  is called a \textit{Poisson submanifold} if \(\Pi|_N \in \Lambda^2 T_N\).
\end{Definition}

Note that this condition is quite restrictive.
For example, if \(M\) is a symplectic manifold, then the only Poisson submanifolds of \(M\) are submanifolds of full dimension, i.e., open submanifolds of \(M\).
Informally, the Poisson submanifolds are unions of symplectic leaves.

Let \(B=B_+\subset G\) be a Borel subgroup of upper triangular matrices and \(B_-\) be a Borel subgroup of lower triangular matrices.
Let \(H\subset G\) denote the Cartan subgroup, which we identify with the subgroup of diagonal matrices.
By \(W\simeq S_N \simeq N(H)/H\) we denote the Weyl group of \(G\).
For any element \(w\in W\), we can assign an element \(\tilde{w}\in G\) which is a lift of \(w\) to the normalizer \(N(H)\) of the torus \(H\).
The element \(\tilde{w}\) is defined up to multiplication by elements of \(H\).
 
\begin{Theorem}[Bruhat decomposition] 
	The group \(G\) has the following decompositions
	\begin{equation}\label{eq:Bruhat}
		G= \bigsqcup_{w \in W} B_- \tilde{w} B_+ = \bigsqcup_{w \in W} B_+ \tilde{w} B_+ =\bigsqcup_{w \in W} B_- \tilde{w} B_-.
	\end{equation}
\end{Theorem}

\begin{Remark}
	The choice of lifts \(\tilde{w}\) is not essential since \(H\subset B_-\) and \(H \subset B_+\).
\end{Remark}

\begin{Remark}
	Let \(w_0=\begin{pmatrix}
		1 & 2 & \dots &N 
		\\
		N & N-1 & \dots & 1 
	\end{pmatrix}\) be the longest element in \(W\). 
	Then \(\tilde{w}_0 B_+\tilde{w}_0^{-1}=B_-\). Hence for any \(w \in W\) we have \(B_- \tilde{w} B_+= \tilde{w}_0 B_+ \widetilde{w_0 w} B_+\). 
	Therefore, the first decomposition~\eqref{eq:Bruhat} is equivalent to the second one. 
	Similarly, they are equivalent to the third one.
	
	The open cell corresponding to \(w=e\) in the first decomposition in~\eqref{eq:Bruhat} is a Gauss (or LDU) decomposition.
\end{Remark}

\begin{Definition}
	For \(w=(w_+,w_-)\in W\times W\) the \emph{double Bruhat cell} is an intersection of Bruhat cells for \(B_+\) and \(B_-\)
	\begin{equation}
		G^w=B_+ \tilde{w}_+B_+ \cap B_- \tilde{w}_-B_-.
	\end{equation}	
\end{Definition}

The formula for the dimension of a double Bruhat cell reads \(\dim G^w=\dim H + l(w_+)+l(w_-)\), where~\(l(u)\) is the length of the element \(u\in W\) (number of inversions).
Two extreme cases are \(G^{(e,e)}=H\) and \(G^{(w_0,w_0)}\) which is an open subset in \(G\).

Double Bruhat cells are important for the Poisson geometry of \(G\).

\begin{Theorem}\label{Th:double Bruhat Poisson}
	For any \(w\in W\times W\) the double Bruhat cell \(G^{w}\) is a Poisson submanifold of \(G\). 
\end{Theorem}

\begin{Remark}
	More generally, by the Semenov--Tian--Shansky theorem \cite{Semenov:1985dressing}, the symplectic leaves on a Poisson--Lie group \(A\) are orbits of the so-called \emph{dressing action} by the dual Poisson--Lie group \(A^*\).
	In our case the dual Poisson--Lie group \(G^*=\{(L^+,L^-)\mid \operatorname{pr}_+L^+\operatorname{pr}_-L^-=e\}\subset B^+\times B^-\), where \(\operatorname{pr}_{\pm}\colon B^\pm \to H\) are natural projections (cf.~Remark~\ref{rem:bialgebra Poisson dual}).
	This explains the appearance of \emph{double} Bruhat cells in Theorem~\ref{Th:double Bruhat Poisson}.
\end{Remark}

See \cite{Hoffmann:2000factorization}, \cite{Kogan:2002symplectic} for the explicit description of symplectic leaves in \(G\).

\begin{Example} 
	Consider the group \(GL(2)\).
	Let us denote coordinates of the generic matrix in \(G\) by \(\begin{pmatrix} a & b \\ c & d \end{pmatrix} \).
	The Poisson brackets~\eqref{eq:PB r mult} between these functions have the form
	\begin{equation}
		\begin{aligned}
			&\{a,b\}=\frac12 a b ,&\;\; &\{a,c\}=\frac12 a c,&\;\;  &\{a,d\}=b c,
			\\
			&\{b,c\}=0,&\;\; &\{b,d\}= \frac12 b d,&\;\;  &\{c,d\}=\frac12 cd.
		\end{aligned}
	\end{equation}
	It is easy to show that functions \(\det=ad-bc\) and \(b /c\) are Casimirs. 	
	
	The Weyl group in this case consists of two elements \(W=\{e,s\}\).
	Hence, there are four double Bruhat cells
	\begin{subequations}\label{eq:Double Bruhat SL(2)}
	\begin{align}
%		\begin{aligned}
		G^{e,e}&=\Big\{\!\begin{pmatrix}
			a & 0 \\ 0 &d
		\end{pmatrix}\!\Big\}=\Big\{\!\begin{pmatrix}
		a & b \\ c &d
		\end{pmatrix}\!\Big| b=0, c=0 \Big\},
		\\
		G^{s,e}&=\Big\{\!\begin{pmatrix}
			a & 0 \\ c &d
		\end{pmatrix}\!\Big| c \neq 0 \Big\} = \Big\{\!\begin{pmatrix}
		a & b \\ c &d
		\end{pmatrix}\!\Big| b=0, c\neq 0 \Big\},
		\\
		G^{e,s}&=\Big\{\!\begin{pmatrix}
		a & b \\ 0 &d
		\end{pmatrix}\!\Big| b \neq 0 \Big\} = \Big\{\!\begin{pmatrix}
		a & b \\ c &d
		\end{pmatrix}\!\Big| b \neq 0, c=0 \Big\},
		\\
		G^{s,s}&=\Big\{\!\begin{pmatrix}
		a & b \\ c &d
		\end{pmatrix}\!\Big| b,c \neq 0 \Big\}.
%		\end{aligned}
	\end{align}
	\end{subequations}	
	It is straightforward to check that these submanifolds are Poisson.
	For example, for \(G^{s,e}\) this means that the Poisson bracket with \(b\) vanishes on a submanifold on which \(b\) vanishes, i.e., \(\{b,f\}|_{b=0}=0\), \(\forall f\).
	
	It is easy to see that the Poisson bracket on \(G^{e,e}\) vanishes, hence it is a 2-parametric set of 0-dimensional symplectic leaves. 
	On the cells \(G^{s,e}, G^{e,s}\), the Poisson bracket is non-zero, hence both of them become a union of 2-dimensional symplectic leaves parametrized by \(\det\).
	Finally, \(G^{s,s}\) is a union of 2-dimensional symplectic leaves parametrized by \(\det\) and \(b/c\).
\end{Example}	
In general, for \(G=GL_N\) the double Bruhat cells can be described by equations and inequalities similarly to~\eqref{eq:Double Bruhat SL(2)}, see \cite[Sec. 4]{Fomin:1999double}.
In order to construct integrable systems more explicitly, we will need some coordinates on the (open subsets of) double Bruhat cells.
We will do this below.

\subsection{Factorization schemes} Now we are going to introduce coordinates on double Bruhat cells \(G^w\), \(w\in W\times W\simeq S_{N}\times S_{N}\).
The group \(S_{N}\) is generated by simple reflections \(s_i=(i,i+1)\), \(1 \le i \le N-1\), which satisfy the braid relations.
We will write \(s_{\bar{1}},\dots, s_{\overline{N-1}}\) for generators of the first factor and \(s_{1},\dots, s_{N-1}\) for generators of the second factor.
We will also use more visible notation \(\bar{s}_i\) for \(s_{\bar{i}}\).
Recall that a \emph{reduced} decomposition of an element \(w \in W\times W\) is a presentation of \(w\) as a product of \(s_i\) of minimal length.
Such a presentation is not unique, but any two reduced decompositions can be connected using the following braid relations
\begin{subequations}\label{eq:braid}
	\begin{align}
		\label{eq:braid 1}
		&s_{i}s_{j}=s_{j}s_{i},\quad \bar{s}_{i}\bar{s}_{j}=\bar{s}_{j}\bar{s}_{i},\qquad j\neq i-1,i,i+1,
		\\\label{eq:braid 2} 
		&s_{i}s_{i+1}s_{i}=s_{i+1}s_{i}s_{i+1},\quad \bar{s}_{i}\bar{s}_{i+1}\bar{s}_{i}=\bar{s}_{i+1}\bar{s}_{i}\bar{s}_{i+1},
		\\
		\label{eq:braid 3}
		&s_i\bar{s}_{j}=\bar{s}_{{j}}s_i.
	\end{align}
\end{subequations}

Let us assume that \(G=SL_N\).
Introduce the following matrices
\begin{subequations}\label{eq:EHF}
	\begin{align}
		&E_i=1+E_{i,i+1}=\exp(E_{i,i+1}),\qquad E_{\bar{i}}=F_i=1+E_{i+1,i}=\exp(E_{i+1,i}), 
		\\
		&H_i(X) =H_{\bar{i}}(X) =\operatorname{diag}(\underbrace{X^{\frac{N-i}{N}},\dots, X^{\frac{N-i}{N}}}_{i}, \underbrace{X^{-\frac{i}{N}},\dots ,X^{-\frac{i}{N}}}_{N-i} ).
	\end{align}	
\end{subequations}
Note that \(H_i(X)\in SL_N\).
Moreover, it is easy to see that \(H_i(X)\) defines a 1-parametric subgroup in \(G\) and this subgroup corresponds to a fundamental coweight.
\footnote{To be precise, \(H_i(X)\) as an element of \(SL_N\) requires a choice of the \(N\)-th root of \(X\). 
On the other hand, it is well defined as an element of \(PGL_N\).}
\begin{Definition}\label{Def:factorization}
	For any reduced word \(w=s_{i_1}s_{i_2}\cdot\dots \cdot s_{i_l}\), \(i_1,\dots, i_l \in \{1,\dots,N-1,\bar{1},\dots,\overline{N-1}\}\) we assign a product
	\begin{equation}\label{eq:factorization}
		\mathbb{L}_{\mathbf{s}}(\mathbf{X})=H_1(X_1)\cdot \dots \cdot H_{N-1}(X_{N-1}) E_{i_1} H_{i_1}(X_N)E_{i_2} H_{i_2}(X_{N+1})\cdot \dots E_{i_l} H_{i_l}(X_{N+l-1}).
	\end{equation}
	This defines a \emph{factorization map} \((\mathbb{C}^*)^{N+l-1}\rightarrow SL_N\).
\end{Definition}
It can be proven that the image of the factorization map \(\mathbb{L}_{\mathbf{s}}\) is contained in the double Bruhat cell \(G^w\).
More precisely
\begin{Theorem}[\cite{Fomin:1999double}]
	The factorization map associated with \(w\in W\times W\) gives a birational isomorphism between \((\mathbb{C}^*)^{N-1+l(w)}\) and the double Bruhat cell \(G^w\).
\end{Theorem}
Due to this theorem, we can view functions \(X_1,\dots, X_{N+l-1}\) as local coordinates on (an open subset of) a double Bruhat cell.
Hence, we can compute the Poisson bracket between them.
We will explain the following result in the next sections.
\begin{Theorem}[\cite{Fock:2006cluster}]
	The Poisson bracket in coordinates \(X_i\) has the form \(\{X_i,X_j\}=\epsilon_{ij}X_iX_j\) for some constants \(\epsilon_{ij}\in \frac12 \mathbb{Z}\).
\end{Theorem}

\begin{Remark}
	Note that the Poisson structure above can be rewritten as follows 
	\begin{equation}
		\{F,G\}=\sum_{i,j} \epsilon_{ij}X_iX_j \frac{\partial F}{\partial X_i}\frac{\partial G}{\partial X_j}= \sum_{i,j} \epsilon_{ij} \frac{\partial F}{\partial \log X_i}\frac{\partial G}{\partial \log X_j}
	\end{equation}
	In other words, this structure is \emph{logarithmically constant}, i.e., constant in coordinates \(x_i=\log X_i\).
	In particular, this proves the Jacobi identity for the Poisson bracket.
\end{Remark}

\begin{Example}\label{Ex:cells SL2}
	The factorization schemes of \(SL(2)\) have the form 
	\begin{subequations}\label{eq:Double Bruhat SL(2) factor}
	\begin{align}
%		\begin{aligned}
			&G^{e,e},\qquad \Big\{H_1(X_1)\Big\}=\Big\{\!\begin{pmatrix}
				X_1^{1/2} & 0 \\ 0 & X_1^{-1/2}
			\end{pmatrix}\!\Big\},
			\\
			&G^{s,e},\qquad \Big\{H_1(X_1)F_1H_1(X_2)\Big\}=\Big\{\!\begin{pmatrix}
				(X_1X_2)^{1/2} & 0 \\ (X_2/X_1)^{1/2} & (X_1X_2)^{-1/2}
			\end{pmatrix}\!\Big\},
			\\
			&G^{e,s},\qquad \Big\{H_1(X_1)E_1H_1(X_2)\Big\}=\Big\{\!\begin{pmatrix}
				(X_1X_2)^{1/2} & (X_1/X_2)^{1/2} \\ 0 & (X_1X_2)^{-1/2}
			\end{pmatrix}\!\Big\},
			\\
%			&G^{s,s},\qquad H_1(X_1)F_1H_1(X_2)E_1H_1(X_3)=\Big\{\!\begin{pmatrix}
%				X_1^{1/2}X_2^{1/2}X_3^{1/2} & X_1^{1/2}X_2^{1/2}X_3^{-1/2} \\ X_1^{-1/2}X_2^{1/2}X_3^{1/2} & X_1^{-1/2}X_3^{-1/2}(X_2^{-1/2}+X_2^{1/2})
%			\end{pmatrix}\!\Big\},
%						\\
			&G^{s,s},\qquad \Big\{H_1(X_1)F_1H_1(X_2)E_1H_1(X_3)\Big\}=\Big\{\!\begin{pmatrix}
				(X_1X_2X_3)^{1/2}& (X_1X_2/X_3)^{1/2} \\ (X_2X_3/X_1)^{1/2} & (X_1X_2X_3)^{-1/2}(1+X_2)
			\end{pmatrix}\!\Big\},
%		\end{aligned}
	\end{align}
	\end{subequations}	
	Note that in the case of $G^{s,s}$ we obtained only the chart with \(a \neq 0\).
	If we change the reduced expression of \(w=\bar{s}_1s_1=s_1\bar{s}_1\), i.e., change the factorization scheme, we get another chart with \(d\neq 0\).
	\begin{equation}
		G^{s,s},\qquad H_1(\tilde{X}_1)E_1H_1(\tilde{X}_2)F_1H_1(\tilde{X}_3)=\Big\{\!\begin{pmatrix}
		(\tilde{X}_1\tilde{X}_2\tilde{X}_3)^{1/2}(1+\tilde{X}_2^{-1})& (\tilde{X}_3\tilde{X}_2/\tilde{X}_1)^{-1/2} \\ (\tilde{X}_1\tilde{X}_2/\tilde{X}_3)^{-1/2} & (\tilde{X}_1\tilde{X}_2\tilde{X}_3)^{-1/2}
	\end{pmatrix}\!\Big\}.
	\end{equation}
	The complement of the union of these two charts has codimension 2 in \(G^{s,s}\). 
\end{Example}

Recall that, in general, different reduced decompositions of \(w\) are connected by braid relations~\eqref{eq:braid}.
Hence, the braid relations correspond to transformations of local coordinates.
Formulas for such transformations emerge from the following relations on matrices~\eqref{eq:EHF}
\begin{subequations}\label{eq:rel EHF}
	\begin{align}\label{eq:rel EHF 1}
		&H_i(X) E_j=E_j H_i(X),\quad H_i(X) F_j=F_j H_i(X),\quad E_iF_j=F_jE_i,\qquad  i\neq j,
		\\ \label{eq:rel EHF 2}
		&H_i(X_1)H_i(X_2)=H_i(X_1X_2),\quad H_i(X_1)H_j(X_2)=H_j(X_2)H_i(X_1),
		\\ \label{eq:rel EHF 3}
		&E_iE_j=E_jE_i,\quad F_iF_j=F_jF_i,\qquad j\neq i{-}1,i,i{+}1,
		\\ \label{eq:rel EHF 4}
		&E_iE_{i+1}H_{i}(X)E_i = H_{i+1}\Big(\frac{1}{1+X^{-1}}\Big) H_{i}(1+X) E_{i+1} E_{i} H_{i+1}\Big(X^{-1}\Big) E_{i+1} H_i\Big(\frac{1}{1+X^{-1}}\Big) H_{i+1}(1+X),
		\\  \label{eq:rel EHF 5}
		&F_iF_{i+1}H_{i}(X)F_i=H_{i}\Big(\frac{1}{1+X^{-1}}\Big) H_{i+1}(1+X) F_{i+1} F_{i} H_{i+1}(X^{-1}) F_{i+1} H_{i+1}\Big(\frac{1}{1+X^{-1}}\Big) H_{i}(1+X),
		\\  \label{eq:rel EHF 6}
		&F_iH_{i}(X)E_i=H_{i}\Big(\frac{1}{1+X^{-1}}\Big) E_i H_{i}(X^{-1}) F_i H_{i}\Big(\frac{1}{1+X^{-1}}\Big) H_{i-1}(1+X) H_{i+1}(1+X).
	\end{align}
\end{subequations}
Here we assumed that \(H_0(X)=H_N(X)=1\).
As we said above, the relations among matrices~\eqref{eq:rel EHF} correspond to braid relations in the Weyl group \(W\times W\).
Namely, relation~\eqref{eq:rel EHF 3} corresponds to~\eqref{eq:braid 1}, relations~\eqref{eq:rel EHF 4} and~\eqref{eq:rel EHF 5} correspond to~\eqref{eq:braid 2}, and
the third relation in~\eqref{eq:rel EHF 1} and~\eqref{eq:rel EHF 6} correspond to~\eqref{eq:braid 3}. 

\begin{Example}\label{Ex: ss}
	The relation between coordinates \(\tilde{\mathbf{X}}\) and \(\mathbf{X}\) in Example~\ref{Ex:cells SL2} for the cell \(G^{s,s}\) has the form 
	\begin{equation}\label{eq:s1s1p mutation}
		\tilde{X}_1=\frac{X_1}{1+X_2^{-1}} ,\; \tilde{X}_2=X_2^{-1},\;\tilde{X}_3=\frac{X_3}{1+X_2^{-1}}.
	\end{equation}
	This can be deduced from the transformation~\eqref{eq:rel EHF 6}.
\end{Example}

\begin{Example}\label{Ex: s123}
	Let us take \(G=SL_3\) and consider the cell \(G^{e,w_0}\).
	Recall that \(w_0\) has two reduced decompositions
	\begin{equation}
		w_0=s_1s_2s_1=s_2s_1s_2
	\end{equation}
	Then using relation~\eqref{eq:rel EHF 5} we get
	\begin{multline}
		\mathbb{L}_{s_1s_2s_1}(X_1,X_2,X_3,X_4,X_5)=
		\\ 
		=H_1(X_1)H_2(X_2)E_1H_1(X_3)E_2H_2(X_4)E_1H_1(X_5)
		\\
		=H_1(X_1)H_2(X_2)\left(H_{2}\Big(\frac{1}{1+X_3^{-1}}\Big) H_{1}(1+X_3) E_{2} E_{1} H_{2}\Big(X_3^{-1}\Big) E_{2} H_1\Big(\frac{1}{1+X_3^{-1}}\Big) H_{2}(1+X_3)\right)
		\\ H_2(X_4)H_1(X_5)
		=H_1(\tilde{X}_1)H_2(\tilde{X}_2)E_2H_2(\tilde{X}_3)E_1H_1(\tilde{X}_4)E_2H_2(\tilde{X}_5)
		\\=
		\mathbb{L}_{s_2s_1s_2}(\tilde{X}_1,\tilde{X}_2,\tilde{X}_3,\tilde{X}_4,\tilde{X}_5),
	\end{multline}	
	where 
	\begin{equation}\label{eq:s1s2s1 mutation}
		\tilde{X}_1=X_1(1+X_3) ,\; \tilde{X}_2=\frac{X_2}{1+X_3^{-1}},\;\tilde{X}_3=X_3^{-1},\; \tilde{X}_4=\frac{X_4}{1+X_3^{-1}}, \; \tilde{X}_5=X_5(1+X_3).
	\end{equation}
	This transformation is a particular example of cluster mutation, which we will discuss below.
\end{Example}

\section{Cluster varieties}
\subsection{Seeds} Informally speaking, \(\mathcal{X}\)-cluster varieties are Poisson varieties with an atlas with nice (Darboux-like) coordinates on each chart and simple (binomial) transformations between charts.
The charts are labeled by so-called seeds.
For the reference about \(\mathcal{X}\)-cluster varieties, see for example \cite{Fock:2006cluster}.

\begin{Definition}
	A \emph{cluster seed} consists of the following data \((I,I_{\mathrm{f}}, \epsilon, \mathbf{X})\), where 
	\begin{itemize}
		\item \(I\) is a finite set. \(I_{\mathrm{f}}\subset I\) is a frozen subset.
		\item \(\epsilon\) is an anti-symmetric matrix with rows and columns labeled by \(I\).
		The matrix element \(\epsilon_{ij}\in \frac12 \mathbb{Z}\), \(\forall i,j \in I\) and, moreover  \(\epsilon_{ij}\in \mathbb{Z}\), if \( i \in I\setminus I_{\mathrm{f}}\) or \(j \in I\setminus I_{\mathrm{f}}\).
		\item \(\mathbf{X}\) is a set of variables labeled by \(I\): \(\mathbf{X}=(X_i\mid i\in I)\).
	\end{itemize}
	
	A \emph{cluster chart} is an algebraic torus \(\mathcal{X}_{\mathsf{s}}=(\mathbb{C}^*)^n\) such that \(\mathbf{X}\) are coordinate functions on it.
	The \emph{cluster Poisson bracket} is defined by the formula
	\begin{equation}\label{eq:PB cluster}
		\{X_i,X_j\}=\epsilon_{ij}X_i X_j, \quad \forall i,j \in I.
	\end{equation}			
\end{Definition}

We will call \((I,I_{\mathrm{f}}, \epsilon)\) \emph{combinatorial data} and \(\mathbf{X}\) \emph{algebraic data}.
The cluster Poisson bracket~\eqref{eq:PB cluster} is quadratic, but in \emph{logarithmic coordinates} \(x_i=\log(X_i)\) we get a constant Poisson bracket \(\{x_i,x_j\}=\epsilon_{ij}\).
Hence, the cluster Poisson bracket can be called \emph{logarithmically constant}.

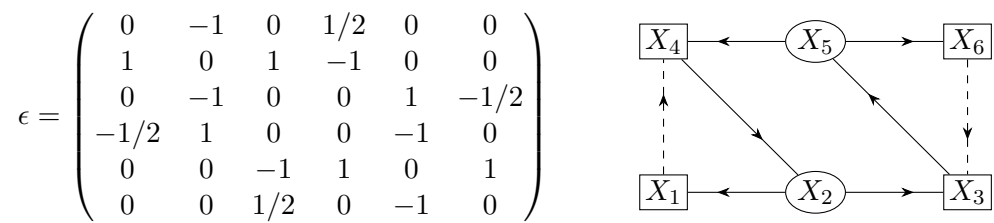
\begin{figure}[h]
	\begin{center}
	\begin{tikzpicture}
		\def\xs{2} 		\def\ys{2}
		
		\node at (-2.5*\xs,0.5*\ys) {	\(\epsilon=\begin{pmatrix} 
				0 & -1 & 0 & 1/2 & 0 &0
				\\
				1 & 0 & 1 & -1 & 0 &0
				\\
				0 & -1 & 0 & 0 & 1 &-1/2
				\\
				-1/2 & 1 & 0 & 0 & -1 &0
				\\
				0 & 0 & -1 & 1 & 0 &1
				\\
				0 & 0 & 1/2 & 0 & -1 &0		
			\end{pmatrix}\)	};
		
		\node[styleNodeFr] (X1) at (0,0) {$X_1$};	
		\node[styleNode] (X2) at (\xs,0) {$X_2$};
		\node[styleNodeFr] (X3) at (2*\xs,0) {$X_3$};
		
		\node[styleNodeFr] (Y1) at (0,\ys) {$X_4$};
		\node[styleNode] (Y2) at (\xs,\ys) {$X_5$};
		\node[styleNodeFr] (Y3) at (2*\xs,\ys) {$X_6$};
		\draw[styleArrow](X2) to  (X1) ;
		\draw[styleArrow](Y2) to  (Y1) ;
		\draw[styleArrow](X2) to  (X3) ;
		\draw[styleArrow](Y2) to  (Y3) ;
		\draw[styleArrow](Y1) to  (X2) ;
		\draw[styleArrow](X3) to  (Y2) ;
		
		\draw[styleArrow, dashed](X1) to  (Y1) ;
		\draw[styleArrow, dashed](Y3) to  (X3) ;
	\end{tikzpicture}
		\caption{\label{Fig:ex} On the left is the matrix \(\epsilon\), and on the right is the corresponding quiver.}		
	\end{center}
\end{figure}

It is convenient to represent graphically combinatorial data using quivers (i.e., oriented graphs).
The set of vertices for the quiver is \(I\).
	Often, the frozen vertices (i.e., ones corresponding to \(I_{\mathrm{f}}\)) are depicted by squares and unfrozen ones are depicted by circles.
If \(\epsilon_{ij}\in \mathbb{Z}_{\ge 0}\), we draw \(\epsilon_{ij}\) solid arrows from \(i\) to \(j\).
If \(\epsilon_{ij}\in \mathbb{Z}_{\ge 0}+\frac{1}{2}\) we draw \((\epsilon_{ij}-\frac12)\) solid arrows and also one dashed arrow between \(i\) and \(j\).
See Fig.~\ref{Fig:ex} for an example.
Here and below, we label the vertex corresponding to \(i \in I \) by the cluster variable \(X_i \in \mathbf{X}\).

\begin{Remark}\label{Rem:seed lattice}
	The combinatorial part of a seed can also be defined as a quadruple \((\Lambda,\mathbf{e}, I, I_{\mathrm{f}})\), where \(\Lambda\) is a lattice (a free abelian group with antisymmetric pairing \((\cdot,\cdot)\colon \Lambda \times \Lambda \to \frac12\mathbb{Z}\)) and \(\mathbf{e}=(e_i\mid i \in I)\) is a set of free generators of \(\Lambda\).
	Then the matrix \(\epsilon\) is a Gram matrix \(\epsilon_{ij}=(e_i,e_j)\).
	As before the set \(I_{\mathrm{f}}\) should satisfy \((e_i,e_j)\in \mathbb{Z}\) for \(i\in I \setminus I_{\mathrm{f}}\) or \(j\in I \setminus I_{\mathrm{f}}\).
	
	Note that the lattice and basis are defined by the matrix \(\epsilon\) up to isomorphism.
	
	Note that the definition of a seed varies among different authors, for example algebraic data (coordinates \(\mathbf{X}\) on the cluster torus \(\mathcal{X}_{\mathsf{s}}\)) are not included in the definition of a seed in \cite{Fock:2006cluster}, \cite{Gross:2015birational}, but are included in \cite{Fomin:2016introduction}.
\end{Remark}

The algebra of functions on the cluster chart \(\mathbb{C}[\mathcal{X}_{\mathsf{s}}]\) is an algebra of Laurent polynomials in variables~\(\mathbf{X}_{\mathsf{s}}\).
In notations above, for any \(\lambda \in \Lambda\) we can assign a monomial \(X_\lambda\), namely, if \(\lambda=\sum n_i e_i\), then \(X_\lambda=\prod X_i^{n_i}\).
The monomials \(\{X_\lambda|\lambda \in \Lambda\}\) form a basis of \(\mathbb{C}[\mathcal{X}_{\mathsf{s}}]\).
The notation \(X_\lambda\) becomes especially useful in a quantum setting, see Sec.~\ref{Sec:quantum}.

\subsection{Mutations} The charts corresponding to seeds are glued by mutations and permutations.
Let \((I,I_{\mathrm{f}},\epsilon,\mathbf{X}), (\tilde{I},\tilde{I}_{\mathrm{f}},\tilde{\epsilon},\tilde{\mathbf{X}}) \) be two seeds.
Let \(\varphi\colon I \to \tilde{I}\) be a bijection which induces a bijection of frozen subsets \(I_{\mathrm{f}} \to \tilde{I}_{\mathrm{f}}\) and satisfies \(\epsilon_{ij}=\tilde{\epsilon}_{\varphi(i)\varphi(j)}\).
Then we can extend this bijection to an isomorphism of seeds by \(\varphi(X_i)=\tilde{X}_{\varphi(i)}\).

For example, for any permutation \(\varphi\colon I \to I\) that preserves \(I_{\mathrm{f}}\) we assign an isomorphism between seeds \((I,I_{\mathrm{f}},\epsilon,\mathbf{X})\) and \( (\tilde{I},\tilde{I}_{\mathrm{f}},\tilde{\epsilon},\tilde{\mathbf{X}}) \), where the matrix \(\tilde{\epsilon}\) is obtained by permutation of columns and rows of matrix~\(\epsilon\) and the cluster variables \(\tilde{\mathbf{X}}\) are obtained by permutation of \(\mathbf{X}\).

The definition of mutation is perhaps a key definition in the cluster theory.

\begin{Definition} \label{Def:mutation}
	\textit{Mutation} at an unfrozen vertex \(k\in I \setminus I_{\mathrm{f}}\) is a transformation of seeds \(\mu_k\colon (I,I_{\mathrm{f}},\epsilon,\mathbf{X}) \rightarrow  (I,I_{\mathrm{f}},\tilde{\epsilon},\tilde{\mathbf{X}}) \) such that
	\begin{equation}\label{eq:mutation rule}
		\tilde{\epsilon}_{ij}=
		\begin{cases} 		
			-\epsilon_{ij}, \quad \text{ if } i=k  \text{ or } j=k  
			\\ 
			\epsilon_{ij}+\frac{\epsilon_{ik}|\epsilon_{kj}|-\epsilon_{jk}|\epsilon_{ki}|}{2},\ \text{ otherwise }
		\end{cases}	
		\quad 
		\tilde{X}_i=\begin{cases} 
			X_k^{-1} \qquad \text{if } k=i 
			\\  
			X_i(1+X_k^{\operatorname{sgn}\epsilon_{ik}})^{\epsilon_{ik}} \quad \text{if } k\neq i 
		\end{cases}.
	\end{equation}
\end{Definition}

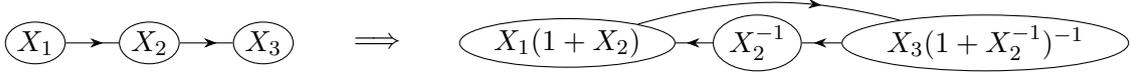
\begin{figure}[h]
	\begin{center}
		\begin{tikzpicture} 
			\def\xs{1}
			\node[styleNode] (A) at (0,0) {$X_1$};	
			\node[styleNode] (B) at (1.5*\xs,0) {$X_2$};
			\node[styleNode] (C) at (3*\xs,0) {$X_3$};
			\draw[styleArrow](A) to   (B) ;
			\draw[styleArrow](B) to   (C) ;
			
			\node at (4.5*\xs,0) {$\Longrightarrow$};
			
			\begin{scope}[shift={(7*\xs,0)}]	
				\node[styleNode] (A) at (0,0) {$X_1(1+X_2)$};	
				\node[styleNode] (B) at (2.5*\xs,0) {$X_2^{-1}$};
				\node[styleNode] (C) at (5.5*\xs,0) {$X_3(1+X_2^{-1})^{-1}$};
				\draw[styleArrow](B) to   (A) ;
				\draw[styleArrow](C) to  (B) ;
				\draw[styleArrow](A) to [bend left=15]    (C) ;
			\end{scope}
		\end{tikzpicture}
	\end{center}
	\caption{\label{Fig:ex mut}Example of mutation in the vertex 2}	
\end{figure}

A simple example of a mutation is given in Fig.~\ref{Fig:ex mut}.
In terms of quivers, the mutation can be stated simply as the following algorithm:
\begin{enumerate}
	\item  Reverse the directions of all arrows incident to the vertex $k$.
	\item For each pair of arrows $i \to k$ and $k \to j$, draw an arrow $ j\to i$ (close 3-cycles).
	\item Delete pairs of arrows of the opposite direction $i \to j$ and $j \to i$ (remove 2-cycles).
\end{enumerate}

In terms of the lattice, the mutation can be given by either of the following two formulas 
\begin{equation}\label{eq:basis mutation}
	\mu_{k,+}(e_{i})=
	\begin{cases} 		
		-e_k, \qquad \text{ if } i=k,
		\\ 
		e_i+\operatorname{max}\big((e_i,e_k),0\big)e_k, \text{ if } i\neq k,
	\end{cases}	
	\quad 
	\mu_{k,-}(e_{i})=
		\begin{cases} 		
			-e_k, \qquad \text{ if } i=k, 
			\\ 
			e_i+\operatorname{max}\big((e_k,e_i),0\big)e_k, \text{ if } i\neq k.
		\end{cases}	
\end{equation}
It is easy to see that bases \(\mu_{k,+}(\mathbf{e})\) and \(\mu_{k,-}(\mathbf{e})\) are connected by a linear transformation that preserves the antisymmetric form on \(\Lambda\). 

The following proposition states that mutation agrees with the cluster Poisson bracket~\eqref{eq:PB cluster}.
\begin{Proposition} 
	Mutation is a Poisson map, i.e., \(\{\tilde{X}_i,\tilde{X}_j\}=\tilde{\epsilon}_{i,j}\tilde{X}_i\tilde{X}_j\).
\end{Proposition}

\begin{Example}\label{Ex:ex ss sss cluster} 
	It is easy to see that the transformation given by formula~\eqref{eq:s1s1p mutation} above corresponds to the mutation of the seed with the quiver depicted on the left in Fig.~\ref{Fig:ex Double Bruhat} at the vertex \(X_2\).
	Similarly, the transformation given by formula~\eqref{eq:s1s2s1 mutation} corresponds to the mutation of the quiver depicted on the right in Fig.~\ref{Fig:ex Double Bruhat} at the vertex \(X_3\).
	Note that edges between frozen vertices are not specified by formula~\eqref{eq:s1s2s1 mutation}; we choose them to agree with the construction of the cluster structure on double Bruhat cells in Section~\ref{Sec:Cluster Bruhat}.
	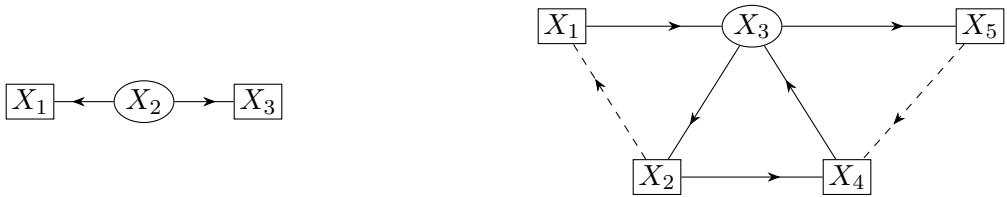
\begin{figure}[h]
		\begin{center}
			\begin{tikzpicture} 
				\def\xs{1}
				\node[styleNodeFr] (X1) at (0,0) {$X_1$};	
				\node[styleNode] (X2) at (1.5*\xs,0) {$X_2$};
				\node[styleNodeFr] (X3) at (3*\xs,0) {$X_3$};
				\draw[styleArrow](X2) to   (X1) ;
				\draw[styleArrow](X2) to   (X3) ;

				\begin{scope}[shift={(7*\xs,-1*\xs)}]	
					\node[styleNodeFr] (X1) at (0,2*\xs) {$X_1$};	
					\node[styleNode] (X3) at (2.5*\xs,2*\xs) {$X_3$};
					\node[styleNodeFr] (X5) at (5.5*\xs,2*\xs) {$X_5$};
					\node[styleNodeFr] (X2) at (1.25*\xs,0) {$X_2$};
					\node[styleNodeFr] (X4) at (3.75*\xs,0) {$X_4$};
					\draw[styleArrow](X1) to   (X3) ;
					\draw[styleArrow](X3) to  (X2) ;
					\draw[styleArrow](X4) to   (X3) ;
					\draw[styleArrow](X3) to  (X5) ;
					\draw[styleArrow](X2) to  (X4) ;
					
					\draw[styleArrow,dashed](X2) to   (X1) ;
					\draw[styleArrow,dashed](X5) to   (X4) ;
				\end{scope}
			\end{tikzpicture}
		\end{center}
		\caption{\label{Fig:ex Double Bruhat}Quivers corresponding to Examples \ref{Ex: ss} and \ref{Ex: s123}}
	\end{figure}
	
\end{Example}

\bigskip 

Let us take a seed \(\mathsf{s}\) and consider the set \(|\mathsf{s}|\) of all seeds (up to permutations) 
connected with \(\mathsf{s}\) via sequences of mutations.
Different mutation sequences may produce the same seed, and such phenomena can be interpreted as relations among cluster mutations.
The most basic relations are given in the following proposition.
\begin{Proposition}	\label{Prop:mutation relations}
	\begin{enumerate}[label=(\alph*)]
	\item \label{item:a}Mutation is an involution, i.e., \(\mu_k\mu_k=\operatorname{id}\).

	\item \label{item:b} If vertices \(i\) and \(j\) are not connected (i.e., \(\epsilon_{ij}=0\)), then mutations in vertices \(i\) and \(j\) commute \(\mu_i\mu_j=\mu_j\mu_i\).
	
	\item \label{item:c} If vertices \(i\) and \(j\) are connected by one arrow  (i.e., \(\epsilon_{ij}=1\)), then mutations in vertices \(i\) and \(j\) satisfy  \(\mu_j\mu_i\mu_j=(i,j)\mu_j\mu_i \), where \((i,j)\) is the transposition.
	\end{enumerate}
\end{Proposition}

Properties \ref{item:a} and \ref{item:b} are straightforward.
The property \ref{item:c} is called the pentagon property.
It is also straightforward, but the accurate proof requires checking several cases; it is easier to perform it in the quantum setting.

\begin{Example}\label{Ex:A2 quiver}
	Let us consider the simplest example of property \ref{item:c}.
	Namely, consider the quiver with two vertices and one edge between them, say from 1 to 2.
	This quiver is called an \(A_2\) quiver since the corresponding non-oriented graph coincides with the \(A_2\) Dynkin diagram.
	We denote the variables corresponding to vertices 1 and 2 by \(X\) and \(Y\), respectively.
	A detailed check of the pentagon identity is given in Fig.~\ref{Fig:ex pentagon}.
	
	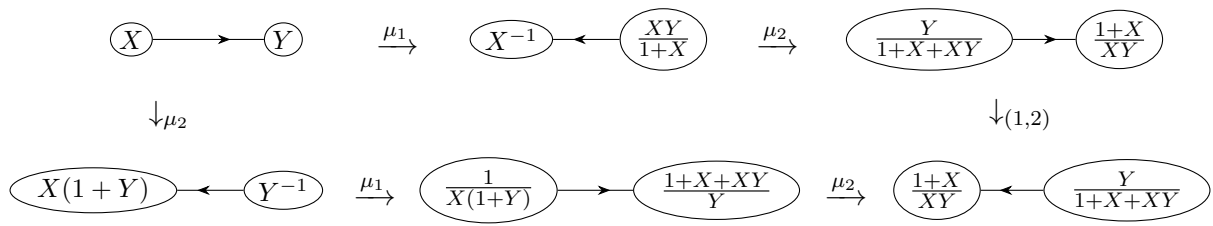
\begin{figure}[h]
		\begin{center}
			\begin{tikzpicture} 
				\def\xs{1}
				
				\begin{scope}
					\node[styleNode] (X) at (-0.5*\xs,0) {\small $X$};	
					\node[styleNode] (Y) at (1.5*\xs,0) {\small $Y$};
					\draw[styleArrow](X) to   (Y) ;
				\end{scope}
				
				\node at (3*\xs,0) {$\xrightarrow{\mu_1}$};
				
				\begin{scope}[shift={(4.5*\xs,0)}]	
					\node[styleNode] (X) at (0,0) {\small $X^{-1}$};	
					\node[styleNode] (Y) at (2*\xs,0) {$\frac{X Y}{1+X}$};
					\draw[styleArrow](Y) to   (X) ;
				\end{scope}
				
				\node at (8*\xs,0) {$\xrightarrow{\mu_2}$};
				
				\begin{scope}[shift={(10*\xs,0)}]	
					\node[styleNode] (X) at (0,0) {$\frac{Y}{1+X+XY}$};	
					\node[styleNode] (Y) at (2.5*\xs,0) {$\frac{1+X}{XY}$};
					\draw[styleArrow](X) to   (Y) ;
				\end{scope}

				\node at (0,-\xs) {$\downarrow_{\mu_2}$};
					
				\begin{scope}[shift={(0,-2*\xs)}]
					\node[styleNode] (X) at (-1*\xs,0) {\small $X(1+Y)$};	
					\node[styleNode] (Y) at (1.5*\xs,0) {\small$ Y^{-1}$};
					\draw[styleArrow](Y) to   (X) ;
				\end{scope}
				
				\node at (2.7*\xs,-2*\xs) {$\xrightarrow{\mu_1}$};
				
				\begin{scope}[shift={(4.2*\xs,-2*\xs)}]	
					\node[styleNode] (X) at (0,0) {$\frac{1}{X(1+Y)}$};	
					\node[styleNode] (Y) at (3*\xs,0) {$\frac{1+X+X Y}{Y}$};
					\draw[styleArrow](X) to   (Y) ;
				\end{scope}
				
				\node at (8.9*\xs,-2*\xs) {$\xrightarrow{\mu_2}$};
				
				\node at (11.2*\xs,-\xs) {$\downarrow_{(1,2)}$};

				\begin{scope}[shift={(10.1*\xs,-2*\xs)}]	
					\node[styleNode] (X) at (0,0) {$\frac{1+X}{XY}$};	
					\node[styleNode] (Y) at (2.5*\xs,0) {$\frac{ Y}{1+X+XY}$};
					\draw[styleArrow](Y) to   (X) ;
				\end{scope}

			\end{tikzpicture}
		\end{center}
		\caption{\label{Fig:ex pentagon}Pentagon for \(A_2\) quiver}	
	\end{figure}
\end{Example}

One can consider the set \(|\mathsf{s}|\) as a set of vertices for a graph, in which two seeds \(\mathsf{s}\) and \(\mathsf{s}'\) are connected by an edge if they differ by one mutation.
The properties~\ref{item:b}, \ref{item:c} mean that this graph contains some 4-gon and 5-gon cycles.
It is instructive to think about this graph as a skeleton of some (multidimensional) polyhedron; then these 4-gon and 5-gon cycles are faces.

\subsection{Cluster varieties}

The \(\mathcal{X}_{|\mathsf{s}|}\) cluster variety is defined as a gluing of \(\mathcal{X}_{\mathsf{s}'}\) for \(\mathsf{s}'\in |\mathsf{s}|\).
By definition, it depends on the class of mutation equivalent seeds \(|\mathsf{s}|\), not on a particular starting representative \(\mathsf{s}\).
See \cite{Fock:2009cluster} for the definition and also \cite{Gross:2015birational} for a more recent treatment.

By \(\mathbb{L}[\mathcal{X}_{|\mathsf{s}|}]\) we denote the algebra of global functions on \(\mathcal{X}_{|\mathsf{s}|}\).
In other words, these are the functions on~\(\mathcal{X}_{\mathsf{s}}\) that are Laurent polynomials in cluster variables \(\mathbf{X}_{\mathsf{s}'}\) for any \(\mathsf{s}'\in |\mathsf{s}|\).
This Laurent property motivates using the letter \(\mathbb{L}\).
This looks like an infinite number of constraints, but it appears that it suffices to check only the seeds neighboring to \(\mathcal{X}_{\mathsf{s}}\).
This property is called \textit{starfish Lemma} or \textit{1-step mutation property}, see \cite[Lem. 3.8]{Gross:2015birational}, \cite{Berenstein:2005cluster}.

\begin{Theorem}
	Let \(\mathsf{s}_i\) be a seed obtained from \(\mathsf{s}\) via mutation in vertex \(i\in I \setminus I_{\mathrm{f}}\).
	Let \(F\) be a function on the initial cluster chart, \(F\in \mathbb{C}[\mathcal{X}_{\mathsf{s}}]\), and assume that for any \(i\in I \setminus I_{\mathrm{f}}\) we have \(F\in \mathbb{C}[\mathcal{X}_{\mathsf{s}_i}]\).
	Then \(F\in \mathbb{C}[\mathcal{X}_{\mathsf{s}'}]\) for any seed \(\mathsf{s}' \in |\mathsf{s}|\).
\end{Theorem}

Geometrically this theorem means that the variety \(\mathcal{X}_{|\mathsf{s}|}\) is isomorphic to \(\mathcal{X}_{\mathsf{s}}\cup\bigcup_{i \in I \setminus I_{\mathrm{f}}} \mathcal{X}_{\mathsf{s}_i}\) up to codimension 2.

As an easy corollary of this theorem, we note 
\begin{Lemma}
	Assume that for a given \(i\in I\) we have \(\epsilon_{ji}\ge 0\) for any \(j \in I\setminus I_{\mathrm{f}}\).
	Then the variable \(X_i\) is a global function.
\end{Lemma}

Perhaps, the most basic example of global functions is provided by \(\mathcal{A}\)-variables.
This is a fundamental notion in the cluster algebra theory, which is dual in some sense to \(\mathcal{X}\)-variables used above.
Assume for simplicity that the adjacency matrix \(\epsilon\) is not degenerate and integer-valued.
Let us define \(A_i\) such that \(X_i=\prod A_j^{\epsilon_{ji}}\).

Note that unless \(\det \epsilon=\pm1\) the functions \(\mathbf{A}\) do not belong to \(\mathbb{C}[\mathcal{X}_{\mathsf{s}}]\), but rather belong to an algebraic extension that includes some roots of variables \(\mathbf{X}\).
Geometrically, the torus \(\mathcal{A}_\mathsf{s}\) is a finite cover of \(\mathcal{X}_\mathsf{s}\).

\begin{Lemma}
	The definition of mutation~\ref{Def:mutation} implies the following transformation of \(\mathcal{A}\)
	\begin{equation}
		\tilde{A}_i=\begin{cases} 
			A_k^{-1}\Big(\prod_{j, \epsilon_{jk}>0}A_j^{\epsilon_{jk}} +
			\prod_{j, \epsilon_{kj}>0}A_j^{\epsilon_{kj}}   \Big) \quad \text{if } k=i, 
			\\  
			A_i \qquad\qquad \text{if } k\neq i.
		\end{cases}
	\end{equation}
\end{Lemma}

The following property is called the Laurent phenomenon.
It was proven in the seminal paper~\cite{Fomin:2002cluster}.
\begin{Theorem}
	For any seed \(\mathsf{s}' \in |\mathsf{s}|\), all \(\mathcal{A}\) cluster variables are Laurent polynomials in initial variables.
\end{Theorem}

\begin{Example}
	Returning to Example~\ref{Ex:A2 quiver} we see that functions \(X_1^{-1}=X^{-1}\)  and \(X_2=Y\) are global.
	These functions are also \(A\) variables \(A_1=Y\), \(A_2=X_1^{-1}\).
\end{Example}

%\bigskip 

We call the group of birational transformations of \(\mathcal{X}_\mathsf{s}\), generated by sequences of mutations (and permutations) which preserve the quiver \(\mathcal{Q}\), the \emph{cluster modular group} \(G_{\mathcal{Q}}\).
The group \(G_{\mathcal{Q}}\) does not depend on the choice of the initial seed \(\mathsf{s}'\in |\mathsf{s}|\) up to (non-canonical) isomorphism.

\section{Clusters and relativistic Toda system}
\label{Sec:Cluster Bruhat} \label{Sec:Rel Toda}
\subsection{Cluster structure on double Bruhat cells} We follow~\cite{Fock:2006cluster} in this section.

\begin{figure}[h]
	\begin{center} 
		\begin{tikzpicture}
			\def\xs{1.5} 
			\def\ys{1}
			
			\begin{scope}
%				\foreach \j in {0,...,3}
%				{
%					\draw[thick, dotted] (-\xs,-\j*\ys) -- (\xs,-\j*\ys);
%				}
				
				\node[styleNodeFr] (X1) at (0,0) {$X_1$};	
				\node[styleNodeFr] (X2) at (-0.5*\xs,-\ys) {$X_2$};
				\node[styleNodeFr] (X3) at (0,-2*\ys) {$X_3$};
				\node[styleNodeFr] (X4) at (0,-3*\ys) {$X_4$};
				\node[styleNodeFr] (X5) at (0.5*\xs,-\ys) {$X$};
				
				\draw[styleArrow](X5) to   (X2) ;
				\draw[styleArrow,dashed](X2) to   (X1) ;
				\draw[styleArrow,dashed](X2) to   (X3) ;
				\draw[styleArrow,dashed](X1) to   (X5) ;
				\draw[styleArrow,dashed](X3) to   (X5) ;
				
			\end{scope}
			
			\begin{scope}[shift={(5*\xs,0)}]	
%				\foreach \j in {0,...,3}
%				{
%					\draw[thick, dotted] (-\xs,-\j*\ys) -- (\xs,-\j*\ys);
%				}
				
				\node[styleNodeFr] (X1) at (0,0) {$X_1$};	
				\node[styleNodeFr] (X2) at (-0.5*\xs,-\ys) {$X_2$};
				\node[styleNodeFr] (X3) at (0,-2*\ys) {$X_3$};
				\node[styleNodeFr] (X4) at (0,-3*\ys) {$X_4$};
				\node[styleNodeFr] (X5) at (0.5*\xs,-\ys) {$X$};
				
				\draw[styleArrow](X2) to   (X5) ;
				\draw[styleArrow,dashed](X1) to   (X2) ;
				\draw[styleArrow,dashed](X3) to   (X2) ;
				\draw[styleArrow,dashed](X5) to   (X1) ;
				\draw[styleArrow,dashed](X5) to   (X3) ;
				
			\end{scope}
		\end{tikzpicture}
	\end{center}
	\caption{\label{Fig:si quiver} Quivers \(\mathcal{Q}_{\bar 2}\) and \(\mathcal{Q}_{2}\) for \(SL_5\)}
\end{figure}
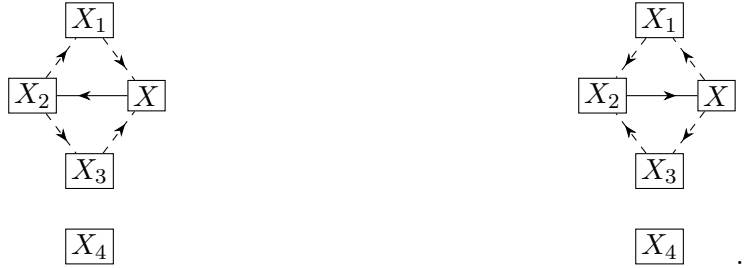

\begin{Lemma}\label{Lem:G si}
	(a) On the double Bruhat cell \(G^{\bar{s}_{i}}\) corresponding to the simple reflection \(\bar{s}_i\), in the parametrization
	\begin{equation}
		\mathbb{L}_{\bar{s}_{i}}(\mathbf{X})=H_1(X_1)\cdot\dots \cdot H_{N-1}(X_{N-1})F_i H_i(X)
	\end{equation}
	the Sklyanin Poisson structure has the form 
	\begin{equation}
		\begin{aligned}
		&\{X,X_i\}=XX_i,\;\; \{X_i,X_{i-1}\}=\frac12 X_iX_{i-1},\;\; \{X_i,X_{i+1}\}=\frac12 X_iX_{i+1},
		\\
		&\{X_{i-1},X\}=\frac12 X_{i-1}X,\;\; \{X_{i+1},X\}=\frac12 X_{i+1}X,
		\end{aligned}
	\end{equation}
	and all other brackets are zero. 
	
	(b) On the double Bruhat cell \(G^{s_i}\) corresponding to the simple reflection \(s_i\), in the parametrization
	\begin{equation}
		\mathbb{L}_{s_i}(\mathbf{X})=H_1(X_1)\cdot\dots \cdot H_{N-1}(X_{N-1})E_i H_i(X)
	\end{equation}
	the Sklyanin Poisson structure has the form 
	\begin{equation}
		\begin{aligned}
			&\{X_i,X\}=X_i X,\;\; \{X_{i-1},X_{i}\}=\frac12 X_{i-1}X_i,\;\; \{X_{i+1},X_{i}\}=\frac12 X_{i+1}X_i,
			\\
			&\{X,X_{i-1}\}=\frac12 XX_{i-1},\;\; \{X,X_{i+1}\}=\frac12 XX_{i+1},
		\end{aligned}
	\end{equation}
	and all other brackets are zero.
\end{Lemma}
For such Poisson brackets, we can assign seeds.
We denote seeds corresponding to cells \(G^{\bar{s}_{i}}\) and \(G^{s_i}\) by \(\mathsf{s}_{\bar{i}}\) and \(\mathsf{s}_{i}\), respectively.
The quivers will be denoted by \(\mathcal{Q}_{\bar{i}}\) and \(\mathcal{Q}_{i}\).
We call them elementary quivers.
Two examples are depicted in Fig.~\ref{Fig:si quiver}.
Note that for \(i=1\) or \(i=N-1\) some of the Poisson brackets above should be ignored, since
there are no \(X_0\) or \(X_N\) in the parametrization.

In order to construct a quiver for generic \(G^w\), we will use amalgamation.

\begin{Definition}
	Assume that we have two seeds \(\mathsf{s}_I=(I,I_{\mathrm{f}},\epsilon^I,\mathbf{X}^I)\) and \(\mathsf{s}_J=(J,J_{\mathrm{f}},\epsilon^J,\mathbf{X}^J)\) and two injections \(L \hookrightarrow I_{\mathrm{f}}\), \(L \hookrightarrow J_{\mathrm{f}}\).
	The \emph{amalgamation} of the seeds \(\mathsf{s}_I\) and \(\mathsf{s}_J\) along \(L\) is the seed \(\mathsf{s}_K=(K,K_{\mathrm{f}},\epsilon^K,\mathbf{X}^K)\) with \(K=I \sqcup_L J\), \(K_{\mathrm{f}}=I_{\mathrm{f}} \sqcup_L J_{\mathrm{f}}\),
	\begin{equation}\label{eq:amalgamation}
		\epsilon_{ij}^K=
		\begin{cases}
			&0,\quad \text{if $i\in I\setminus L$, $j\in J \setminus L$ or vice versa},
			\\
			&\epsilon_{ij}^I,\quad \text{if $i\in I\setminus L$, $j\in I$ or vice versa},
			\\
			&\epsilon_{ij}^J,\quad \text{if $i\in J\setminus L$, $j\in J$ or vice versa},
			\\
			&\epsilon_{ij}^I+\epsilon_{ij}^J,\quad \text{if $i,j\in L$},
		\end{cases}\qquad 
				X_i^K=
		\begin{cases}
			&X_i^I,\quad \text{if $i\in I\setminus L$},
			\\
			&X_i^J,\quad \text{if $i\in J\setminus L$},
			\\
			&X_i^IX_i^J,\quad \text{if $i\in L$}.
		\end{cases}
	\end{equation}
\end{Definition}
\begin{Remark}
	If for some \(i\in L\) we have \(\epsilon_{ij}^K \in \mathbb{Z}\), \(\forall j\) then we can unfreeze \(i\).
\end{Remark}

\begin{Proposition}[\cite{Fock:2006cluster}]
	(a) Amalgamation is a Poisson map \(\mathcal{X}_{\mathsf{s}_I}\times \mathcal{X}_{\mathsf{s}_J}\to \mathcal{X}_{\mathsf{s}_K}\).
	
	(b) Amalgamation commutes with mutation (at unfrozen vertices).
\end{Proposition}

\begin{Theorem}[\cite{Fock:2006cluster}] \label{Th:Double Bruhat cluster} 
	Let \(w \in W \times W\) with reduced expression \(w=s_{i_1}\cdot\dots \cdot s_{i_l}\).
	Consider a seed with the combinatorial data given by the amalgamation of seeds \(\mathsf{s}_{i_1}, \mathsf{s}_{i_2},\dots, \mathsf{s}_{i_l}\) and cluster variables \(\mathbf{X}\) given by factorization coordinates \eqref{eq:factorization}.
	\begin{enumerate}[label=(\alph*)]	
		\item The cluster Poisson bracket coincides with the Sklyanin bracket.
		
		\item Refactorization using relations~\eqref{eq:rel EHF} corresponds to mutations between cluster seeds or trivial transformations.
	\end{enumerate}	
\end{Theorem}

Let us describe this amalgamation explicitly.
We depict quivers \(\mathcal{Q}_{i_1}, \mathcal{Q}_{i_2}, \dots, \mathcal{Q}_{i_l}\) consecutively on the \(N-1\) parallel lines (cf.~Fig.~\ref{Fig:si quiver}).
Then we glue \(N-1\) vertices  on the right boundary of \(\mathcal{Q}_{i_j}\) with \(N-1\) vertices on the left boundary of \(\mathcal{Q}_{i_{j+1}}\) for all \(1\le j\le l-1\).
The vertices that do not belong to the right or left boundary of the resulting quiver can be unfrozen.

Such an amalgamation rule is motivated by the relations~\eqref{eq:rel EHF} between \(H_i,E_i,F_i\) used before.
For example, relation \(H_i(x)H_i(y)=H_i(xy)\) corresponds to the multiplication of variables under amalgamation in the definition~\eqref{eq:amalgamation}.

\begin{Example} 
	Let us revisit examples \ref{Ex: ss}, \ref{Ex: s123}, \ref{Ex:ex ss sss cluster}.
	In the first of these examples we have \(G=SL_2\), \(w=\bar{s}_{1}s_{1}\), and the factorization scheme
	\begin{equation}\label{eq:factorization ss}
		H_1(X_1)F_1H_1(X_2)E_1H_1(X_3).
	\end{equation}
	Then, according to the algorithm above, the seed is amalgamated from two seeds.
	We denote the variables in the factorization schemes as follows:
	\begin{equation}
		H_1(X_1)F_1H_1(X_2'),\qquad 
		H_1(X_2'')E_1H_1(X_3).
	\end{equation}
	The expression~\eqref{eq:factorization ss} can be obtained as a product of these two factors.
	Then we get the relation among the variables, \(X_2=X_2'X_2''\).
	The quivers for two elementary seeds and the amalgamated seed are drawn in Fig.~\ref{Fig:amalgamation example ss}.
	\begin{figure}[h]
		\begin{center}
			\begin{tikzpicture}
				\def\xs{1.5} 
				\def\ys{1.5}
%				\foreach \j in {0,...,1}
%				{
%					\draw[thick, dotted] (-\xs,\j*\ys) -- (6*\xs,\j*\ys);
%				}
				\begin{scope}[shift={(0*\xs,0)}]	
					
					\node[styleNodeFr] (X1) at (-0.5*\xs,0) {$X_1$};	
					\node[styleNodeFr] (X2) at (1*\xs,0) {$X_2'$};
					
					\draw[styleArrow](X2) to   (X1) ;
					
				\end{scope}
				
				\begin{scope}[shift={(2.5*\xs,0)}]	
					
					\node[styleNodeFr] (X2) at (-0.5*\xs,0) {$X_2''$};	
					\node[styleNodeFr] (X3) at (1*\xs,0) {$X_3$};
					
					\draw[styleArrow] (X2) to   (X3) ;
				
				\end{scope}

				\begin{scope}[shift={(6*\xs,0)}]	
					\node[styleNodeFr] (X1) at (0,0) {$X_1$};	
					\node[styleNode] (X2) at (1.5*\xs,0) {$X_2$};
					\node[styleNodeFr] (X3) at (3*\xs,0) {$X_3$};
					\draw[styleArrow](X2) to   (X1) ;
					\draw[styleArrow](X2) to   (X3) ;

				\end{scope}
			\end{tikzpicture}
			\caption{\label{Fig:amalgamation example ss} \(G=SL_2\), \(w=\bar{s}_{1}s_{1}\), left: two elementary quivers, right: amalgamated quiver.}
		\end{center}
	\end{figure}
	
	Now consider Example \ref{Ex: s123}.
	Here we have \(G=SL_3\), \(w=s_1s_2s_{1}\), and the factorization scheme
	\begin{equation}\label{eq:factorization s123}
		H_1(X_1)H_2(Y_1)E_1H_1(X_2)E_2H_2(Y_2)E_1H_1(X_3).
	\end{equation}
	The seed is now amalgamated from three basic ones.
	Denote the variables in the corresponding factorization schemes as
	\begin{equation}
		H_1(X_1)H_2(Y_1')E_1H_1(X_2'),\qquad 
		H_1(X_2'')H_2(Y_1'')E_2H_2(Y_2'),\qquad 
		H_1(X_2''')H_2(Y_2'')E_1H_1(X_3).
	\end{equation}
	The expression~\eqref{eq:factorization s123} can be obtained as a product of these three factors.
	Then we get relations between variables:
	\begin{equation}
		X_2=X_2'X_2''X_2''', \quad Y_1=Y_1'Y_1'',\quad Y_2=Y_2'Y_2''.
	\end{equation}
	The quivers for three elementary seeds and for the amalgamated seed are drawn in Fig.~\ref{Fig:amalgamation example s123}. 
		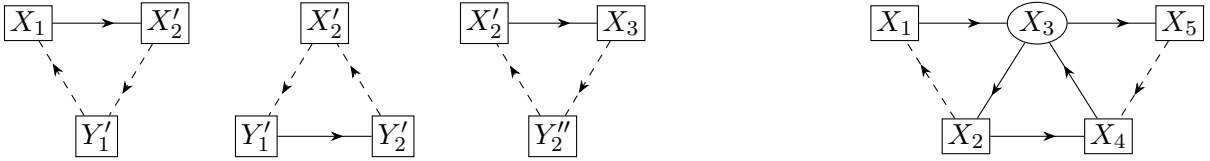
\begin{figure}[h]
		\begin{center}
			\begin{tikzpicture}
				\def\xs{1.5} 
				\def\ys{1.5}
%				\foreach \j in {0,...,1}
%				{
%					\draw[thick, dotted] (-\xs,\j*\ys) -- (6*\xs,\j*\ys);
%				}
				\begin{scope}[shift={(0*\xs,0)}]	
					
					\node[styleNodeFr] (X1) at (-0.6*\xs,0) {$X_1$};	
					\node[styleNodeFr] (Y1) at (0,-\ys) {$Y_1'$};
					\node[styleNodeFr] (X2) at (0.6*\xs,0) {$X_2'$};
					
					\draw[styleArrow](X1) to   (X2) ;
					\draw[styleArrow,dashed](X2) to   (Y1) ;
					\draw[styleArrow,dashed](Y1) to   (X1) ;
					
				\end{scope}
				
				\begin{scope}[shift={(2*\xs,0)}]	
					\node[styleNodeFr] (X1) at (0,0) {$X_2''$};	
					\node[styleNodeFr] (Y1) at (-0.6*\xs,-\ys) {$Y_1''$};
					\node[styleNodeFr] (Y2) at (0.6*\xs,-\ys) {$Y_2'$};
					
					\draw[styleArrow](Y1) to   (Y2) ;
					\draw[styleArrow,dashed](Y2) to   (X1) ;
					\draw[styleArrow,dashed](X1) to   (Y1) ;

				\end{scope}
				\begin{scope}[shift={(4*\xs,0)}]	
					
					\node[styleNodeFr] (X2) at (-0.6*\xs,0) {$X_2'''$};	
					\node[styleNodeFr] (Y2) at (0,-\ys) {$Y_2''$};
					\node[styleNodeFr] (X3) at (0.6*\xs,0) {$X_3$};
					
					\draw[styleArrow](X2) to   (X3) ;
					\draw[styleArrow,dashed](X3) to   (Y2) ;
					\draw[styleArrow,dashed](Y2) to   (X2) ;
					
				\end{scope}

				\begin{scope}[shift={(7*\xs,0)}]	
					\node[styleNodeFr] (X1) at (0,0) {$X_1$};	
					\node[styleNode] (X3) at (1.25*\xs,0) {$X_2$};
					\node[styleNodeFr] (X5) at (2.5*\xs,0) {$X_3$};
					\node[styleNodeFr] (X2) at (0.625*\xs,-\ys) {$Y_1$};
					\node[styleNodeFr] (X4) at (1.875*\xs,-\ys) {$Y_2$};
					\draw[styleArrow](X1) to   (X3) ;
					\draw[styleArrow](X3) to  (X2) ;
					\draw[styleArrow](X4) to   (X3) ;
					\draw[styleArrow](X3) to  (X5) ;
					\draw[styleArrow](X2) to  (X4) ;
					
					\draw[styleArrow,dashed](X2) to   (X1) ;
					\draw[styleArrow,dashed](X5) to   (X4) ;
					
				\end{scope}
			\end{tikzpicture}
			\caption{\label{Fig:amalgamation example s123} \(G=SL_3\), \(w=s_1s_2s_{1}\), left: three elementary quivers, right: amalgamated quiver.}
		\end{center}
	\end{figure}

\end{Example}

\begin{proof}[Idea of the proof of Theorem \ref{Th:Double Bruhat cluster}]
	(a) Induction on \(l\).
	The base is given by \(l=1\).
	This is a direct computation which constitutes the proof of Lemma \ref{Lem:G si}.
	The induction step follows from the base and the Poisson--Lie multiplication property.
	
	(b) Comparison of the formulas for mutation~\eqref{eq:mutation rule} and~\eqref{eq:rel EHF}.
	More explicitly, relations~\eqref{eq:rel EHF 4}, \eqref{eq:rel EHF 5}, \eqref{eq:rel EHF 6} correspond to mutations at vertices with variables \(X\), and relation~\eqref{eq:rel EHF 3} and the third relation in~\eqref{eq:rel EHF 1} correspond to braid transformations which do not transform the seed.
\end{proof}

%\newpage
\subsection{Relativistic Toda systems}
Now we have constructed multiplicative spaces with Poisson structures and introduced coordinates with logarithmically constant brackets.
In order to have an integrable system, it remains to construct a system of commuting Hamiltonians.

Recall the construction of the open Toda system in Example \ref{Ex:Toda}.
The phase space consists of
tridiagonal matrices \(L\).
In more invariant terms, such matrices are sums of generators of the Cartan
subalgebra and generators corresponding to simple roots (positive and negative).
Now we are going to consider
the multiplicative analog of this.
We follow \cite{Fock:1997note}, \cite{Fock:2016}.

\begin{Definition}\label{Def:Coxeter}
	An element \(c \in W\) is called a Coxeter element if it is a product of all simple reflections taken once.
\end{Definition}

It can be proven that all Coxeter elements are conjugated.
In \(G = SL_N\) (which is our running example),
we have \(W = S_N\) and \(c\) is a cycle of length \(N\). 

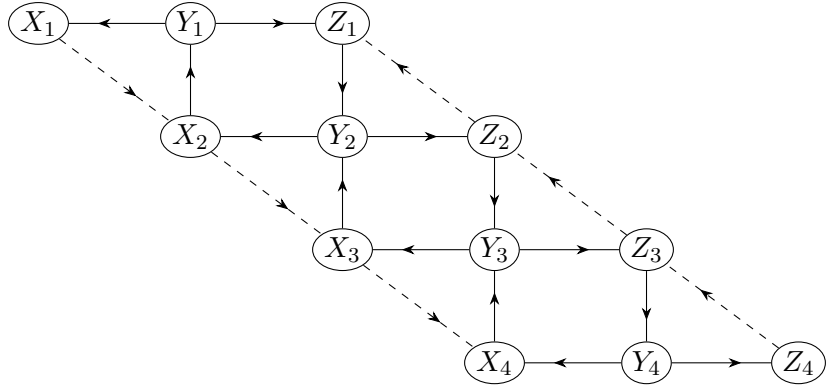
\begin{figure}[h]
	\begin{center}
		\begin{tikzpicture}
			\def\xs{2} 		
			\def\ys{1.5}	
			\def\NN{4} 
			\pgfmathtruncatemacro{\Nplusone}{\NN+1}
			
%			\foreach \j in {1,...,\NN}
%			{
%				\draw[ultra thin] (0.5*\xs,-\j*\ys) -- (4*\xs,-\j*\ys);
%			}

			\foreach \j in {1,...,\NN}
			{
				\pgfmathtruncatemacro{\jminusone}{\j-1}	
				\pgfmathtruncatemacro{\jplusone}{\j+1}	
				\node[styleNode] (X\j) at(\jminusone*\xs,-\j*\ys){$X_{\j}$};
				\node[styleNode] (Y\j) at(\j*\xs,-\j*\ys){$Y_{\j}$};
				\node[styleNode] (Z\j) at(\jplusone*\xs,-\j*\ys){$Z_{\j}$};
				
			}

			\foreach \j in {2,...,\NN}
			{
				\pgfmathtruncatemacro{\jminusone}{\j-1}	
				\draw[styleArrow](Y\j) to (X\j);
				\draw[styleArrow](Y\j) to (Z\j);
				
				\draw[styleArrow](X\j) to (Y\jminusone);
				\draw[styleArrow, dashed](X\jminusone) to (X\j);
				
				\draw[styleArrow](Z\jminusone) to (Y\j);
				
				\draw[styleArrow, dashed](Z\j) to (Z\jminusone);
			}
			\draw[styleArrow](Y1) to (X1);
			\draw[styleArrow](Y1) to (Z1);

		\end{tikzpicture}
		\caption{\label{Fig:Gcc}Quiver for the double Bruhat cell of \(SL_5\).}
	\end{center}
\end{figure}

The double Bruhat cell \(G^{\bar{c},c}\) is called a \emph{Coxeter cell}.
Let us consider
\(w=\bar{s}_{1}s_1\bar{s}_{2}s_2\cdot \dots \bar{s}_{N-1}s_{N-1}\).
We have the following factorization scheme
\begin{multline}\label{eq:factorization Coxeter}
	\mathbb{L}(\mathbf{X},\mathbf{Y},\mathbf{Z})=H_1(X_1)\cdot\ldots \cdot H_{N-1}(X_{N-1})\cdot \big(F_1 H_1(Y_1) E_1 H_1(Z_1)\big) \\ \cdot \big( F_2 H_2(Y_2) E_2 H_2(Z_2)\big)\cdot \ldots \cdot \big(F_{N-1} H_{N-1}(Y_{N-1}) E_{N-1} H_{N-1}(Z_{N-1})\big).
\end{multline}
The corresponding quiver is obtained by the consecutive amalgamation of seeds \(\mathsf{s}_{\bar{1}}, \mathsf{s}_{1}, \dots, \mathsf{s}_{\overline{N-1}}, \mathsf{s}_{N-1}\).
It is depicted in Fig.~\ref{Fig:Gcc}.

The system of commuting Hamiltonians is constructed similarly to the additive case.

\begin{Proposition}
	Let \(L=\mathbb{L}(\mathbf{X},\mathbf{Y},\mathbf{Z})\) and \(H_k=\operatorname{Tr}L^k\).
	Then \(\{H_k,H_m\}=0\).
\end{Proposition}
\begin{proof}
We use the same notations as above, namely \(L_1=L\otimes 1\), \(L_2=1\otimes L\), and \(\operatorname{Tr}_{12}\) is 
the trace of an operator on \(\mathbb{C}^N\otimes \mathbb{C}^N\).
We have
	\begin{multline}
		\{H_k,H_m\}=\{\operatorname{Tr}L^k,\operatorname{Tr}L^m\}=\operatorname{Tr}_{12}\{L_1^k,L_2^m\}
		=\operatorname{Tr}_{12}\sum_{i=1}^k\sum_{j=1}^m L_1^{i-1}L_2^{j-1} \{L_1,L_2\} L_1^{k-i}L_2^{m-j}
		\\
		=\operatorname{Tr}_{12}\sum_{i=1}^k\sum_{j=1}^m L_1^{i-1}L_2^{j-1} [r,L_1L_2] L_1^{k-i}L_2^{m-j} = \operatorname{Tr}_{12}[r,L_1^kL_2^m] =0.
	\end{multline}
\end{proof}

Since \(L\in SL_N\), we have \(N-1\) algebraically independent Hamiltonians \(H_1,\dots, H_{N-1}\).
On the other hand, the Hamiltonians are invariant under conjugation, hence we can move all \(\mathbf{Z}\) in formula~\eqref{eq:factorization Coxeter} to the left.
In other words \(\mathbb{L}(\mathbf{X},\mathbf{Y},\mathbf{Z})\) is conjugated to
\begin{equation}
	H_1(X_1Z_1)\cdot\ldots \cdot H_{N-1}(X_{N-1}Z_{N-1})\cdot \big(F_1 H_1(Y_1) E_1\big) \cdot \big( F_2 H_2(Y_2) E_2 \big)\cdot \ldots \cdot \big( F_{N-1} H_{N-1}(Y_{N-1}) E_{N-1} \big)
\end{equation}
Hence the Hamiltonians depend only on the product \(X_iZ_i\), not on \(X_i\) and \(Z_i\) themselves.
One more way to say it is that the Hamiltonians descend to the quotient \(G^{\bar{c},c}/\operatorname{Ad}H\).
We claim that this quotient is the phase space of the integrable system.

For this, we need to define a natural Poisson structure on the quotient. 
Note that the \(r\)-matrix~\eqref{eq:r matrix} is invariant under conjugation by \(H\).
Hence, the Poisson bivector defined by the Sklyanin bracket~\eqref{eq:PB r mult} vanishes on \(H\).
Therefore, left and right translations by elements of \(H\) preserve the Poisson structure on \(G\) due to the Poisson--Lie property.
In particular, the Poisson structure on \(G^{\bar{c},c}\) is invariant under the adjoint action of \(H\), and we get a natural Poisson structure on the quotient.

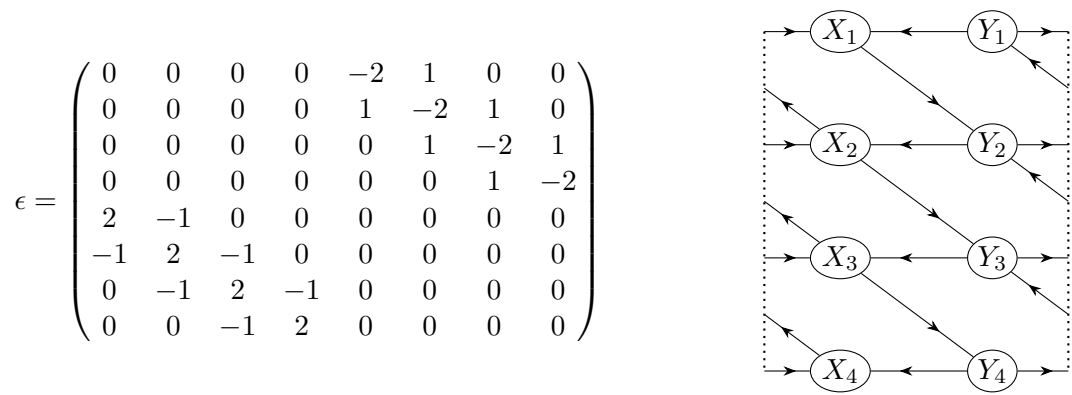
\begin{figure}[h]
	\begin{center}
		\begin{tikzpicture}
			\def\xs{2} 		
			\def\ys{1.5}	
			\def\NN{4} 
			\pgfmathtruncatemacro{\Nplusone}{\NN+1}
			
			\node at ($(-\Nplusone*\xs/2,-\Nplusone*\ys/2)$) {	\(\epsilon=\begin{pmatrix} 
					0 & 0 & 0 & 0 & -2 & 1 &0 &0
					\\
					0 & 0 & 0 & 0 & 1 & -2 &1 &0
					\\
					0 & 0 & 0 & 0 & 0 & 1 & -2 &1
					\\
					0 & 0 & 0 & 0 & 0 & 0 &1 &-2
					\\
					2 & -1 &0 &0 & 0 & 0 & 0 & 0
					\\
					-1 & 2 &-1 &0 & 0 & 0 & 0 & 0
					\\
					0 & -1 & 2 &-1 & 0 & 0 & 0 & 0
					\\
					0 & 0 &-1 &2 & 0 & 0 & 0 & 0					
				\end{pmatrix}\)	};
			
			%vertices
			\foreach \j in {1,...,\NN}
			{
				\node[styleNode] (X\j) at(\xs,-\j*\ys){$X_{\j}$};
				\node[styleNode] (Y\j) at(2*\xs,-\j*\ys){$Y_{\j}$};
			}
			
			%arrows
			\foreach \j in {2,...,\NN}
			{
				\pgfmathtruncatemacro{\jminusone}{\j-1}				\draw[styleArrow](Y\j) to (X\j);
				\draw[styleArrow](X\jminusone) to (Y\j);
				\draw[styleArrow] ($(X\j)-(0.5*\xs,0)$) to (X\j);
				\draw[styleArrow] (Y\j) to ($(Y\j)+(0.5*\xs,0)$) ;
				\draw[styleArrow] ($(Y\jminusone)+(0.5*\xs,-0.5*\ys)$) to (Y\jminusone);
				\draw[styleArrow] (X\j) to ($(X\j)+(-0.5*\xs,0.5*\ys)$);
				
			}
			%arrows near boundary
			\draw[styleArrow](Y1) to (X1);
			\draw[styleArrow] ($(X1)-(0.5*\xs,0)$) to (X1);
			\draw[styleArrow] (Y1) to ($(Y1)+(0.5*\xs,0)$);
			
			%boundary of cylinder
			\draw[thick, dotted] (0.5*\xs,-\ys) -- (0.5*\xs,-\NN*\ys)  ;
			\draw[thick, dotted] (2.5*\xs,-\ys) -- (2.5*\xs,-\NN*\ys)  ;
		\end{tikzpicture}
		\caption{\label{Fig:Toda open} On the left is the matrix \(\epsilon\), and on the right is the quiver of the open relativistic Toda system for \(N=5\). The quiver is drawn on a cylinder.}
	\end{center}
\end{figure}

The local coordinates on the quotient are \(Y_1,\dots, Y_{N-1}\) and \(X_1Z_1,\dots, X_{N-1}Z_{N-1}\).
In cluster terms, it means that we take an amalgamation of the left and right boundaries of the quiver in Fig.~\ref{Fig:Gcc}.
After this operation, we can unfreeze the obtained vertices.
The corresponding quiver and adjacency matrix are depicted in Fig.~\ref{Fig:Toda open}.

In more invariant terms, the adjacency matrix has the form \(\epsilon=\begin{pmatrix}
	0 & -C \\ C &0 
\end{pmatrix}\), where \(C\) is the Cartan matrix of the \(A_{N-1}\) root system. 
In particular, we see that this matrix is non-degenerate, so the Poisson bracket on \(G^{\bar{c},c}/\operatorname{Ad}H\) is non-degenerate. 
It can be proved that Hamiltonians \(H_1,\dots, H_{N-1}\) are algebraically independent on this space. 
Moreover, they can be identified (see \cite{Fock:1997note}, \cite{Marshakov:2013}) with the Hamiltonians of the open \(SL_N\) relativistic Toda system introduced by Ruijsenaars \cite{Ruijsenaars:1990Toda}.
\begin{multline}
	H=\operatorname{Tr}(\mathbb{L}+\mathbb{L}^{-1})
	=2\Big(\cosh p_1 \sqrt{1+\re^{q_{1,2}}}
	\\
	\,+ \sum_{i=2}^{N-1} \cosh p_i  \sqrt{1+\re^{q_{i,i+1}}}\sqrt{1+\re^{q_{i-1,i}}}
	\,+\cosh p_N \sqrt{1+\re^{q_{N-1,N}}}\Big)
\end{multline}
where \(q_i\) and \(p_i\) are Darboux coordinates, \(\cosh(p)=(\re^p+\re^{-p})/2\), \(q_{i,j}=q_i-q_j\).

\begin{Remark}\label{Rem:4d Coulomb}
	It was observed recently that the phase space of the relativistic open Toda system appears as a Coulomb branch
    for the pure \(SL(N)\) 4d (on \(\mathbb{R}^3\times S^1\)) supersymmetric gauge theory \cite{Braverman:2019coulomb}, \cite{Finkelberg:2019multiplicative}.
    Moreover, the Coulomb branches of any 4d quiver gauge theory can be, in some sense, constructed from open Toda systems \cite{Schrader:2019k}.
\end{Remark}

Note that most of the constructions above can be stated in more invariant terms, namely, in terms of root data.
For example, in the formula for the Toda Hamiltonian~\eqref{eq:Toda Ham}, one can easily see a sum over the simple roots of \(A_{N-1}\).
In the formula for the \(r\)-matrix~\eqref{eq:r matrix} we see a sum over all positive roots of \(A_{N-1}\).
The definitions of a Coxeter element and a Coxeter double Bruhat cell make sense for any root system.
And finally, the adjacency matrix for the quiver of the open relativistic Toda system \(\epsilon=\begin{pmatrix}
	0 & -C \\ C &0
\end{pmatrix}\) is well defined for any simply-laced Lie algebra.

\section{Plabic graphs}\label{Sec:plabic}

One can reformulate the constructions above using another combinatorial tool: \emph{plabic graphs}.
The word plabic is an abbreviation for ``planar bicolored''.
It was introduced by Postnikov~\cite{Postnikov2006total}.
\begin{Definition}
	A \emph{plabic} graph \(\Gamma\) is a graph drawn on an oriented surface \(\Sigma\) (possibly with boundary) such that its vertices are colored black and white and the connected components of \(\Sigma \setminus \Gamma\) are contractible.
\end{Definition}
Connected components of \(\Sigma \setminus \Gamma\) are called faces of \(\Gamma\).
For a given plabic graph \(\Gamma\), one can define the dual quiver \(\mathcal{Q}\).

\begin{Definition}\label{Def:quiver from plabic}
	The vertices of \(\mathcal{Q}\) correspond to the faces of \(\Gamma\).
	Arrows of \(\mathcal{Q}\) correspond to the edges of \(\Gamma\) that connect vertices of different colors.
	Namely, if such an edge \(e\) separates faces \(f_1\) and \(f_2\), then the corresponding arrow \(a\) of \(\mathcal{Q}\) connects \(f_1\) and \(f_2\), and is oriented in such a way that the intersection of \(e\) and \(a\) is positive, where \(e\) is oriented from black to white.
\end{Definition}
Equivalently, one can say that the orientation of the quiver \(\mathcal{Q}\) is chosen such that edges go clockwise around black vertices of \(\Gamma\) and counterclockwise around white vertices of \(\Gamma\).

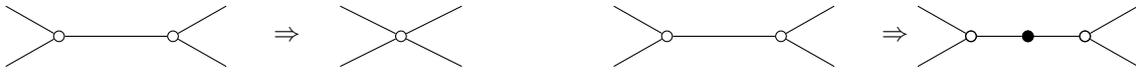
\begin{figure}[h]
	\centering
	\begin{tikzpicture}[font= \small]
		\def\xs{1}
		\def\ys{0.8}
		
		\begin{scope}[shift={(0,0)}]
			
			\node[circle, draw, inner sep=0pt, minimum size=4pt] (left) at (-\xs,0) {};
			
			\node[circle, draw, inner sep=0pt, minimum size=4pt] (right) at (0.5*\xs,0) {};

			\draw[] (left) -- (right);
			
			\draw[] (left) -- ++ ($(-0.7*\xs,0.5*\ys)$);
			\draw[] (left) -- ++ ($(-0.7*\xs,-0.5*\ys)$);
			
			\draw[] (right) -- ++ ($(0.7*\xs,0.5*\ys)$);
			\draw[] (right) -- ++ ($(0.7*\xs,-0.5*\ys)$);
			
		\end{scope}
		\node at (1.5*\xs+0.5,0) {$\Leftrightarrow$};
		
		%	after spider
		\begin{scope}[shift={(3.5*\xs,0)}]	
			\node[circle, draw, inner sep=0pt, minimum size=4pt] (center) at (0,0) {};
			
			% Draw lines starting from the edge of the circle
			\draw (center) -- ++(-0.8*\xs,0.5*\ys);
			\draw (center) -- ++(-0.8*\xs,-0.5*\ys);
			\draw (center) -- ++(0.8*\xs,0.5*\ys);
			\draw (center) -- ++(0.8*\xs,-0.5*\ys);

		\end{scope}	
		
		\begin{scope}[shift={(8*\xs,0)}]

			\node[circle, draw, inner sep=0pt, minimum size=4pt] (left) at (-\xs,0) {};
			
			\node[circle, draw, inner sep=0pt, minimum size=4pt] (right) at (0.5*\xs,0) {};

			\draw[] (left) -- (right);
			
			\draw[] (left) -- ++ ($(-0.7*\xs,0.5*\ys)$);
			\draw[] (left) -- ++ ($(-0.7*\xs,-0.5*\ys)$);
			
			\draw[] (right) -- ++ ($(0.7*\xs,0.5*\ys)$);
			\draw[] (right) -- ++ ($(0.7*\xs,-0.5*\ys)$);
		\end{scope}
		\node at (9.5*\xs+0.5,0) {$\Leftrightarrow$};
		
		%	after spider
		\begin{scope}[shift={(12*\xs,0)}]	
			
			\node[circle, draw, inner sep=0pt, minimum size=4pt] (left) at (-\xs,0) {};
			
			\node[circle, draw, inner sep=0pt, minimum size=4pt] (right) at (0.5*\xs,0) {};		
			
			\node[circle, draw, fill=black, inner sep=0pt, minimum size=4pt] (center) at (-0.25*\xs,0) {};		
			
			\draw[] (left) to (center) to (right);
			
			\draw[] (left) -- ++ ($(-0.7*\xs,0.5*\ys)$);
			\draw[] (left) -- ++ ($(-0.7*\xs,-0.5*\ys)$);
			
			\draw[] (right) -- ++ ($(0.7*\xs,0.5*\ys)$);
			\draw[] (right) -- ++ ($(0.7*\xs,-0.5*\ys)$);

			\draw ($(0.5*\xs,0)$) circle  (2pt);
			\draw ($(-\xs,0)$) circle  (2pt);
			\draw[fill] ($(-0.25*\xs,0)$) circle  (2pt);

		\end{scope}	
	\end{tikzpicture}	
	\caption{\label{Fig:contr 2 alent}On the left, edge contraction; on the right, insertion of a 2-valent vertex}
\end{figure}

The plabic graph \(\Gamma\) is not assumed to be bipartite, but one can make it bipartite using either of the transformations given in Fig.~\ref{Fig:contr 2 alent} (contraction/uncontraction of an edge and insertion/removal of a 2-valent vertex).
It is easy to see that contraction does not change the quiver, while insertion of a 2-valent vertex adds a cycle of length 2, preserving the adjacency matrix of the quiver \(\mathcal{Q}\).

For non-compact \(\Sigma\) we will also allow \emph{semi-infinite} edges of \(\Gamma\), i.e., edges with one endpoint at a vertex of \(\Gamma\) and the other at infinity.
Vertices that correspond to unbounded faces will be frozen.
The arrows corresponding to the semi-infinite edges will be dashed, with the same rule for orientation.
See an example in Fig.~\ref{Fig:examples plabic} below.

Now for any \(w \in W \times W\) with reduced expression \(w=s_{i_1}\cdot\ldots \cdot s_{i_l}\) we construct a plabic graph \(\Gamma_{\mathbf{i}}\), where \(\mathbf{i}=(i_1,\dots,i_l)\).
Actually, the construction is more general; it also defines plabic graphs (and their quivers via Definition \ref{Def:quiver from plabic}) for non-reduced decompositions.
Such graphs are sometimes called \emph{plabic fences}.

\begin{Definition}\label{Def:Gamma i}
	Let \(\mathbf{i}=(i_1,\dots,i_l)\) be a word in the alphabet \(1,\dots,N-1,\bar{1},\dots, \overline{N-1}\).
	The graph \(\Gamma_{\mathbf{i}}\) consists of infinite horizontal lines and finite vertical segments.
	The \(N\) horizontal lines are given by equations \(y=-i\), \(i=1,\dots,N\).
	The vertical segments are in correspondence with the letters \(i_j\) in the word \(\mathbf{i}\).
	The segment corresponding to \(i_j\) goes from line \(y=-|i_j|\) to \(y=-|i_j|-1\) and has a white vertex above and a black vertex below if \(i_j \in \{1,\dots,N-1\}\) and a black vertex above and a white vertex below if \(i_j \in \{\bar{1},\dots,\overline{N-1}\}\).
	The order of the vertical segments agrees with the order of factors in the word \(\mathbf{i}\).
\end{Definition}

Examples of the plabic graphs \(\Gamma_{\mathbf{i}}\) and corresponding quivers are given in Fig.~\ref{Fig:examples plabic}.
The quivers there are depicted in blue in order to distinguish them from the plabic graphs.
\begin{figure}[h]
	\centering
	\begin{tikzpicture}
		\def\xs{1.5}
		\def\ys{1.5}
		
		\begin{scope}[shift={(0,0)}]
			
			\node[circle, draw, inner sep=0pt, minimum size=4pt] (d1) at (-\xs,-2*\ys) {};
			
			\node[circle, draw, fill=black,  inner sep=0pt, minimum size=4pt] (u1) at (-\xs,-\ys) {};		
			
			\node[circle, draw, fill=black, inner sep=0pt, minimum size=4pt] (d2) at (0*\xs,-2*\ys) {};
			
			\node[circle, draw,   inner sep=0pt, minimum size=4pt] (u2) at (0*\xs,-\ys) {};

			\draw[] (d1) -- (u1);
			\draw[] (d2) -- (u2);	
			
			%			sources and targets			
			
			\node (tau2) at (-2*\xs,-2*\ys) {};			
			\node (tau1) at (-2*\xs,-\ys) {};			
			\node (sigma2) at (1*\xs,-2*\ys) {};	
			\node (sigma1) at (1*\xs,-\ys) {};
			
			\draw (sigma1) to (u2) to (u1) to (tau1);
			\draw (sigma2) to (d2) to (d1) to (tau2);
			
			\node[styleNodeFr,blue] (x1) at (-1.5*\xs,-1.5*\ys) {$X_1$};
			\node[styleNode, blue] (x2) at (-0.5*\xs,-1.5*\ys) {$X_2$};
			\node[styleNodeFr,blue] (x3) at (0.5*\xs,-1.5*\ys) {$X_3$};
			
			\draw[special arrow=0.8, blue] (x2) to (x1); 
			\draw[special arrow=0.8, blue] (x2) to (x3);

		\end{scope}
		\begin{scope}[shift={(5*\xs,0.5*\ys)}]
			
			\node[circle, draw, fill=black, inner sep=0pt, minimum size=4pt] (d1) at (-\xs,-2*\ys) {};
			
			\node[circle, draw, inner sep=0pt, minimum size=4pt] (u1) at (-\xs,-\ys) {};		
			
			\node[circle, draw, fill=black, inner sep=0pt, minimum size=4pt] (d2) at (0*\xs,-3*\ys) {};
			
			\node[circle, draw,   inner sep=0pt, minimum size=4pt] (u2) at (0*\xs,-2*\ys) {};
			
			\node[circle, draw, fill=black, inner sep=0pt, minimum size=4pt] (d3) at (1*\xs,-2*\ys) {};
			
			\node[circle, draw,   inner sep=0pt, minimum size=4pt] (u3) at (1*\xs,-1*\ys) {};

			\draw[] (d1) -- (u1);
			\draw[] (d2) -- (u2);	
			\draw[] (d3) -- (u3);	
			%			sources and targets			
			\node (tau3) at (-2*\xs,-3*\ys) {};	
			\node (tau2) at (-2*\xs,-2*\ys) {};			
			\node (tau1) at (-2*\xs,-\ys) {};			
			\node (sigma3) at (2*\xs,-3*\ys) {};	
			\node (sigma2) at (2*\xs,-2*\ys) {};	
			\node (sigma1) at (2*\xs,-\ys) {};
			
			\draw (sigma1) to (u3) to (u1) to (tau1);
			\draw (sigma2) to (d3) to (u2) to (d1) to (tau2);
			\draw (sigma3) to (d2) to (tau3);

			\node[styleNodeFr,blue] (x1) at (-1.5*\xs,-1.5*\ys) {$X_1$};
			\node[styleNodeFr,blue] (x2) at (-0.8*\xs,-2.5*\ys) {$X_2$};
			\node[styleNode,blue] (x3) at (0*\xs,-1.5*\ys) {$X_3$};
			\node[styleNodeFr,blue] (x4) at (0.8*\xs,-2.5*\ys) {$X_4$};
			\node[styleNodeFr,blue] (x5) at (1.5*\xs,-1.5*\ys) {$X_5$};
			
			\draw[special arrow=0.8, blue] (x1) to (x3); 
			\draw[special arrow=0.8, blue] (x3) to (x2); 
			\draw[special arrow=0.8, blue] (x3) to (x5); 
			\draw[special arrow=0.8, blue] (x4) to (x3); 
			\draw[special arrow=0.8, blue] (x2) to (x4); 
			\draw[special arrow=0.8, blue, dashed] (x2) to[bend left=20] (x1); 
			\draw[special arrow=0.8, blue,dashed] (x5) to[bend left=20] (x4); 
			
		\end{scope}
		
	\end{tikzpicture}		
	\caption{\label{Fig:examples plabic} Plabic graphs and quivers, left: \(G=SL_2\), \(w=\overline{s}_1s_1\), right: \(G=SL_3\), \(w=s_1s_2s_1\)}
\end{figure}
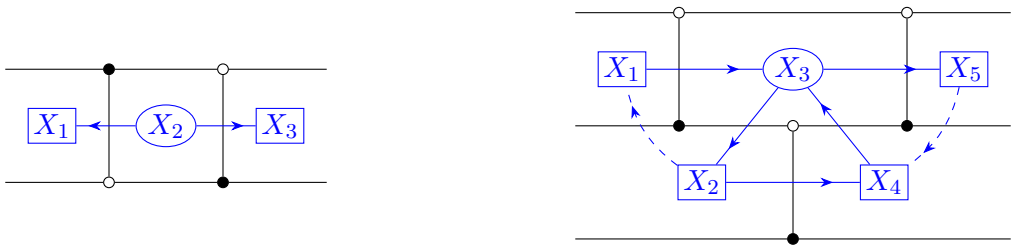

It is straightforward to see that quivers corresponding to such networks agree with the ones given by amalgamation of seeds \(\mathsf{s}_{i_1}, \mathsf{s}_{i_2},\dots, \mathsf{s}_{i_l}\).
In particular, the quivers given in Fig.~\ref{Fig:examples plabic} coincide with the ones in Fig.~\ref{Fig:amalgamation example ss} and  Fig.~\ref{Fig:amalgamation example s123}.

The following lemma is an analog of Theorem~\ref{Th:Double Bruhat cluster}(b) in terms of plabic graphs.

\begin{figure}[h]
	\centering
	\begin{tikzpicture}[font= \small]
		\def\xs{1.5}
		\def\ys{1.0}
		
		%	before spider
		\begin{scope}[shift={(0,0)}]
			
			\node[circle, draw, inner sep=0pt, minimum size=4pt] (w1) at (-\xs,-\ys) {};
			\node[circle, draw,  inner sep=0pt, minimum size=4pt] (w2) at (\xs,\ys) {};

			\node[circle, draw, fill=black, inner sep=0pt, minimum size=4pt] (b1) at (\xs,-\ys) {};
			\node[circle, draw, fill=black, inner sep=0pt, minimum size=4pt] (b2) at (-\xs,\ys) {};

%			\draw ($(-\xs,-\ys)$) circle (2pt);
%			\draw ($(\xs,\ys)$) circle (2pt);
%			\draw[fill] ($(\xs,-\ys)$) circle  (2pt);
%			\draw[fill] ($(-\xs,\ys)$) circle  (2pt);

			% Draw lines with different colors
			\draw[] (w1) -- (b1) -- (w2) -- (b2) -- (w1);
			
			\draw[] (w1) -- ++ (-0.5*\xs,-0.5*\ys);
			\draw[] (b1) -- ++ (0.5*\xs,-0.5*\ys);
			\draw[] (b2) -- ++ (-0.5*\xs,0.5*\ys);
			\draw[] (w2) -- ++ (0.5*\xs,0.5*\ys);

			\node[blue] at (0,0) {$X$};
			\node[blue] at (1.5*\xs,0) {$X_3$};
			\node[blue] at (-1.5*\xs,0) {$X_1$};
			\node[blue] at (0,1.5*\ys) {$X_2$};
			\node[blue] at (0,-1.5*\ys) {$X_4$};
			
		\end{scope}
		
		\node at (1.5*\xs+2,0) {$\Rightarrow$};
		
		%	after spider
		\begin{scope}[shift={(3*\xs+4,0)}]
			
			\node[circle, draw, inner sep=0pt, minimum size=4pt] (w1) at (-\xs,-\ys) {};
			\node[circle, draw,  inner sep=0pt, minimum size=4pt] (w2) at (\xs,\ys) {};
			
			\node[circle, draw, fill=black, inner sep=0pt, minimum size=4pt] (b1) at (\xs,-\ys) {};
			\node[circle, draw, fill=black, inner sep=0pt, minimum size=4pt] (b2) at (-\xs,\ys) {};
			
			\node[circle, draw, inner sep=0pt, minimum size=4pt] (w3) at (-0.5*\xs,0.5*\ys) {};
			\node[circle, draw,  inner sep=0pt, minimum size=4pt] (w4) at (0.5*\xs,-0.5*\ys) {};
			
			\node[circle, draw, fill=black, inner sep=0pt, minimum size=4pt] (b3) at (-0.5*\xs,-0.5*\ys) {};
			\node[circle, draw, fill=black, inner sep=0pt, minimum size=4pt] (b4) at (0.5*\xs,0.5*\ys) {};

			\draw[] (w3) -- (b3) -- (w4) -- (b4) -- (w3);

			\draw[] (b3) -- (w1) -- ++ (-0.5*\xs,-0.5*\ys);
			\draw[] (w4) -- (b1) -- ++ (0.5*\xs,-0.5*\ys);
			\draw[] (w3) -- (b2) -- ++ (-0.5*\xs,0.5*\ys);
			\draw[] (b4) -- (w2) -- ++ (0.5*\xs,0.5*\ys);

			\node[blue] at (0,0) {$X^{-1}$};
			\node[blue] at (1.5*\xs,0) {$X_3(1+X^{-1})^{-1}$};
			\node[blue] at (-1.5*\xs,0) {$X_1(1+X^{-1})^{-1}$};
			\node[blue] at (0,1.5*\ys) {$X_2(1+X)$};
			\node[blue] at (0,-1.5*\ys) {$X_4(1+X)$};
			
		\end{scope}
	\end{tikzpicture}
	\caption{\label{fi:face4}
		4-gon face mutation (spider move)
	}
\end{figure}
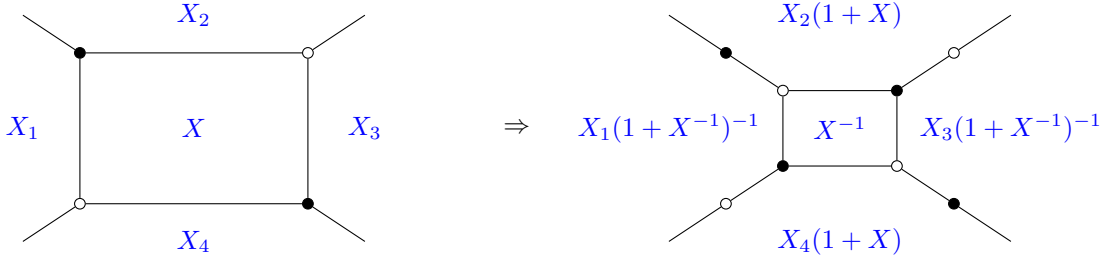

\begin{Lemma}\label{Lemma:braid plabic}
	(a) Transformations of reduced decompositions given by~\eqref{eq:braid 1} and~\eqref{eq:braid 3} for \(i\neq j\) correspond to an isotopy of a plabic graph.
	
	(b) Transformations of reduced decompositions given by~\eqref{eq:braid 2} and~\eqref{eq:braid 3} for \(i=j\) correspond to the transformation of a plabic graph given in Fig.~\ref{fi:face4} (up to contractions of edges and removal of 2-valent vertices).
\end{Lemma}
Such transformations are assigned to 4-gon faces and are called 4-gon mutations or spider moves.
Recall that the variables are assigned to the vertices of the quiver, so in the plabic graph description, the variables are assigned to the \emph{faces}.
The transformation of variables for a spider move is also shown in Fig.~\ref{fi:face4}, which is a particular case of the formulas~\eqref{eq:mutation rule}.
Note that not every quiver mutation corresponds to a spider move, since quivers can have vertices of valency greater than 4.

\begin{figure}[h]
	\centering
	\begin{tikzpicture}
		\def\xs{1.25}
		\def\ys{1}
		
		\begin{scope}[shift={(0,0)}]
			
			\node[circle, draw, inner sep=0pt, minimum size=4pt] (d1) at (-\xs,-3*\ys) {};
			
			\node[circle, draw, fill=black,  inner sep=0pt, minimum size=4pt] (u1) at (-\xs,-2*\ys) {};		
			
			\node[circle, draw, fill=black, inner sep=0pt, minimum size=4pt] (d2) at (0*\xs,-3*\ys) {};
			
			\node[circle, draw,   inner sep=0pt, minimum size=4pt] (u2) at (0*\xs,-2*\ys) {};

			\draw[] (d1) -- (u1);
			\draw[] (d2) -- (u2);	
			
			%			sources and targets			
			
			\node (tau4) at (-2*\xs,-4*\ys) {};
			\node (tau3) at (-2*\xs,-3*\ys) {};			
			\node (tau2) at (-2*\xs,-2*\ys) {};
			\node (tau1) at (-2*\xs,-1*\ys) {};		
			\node (sigma4) at (1*\xs,-4*\ys) {};	
			\node (sigma3) at (1*\xs,-3*\ys) {};	
			\node (sigma2) at (1*\xs,-2*\ys) {};
			\node (sigma1) at (1*\xs,-\ys) {};
			
			\draw (sigma1) to  (tau1);
			\draw (sigma2) to (u2) to (u1) to (tau2);
			\draw (sigma3) to (d2) to (d1) to (tau3);
			\draw (sigma4) to  (tau4);

		\end{scope}
		\begin{scope}[shift={(5*\xs,0)}]
			
			\node[circle, draw, inner sep=0pt, minimum size=4pt] (d1) at (-\xs,-3*\ys) {};
			
			\node[circle, draw, fill=black,  inner sep=0pt, minimum size=4pt] (u1) at (-\xs,-2*\ys) {};		
			
			\node[circle, draw, fill=black, inner sep=0pt, minimum size=4pt] (d2) at (0*\xs,-3*\ys) {};
			
			\node[circle, draw,   inner sep=0pt, minimum size=4pt] (u2) at (0*\xs,-2*\ys) {};
			
			\node[circle, draw, fill=black, inner sep=0pt, minimum size=4pt] (d3) at (-0.7*\xs,-2.8*\ys) {};
			
			\node[circle, draw,   inner sep=0pt, minimum size=4pt] (u3) at (-0.7*\xs,-2.2*\ys) {};		
			
			\node[circle, draw,  inner sep=0pt, minimum size=4pt] (d4) at (-0.3*\xs,-2.8*\ys) {};
			
			\node[circle, draw, fill=black,  inner sep=0pt, minimum size=4pt] (u4) at (-0.3*\xs,-2.2*\ys) {};

			\draw[] (d4) -- (u4);
			\draw[] (d3) -- (u3);	
			
			%			sources and targets			
			
			\node (tau4) at (-2*\xs,-4*\ys) {};
			\node (tau3) at (-2*\xs,-3*\ys) {};			
			\node (tau2) at (-2*\xs,-2*\ys) {};
			\node (tau1) at (-2*\xs,-1*\ys) {};		
			\node (sigma4) at (1*\xs,-4*\ys) {};	
			\node (sigma3) at (1*\xs,-3*\ys) {};	
			\node (sigma2) at (1*\xs,-2*\ys) {};
			\node (sigma1) at (1*\xs,-\ys) {};
			
			\draw (sigma1) to  (tau1);
			\draw (sigma2) to (u2) to (u4) to (u3) to (u1) to (tau2);
			\draw (sigma3) to (d2) to (d4) to (d3) to (d1) to (tau3);
			\draw (sigma4) to  (tau4);				
			
		\end{scope}

		\begin{scope}[shift={(10*\xs,0)}]
			
			\node[circle, draw, fill=black,  inner sep=0pt, minimum size=4pt] (d1) at (-\xs,-3*\ys) {};
			
			\node[circle, draw,  inner sep=0pt, minimum size=4pt] (u1) at (-\xs,-2*\ys) {};		
			
			\node[circle, draw,  inner sep=0pt, minimum size=4pt] (d2) at (0*\xs,-3*\ys) {};
			
			\node[circle, draw, fill=black,  inner sep=0pt, minimum size=4pt] (u2) at (0*\xs,-2*\ys) {};

			\draw[] (d1) -- (u1);
			\draw[] (d2) -- (u2);	
			
			%			sources and targets			
			
			\node (tau4) at (-2*\xs,-4*\ys) {};
			\node (tau3) at (-2*\xs,-3*\ys) {};			
			\node (tau2) at (-2*\xs,-2*\ys) {};
			\node (tau1) at (-2*\xs,-1*\ys) {};		
			\node (sigma4) at (1*\xs,-4*\ys) {};	
			\node (sigma3) at (1*\xs,-3*\ys) {};	
			\node (sigma2) at (1*\xs,-2*\ys) {};
			\node (sigma1) at (1*\xs,-\ys) {};
			
			\draw (sigma1) to  (tau1);
			\draw (sigma2) to (u2) to (u1) to (tau2);
			\draw (sigma3) to (d2) to (d1) to (tau3);
			\draw (sigma4) to  (tau4);

		\end{scope}

	\end{tikzpicture}		
	\caption{\label{Fig:example braid plabic} Transformation of plabic graphs corresponding to \(G=SL_4\), \(\bar{s}_2s_2=s_2\bar{s}_2\) }
\end{figure}
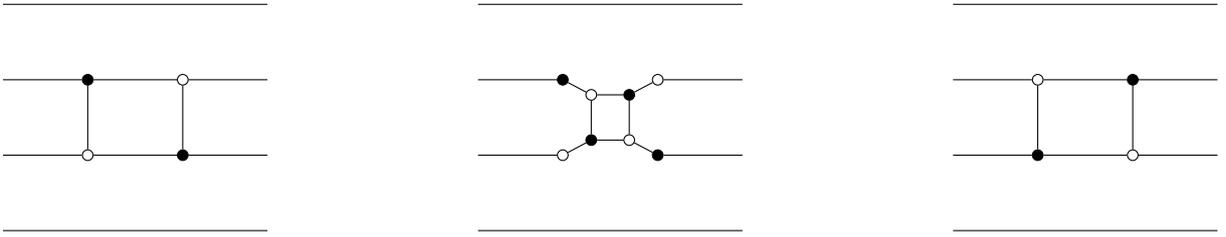

The proof of Lemma~\ref{Lemma:braid plabic} is straightforward.
See Fig.~\ref{Fig:example braid plabic} for an example.
In terms of a plabic graph, the first step is a spider move, while the second step is a removal of 2-valent vertices.

Finally, let us explain the meaning of the factorization schemes~\eqref{eq:factorization} in this combinatorial setting.
Let us orient edges in the plabic graph \(\Gamma_{\mathbf{i}}\) such that all horizontal lines go from right to left and all vertical edges go from black to white vertices (this is an example of perfect orientation from \cite{Postnikov2006total}).
Let us add (infinitely remote) boundary vertices to the network, namely source vertices \(\sigma_i=(+\infty,-i)\) and target vertices \(\tau_i=(-\infty,-i)\), \(1\le i \le N\).
For any oriented path \(p\) from \(\sigma_j\) to \(\tau_i\), let \(\wt(p)\) be the product of variables assigned to faces \emph{below} the path.
The transfer matrix $\tilde{T}$ assigned to a network is an \(N\times N\) matrix with elements
\begin{equation}\label{eq:T=Transfer matrix}
	\tilde{T}_{i,j}=\sum_{p \colon \sigma_j \to \tau_i} \wt(p),
\end{equation}
where the summation runs over paths from \(\sigma_j\) to \(\tau_i\).
Let us define the \emph{normalized transfer matrix} as \(T=(\det{\tilde{T}})^{-1/N}\tilde{T}\).
Clearly we have \(T\in SL_N\).

For example, let us take \(G=SL_3\) and \(w=\bar{s}_1s_1\bar{s}_2s_2\).
The corresponding network is depicted in Fig.~\ref{Fig:network}.
\begin{figure}[h]
	\centering
	\begin{tikzpicture}
		\def\xs{1.5}
		\def\ys{1.5}

			\node[circle, draw, inner sep=0pt, minimum size=4pt] (d1) at (-\xs,-2*\ys) {};
			
			\node[circle, draw, fill=black,  inner sep=0pt, minimum size=4pt] (u1) at (-\xs,-\ys) {};		
			
			\node[circle, draw,  fill=black,  inner sep=0pt, minimum size=4pt] (d2) at (0*\xs,-2*\ys) {};
			
			\node[circle, draw,  inner sep=0pt, minimum size=4pt] (u2) at (0*\xs,-1*\ys) {};
			
			\node[circle, draw,  inner sep=0pt, minimum size=4pt] (d3) at (1*\xs,-3*\ys) {};
			
			\node[circle, draw, fill=black, inner sep=0pt, minimum size=4pt] (u3) at (1*\xs,-2*\ys) {};

			\node[circle, draw, fill=black, inner sep=0pt, minimum size=4pt] (d4) at (2*\xs,-3*\ys) {};

			\node[circle, draw,   inner sep=0pt, minimum size=4pt] (u4) at (2*\xs,-2*\ys) {};

			\draw[special arrow=0.7] (u1) -- (d1);
			\draw[special arrow=0.7] (d2) -- (u2);	
			\draw[special arrow=0.7] (u3) -- (d3);	
			\draw[special arrow=0.7] (d4) -- (u4);	
%			sources and targets			
			\node (tau3) at (-2*\xs,-3*\ys) {$\tau_3$};	
			\node (tau2) at (-2*\xs,-2*\ys) {$\tau_2$};			
			\node (tau1) at (-2*\xs,-\ys) {$\tau_1$};			
			\node (sigma3) at (3*\xs,-3*\ys) {$\sigma_3$};	
			\node (sigma2) at (3*\xs,-2*\ys) {$\sigma_2$};	
			\node (sigma1) at (3*\xs,-\ys) {$\sigma_1$};
			
			\draw[special arrow=0.7] (sigma1) to (u2);
			\draw[special arrow=0.7] (u2) to (u1);
			\draw[special arrow=0.7] (u1) to (tau1);
			\draw[special arrow=0.7] (sigma2) to (u4);
			\draw[special arrow=0.7] (u4) to (u3);
			\draw[special arrow=0.7] (u3) to (d2); 
			\draw[special arrow=0.7] (d2) to (d1);
			\draw[special arrow=0.7] (d1) to (tau2);
			\draw[special arrow=0.7] (sigma3) to (d4);
			\draw[special arrow=0.7](d4) to (d3);
			\draw[special arrow=0.7] (d3) to (tau3);

			\node[blue] (x1) at (-1.5*\xs,-1.5*\ys) {$X_1$};
			\node[blue] (x2) at (-0.5*\xs,-2.5*\ys) {$X_2$};
			\node[blue] (x3) at (-0.5*\xs,-1.5*\ys) {$Y_1$};
			\node[blue] (x4) at (1.5*\xs,-2.5*\ys) {$Y_2$};
			\node[blue] (x5) at (1.5*\xs,-1.5*\ys) {$Z_1$};
			\node[blue] (x6) at (2.5*\xs,-2.5*\ys) {$Z_2$};

	\end{tikzpicture}		
	\caption{\label{Fig:network} Network corresponding to  \(G=SL_3\), \(w=\overline{s}_1s_1\overline{s}_2s_2\). }
\end{figure}
The corresponding transfer matrix is equal to 
\begin{equation}
	T=(X_1Y_1Z_1)^{-1/3}(X_2Y_2Z_2)^{-2/3}\begin{pmatrix}
		 X_1X_2Y_1Y_2Z_1Z_2 & X_1X_2Y_1Y_2Z_2 & X_1X_2Y_1Y_2
		 \\ 
		 X_2Y_1Y_2Z_1Z_2 & X_2Y_2Z_2(1+Y_1) & X_2Y_2 (1+Y_1)
		 \\
		 0& Y_2Z_2 & 1+Y_2
	\end{pmatrix}.
\end{equation}
\begin{Lemma}\label{Lem:L=T}	
	For any reduced expression \(w=s_{i_1}\cdot\ldots \cdot s_{i_l}\), the normalized transfer matrix constructed from the network \(\Gamma_{\mathbf{i}}\) is equal to the image of the factorization map \(T=\mathbb{L}_{\mathbf{s}}(\mathbf{X})\).
\end{Lemma}
It is straightforward to show this by induction on \(l(w)\).

\section{Moduli spaces of framed local systems} \label{Sec:FG}

We follow \cite{Fock:2006moduli}, \cite{Goncharov:2017ideal}, \cite{Goncharov:2019quantum} in this section. 

\subsection{Varieties \(\mathcal{X}_{G, \Sigma}\) and \(\mathcal{P}_{G, \Sigma}\)} Let \(\Sigma\) be an oriented surface with punctures and marked points on its boundary. 
We require that any boundary component contains at least one marked point, and there is at least one puncture or marked point. 

By a \(G\)-local system we mean a principal \(G\)-bundle with a flat connection. 
For a given local system and a curve \(\gamma\) from \(p_1\) to \(p_2\), we can compute the parallel transport map \(T_\gamma\) connecting the fibers over these points. 
If the trivializations of the fibers are given, then \(T_\gamma\) can be viewed as an element of \(G\). 
In the case of a closed path \(\gamma\), we have an automorphism of the fiber over the point \(p_1=p_2\). 
For a given trivialization of the fiber, this automorphism is given by the action of a monodromy element \(M_\gamma\in G\). 

We will concentrate on the cases \(G=SL_N\) and \(G=PGL_N\). 
In these cases, instead of principal \(G\)-bundles one can consider rank \(N\) vector bundles and projective bundles with flat connections, respectively.
These groups are related by the finite central quotient \(PGL_N\simeq SL_N/\mu_N\), where \(\mu_N=\{z\in \mathbb{C}^\times\mid z^N=1\}\). 
Locally, the corresponding moduli spaces of local systems differ by a finite cover. 
It appears that \(PGL_N\)-local systems are more suited for the cluster description. 
We mainly stick to the choice \(G=PGL_N\) below; however, we occasionally pass to \(SL_N\)-lifts when matrix elements, traces, or characteristic polynomials are needed.

\begin{Definition}\label{Def:X G,S}
	A framing of a \(G\)-local system on \(\Sigma\) is the choice of a flat section of the \(B\)-reduction of the local system on a small neighborhood of any puncture or marked point.
	The moduli space of the framed \(G\)-local systems on \(\Sigma\) is denoted by \(\mathcal{X}_{G,\Sigma}\).
\end{Definition}

Recall that a (complete) flag in \(\mathbb{C}^N\) is a sequence of subspaces 
\(
	0 = F_0 \subset F_1 \subset \dots \subset F_N = \mathbb{C}^N
\) such that \(\dim F_k = k\).
For any flag \(F\), there exists a unique Borel subgroup in \(G\) that preserves \(F\).
In more elementary terms, the choice of framing for a puncture \(p \in \Sigma\) is a choice of a complete flag that is invariant under monodromy around \(p\).
Assume that the monodromy is generic, namely, its matrix has \(N\) eigenvectors with different eigenvalues.
Then there exist \(N!\) invariant flags.
In particular, if there are no boundary components on \(\Sigma\), then \(\mathcal{X}_{G, \Sigma}\) is (at generic points) an \(N!^{\text{number of punctures}}\) covering of the moduli space of local systems on \(\Sigma\).

On the other hand, in the neighborhood of the marked points on the boundary, the local system can be trivialized, therefore, there are continuous families (namely \(G/B\)) of framing choices.

Let \(F, F'\) be two flags in \(\mathbb{C}^N\) in general position.
The latter implies that the intersection
\(L_i = F_i \cap F'_{N + 1 - i}\) is 1-dimensional for all \(i\). 
The \emph{pinning} over \((F, F')\) is a choice of vectors \(v_i \in L_i\), \(v_i \neq 0\) 
up to a total rescaling \((v_1, \dots, v_N) \mapsto (\lambda v_1, \dots, \lambda v_N)\). 
There is a natural free and transitive action of the Cartan subgroup \(H \subset PGL_N\) 
	on the set of pinnings over \((F,F')\). 
In particular, the choice of pinning depends on \(N - 1 = \operatorname{rk} G\) parameters.

\begin{Definition}
	By \(\mathcal{P}_{G, \Sigma}\) we denote the moduli space of framed \(G\)-local systems on \(\Sigma\) with the choice of pinning for each boundary segment of \(\Sigma\).
\end{Definition}

More explicitly, we can trivialize the local system near any segment \(AA'\) on the boundary of~\(\Sigma\).
The framing gives a pair of flags \(F, F'\) corresponding to marked points \(A, A'\); using trivialization, we can think of them as flags in the same \(\mathbb{C}^N\).
The choice of pinning upgrades this to the choice of a projective basis $(v_1,\dots,v_N)$ in \(\mathbb{C}^N\) assigned to the segment\footnote{\label{foot:two bases} More precisely, there are two bases $(v_1,\dots,v_N)$ and $(v_N,\dots,v_1)$ that should be considered on an equal footing.}.
This allows us to compute parallel transports from one boundary segment to another.
Such parallel transports are also called Wilson lines.

If \(\Sigma\) has no boundary components, then \(\mathcal{X}_{G,\Sigma}=\mathcal{P}_{G,\Sigma}\).
In general,
\begin{equation}
	\dim \mathcal{P}_{G,\Sigma}=\dim \mathcal{X}_{G,\Sigma}+(N-1) (\text{number of boundary segments}).
\end{equation} 
Sometimes it is also convenient to consider moduli spaces that are intermediate between \(\mathcal{X}_{G, \Sigma}\) and \(\mathcal{P}_{G, \Sigma}\), namely when the pinnings are assigned only to some of the boundary segments.

\begin{Theorem}[\cite{Fock:2006moduli},\cite{Goncharov:2019quantum}] \label{Th:P G,S}
	The varieties \(\mathcal{P}_{G, \Sigma}\), \(\mathcal{X}_{G, \Sigma}\) (and all intermediate ones) have a natural cluster structure.
\end{Theorem}

\begin{figure}[h]
	\centering
	\includegraphics[]{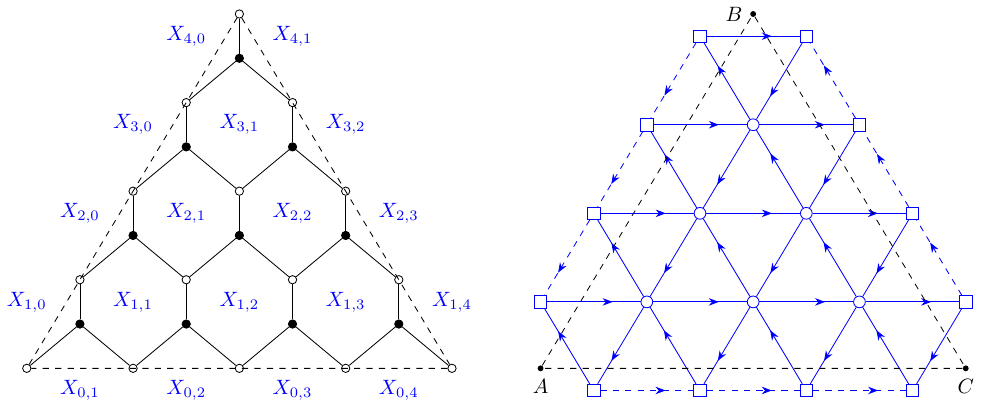}
	\caption{\label{Fig:triangle} $N=5$, left: plabic graph with variables, right: quiver}
\end{figure}

Let us construct some seeds for these cluster structures.
Consider an ideal triangulation \(\mathcal{T}\) of \(\Sigma\), that is, a triangulation with vertices at the punctures and marked points on the boundary.
For any triangle \(ABC\), we assign a plabic graph as on the left in Fig.~\ref{Fig:triangle}.
The corresponding quiver is depicted on the right in Fig.~\ref{Fig:triangle}.

The seed \(\mathsf{s}_\mathcal{T}\) for \(\mathcal{P}_{G, \Sigma}\) is obtained via amalgamation of the seeds corresponding to triangles in \(\mathcal{T}\).
The corresponding cluster chart will be denoted by \(\mathcal{X}_\mathcal{T}=\mathcal{X}_{\mathsf{s}_\mathcal{T}}\).
In this construction, for each segment on the boundary, we assign \(N-1\) frozen vertices.
For instance, for the triangle in Fig.~\ref{Fig:triangle}, to the segment \(AB\) we assign variables \(X_{1,0}, X_{2,0}, X_{3,0}, X_{4,0}\).
These \(N-1\) variables encode the choice of the pinning.
In particular, if we want to remove it from the moduli space data, we remove the corresponding variables.

\subsection{Local system in cluster coordinates} Let us now relate this combinatorial construction to the local systems.
Note that if we exclude the bottom \(N-1\) faces in the triangular plabic graph in Fig.~\ref{Fig:triangle}, we will get the plabic graph  \(\Gamma_{\mathbf{i}_0}\), where \(\mathbf{i}_0\) is the word corresponding to a reduced decomposition of \(w_0\) given by
\begin{equation}
	w_0=\left(s_{N-1}s_{N-2}\cdots s_2s_1\right)\left(s_{N-1}s_{N-2}\cdots s_2\right)\cdots \left(s_{N-1}s_{N-2}\right)\left(s_{N-1}\right).
\end{equation}
This allows us to define a transfer matrix that geometrically corresponds to the parallel transport from the side \(BC\) to the side \(BA\).
The corresponding formula reads
\begin{multline}
	T_{BC,BA}=\mathbb{L}_{w_0}(\mathbf{X})=	H_4(X_{1,0})H_3(X_{2,0}) H_2(X_{3,0}) H_1(X_{4,0})\; E_4 E_3E_2E_1
	\\ H_4(X_{1,1}) H_3(X_{2,1}) H_2(X_{3,1}) \; E_4 E_3E_2\; H_4(X_{1,2}) H_3(X_{2,2})
	\\
	 E_4 E_3 \; H_4(X_{1,3}) \; E_4 \;
	 H_4(X_{1,4})H_3(X_{2,3}) H_2(X_{3,2})H_1(X_{4,1}).
\end{multline}
Similarly, one can define parallel transports from \(AB\) to the side \(AC\) and from \(CA\) to the side \(CB\).
For example
\begin{multline}
	T_{AB,AC}=H_4(X_{0,4})H_3(X_{0,3}) H_2(X_{0,2}) H_1(X_{0,1})\; E_4 E_3E_2E_1 
	\\ H_4(X_{1,3}) H_3(X_{1,2}) H_2(X_{1,1}) \; E_4 E_3E_2\; H_4(X_{2,2}) H_3(X_{2,1})
	\\
	E_4 E_3 \; H_4(X_{3,1}) \; E_4 \;
	H_4(X_{4,0})H_3(X_{3,0}) H_2(X_{2,0})H_1(X_{1,0}).
\end{multline}

We will need a slight modification of the factorization formula~\eqref{eq:factorization}.
Let  \(w=s_{i_1}s_{i_2}\cdot\dots \cdot s_{i_l}\) be a reduced decomposition, and assume that for any \(j\in \{1,\dots,N-1\}\) the letter \(j\) or the letter \(\bar{j}\) appears in the list \((i_1,\dots, i_l)\).
Then \(\overline{\mathbb{L}}_{\mathbf{s}}(\mathbf{X})\) is defined by removing from \(\mathbb{L}_{\mathbf{s}}(\mathbf{X})\) the factors corresponding to frozen variables.
Hence the expression \(\overline{\mathbb{L}}_{\mathbf{s}}(\mathbf{X})\) depends only on \(l(w)-(N-1)\) variables, contrary to \(\mathbb{L}_{\mathbf{s}}(\mathbf{X})\) that depends on \(l(w)+N-1\) variables.
Using these notations, we can write
\begin{equation}\label{eq:T BC BA}
	T_{BC,BA}=\left(\prod H_i(X_{N-i,0})\right) \; \overline{\mathbb{L}}_{w_0}(\mathbf{X})\;
	\left(\prod H_i(X_{N-i,i})\right).
\end{equation}

\begin{figure}[h]
	\centering
	\includegraphics[]{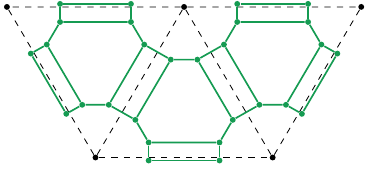}
	\caption{\label{Fig:rectangular-hexagonal graph} Rectangular-hexagonal graph constructed from triangulation}
\end{figure}

Consider an additional graph with hexagonal faces inside each triangle and rectangles around each side of the triangulation, see Fig.~\ref{Fig:rectangular-hexagonal graph}.
We usually depict this graph in green.

We will encode the local system in terms of parallel transports on this graph.
Namely, to each vertex of this graph, we assign a projective basis in the fiber, and for any edge, we assign an element in \(PGL_N\) corresponding to the parallel transport.
We already saw projective bases assigned to the boundary segments with pinning above.

\begin{figure}[h]
	\centering
	\includegraphics[]{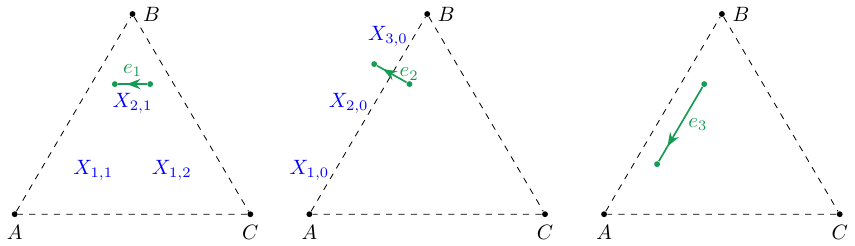}
	\caption{\label{Fig:edges} Three types of edges of a rectangular-hexagonal graph. The cluster variables are given for $N=4$. The corresponding parallel transport matrices are~\eqref{eq:T1 T2 T3}}
\end{figure}

Note that there are three types of edges in this rectangular-hexagonal graph: ones near the vertices of the triangulation, ones orthogonal to the sides of the triangulation, and ones parallel to the sides of the triangulation; see Fig.~\ref{Fig:edges}.
To each of these edges \(e_i\) we assign a parallel transport element \(T_i\in PGL_N\) as follows:
\begin{subequations}\label{eq:T1 T2 T3}
	\begin{align}
		&T_1=\overline{\mathbb{L}}_{w_0}(\mathbf{X})=E_3E_2E_1  H_3(X_{1,2}) H_2(X_{1,1}) E_3E_2  H_3(X_{2,1}) E_3, \label{eq:T1}
		\\
		&T_2=\prod\nolimits_{i=1}^{N-1} H_i(X_{N-i,0})=H_3(X_{1,0})H_2(X_{2,0})H_1(X_{3,0}), \label{eq:T2}
		\\
		&T_3=S=\sum\nolimits_{i=1}^N (-1)^{i-1}E_{N+1-i,i}. \label{eq:S}
	\end{align}
\end{subequations}
Here we give both the generic formula and the explicit formula for \(G=PGL_4\) and variables inside the triangle as in Fig.~\ref{Fig:edges}.
By definition, the parallel transport for any path in a rectangular-hexagonal graph is a product of transports along the edges.
In particular, the parallel transport $T_{BC, BA}$ given in formula~\eqref{eq:T BC BA} above now corresponds to the path that consists of one \(e_2\) edge, one \(e_1\) edge, and one more \(e_2\) edge.

The element \(S\) defined in~\eqref{eq:S} is a matrix with \(1\) and \(-1\) alternating on the secondary diagonal.
This element is a lift of \(w_0\in S_N\) to the group \(PGL_N\).
Its action corresponds to reordering the elements of the basis in \(\mathbb{C}^N\), cf.~footnote~\ref{foot:two bases}.
Conjugation by \(S\) acts as
\begin{equation}\label{eq:S conjugation}
	S E_iS^{-1}=F_{N-i}^{-1}, \qquad S F_iS^{-1}=E_{N-i}^{-1}, \qquad S H_i(X)S^{-1}=H_{N-i}(X)^{-1}.
\end{equation}
\begin{Lemma}\label{Lem:faces trivial}
	The parallel transport around any contractible cycle in a rectangular-hexagonal graph is trivial.
\end{Lemma}

\begin{figure}[h]
	\centering
	\includegraphics[]{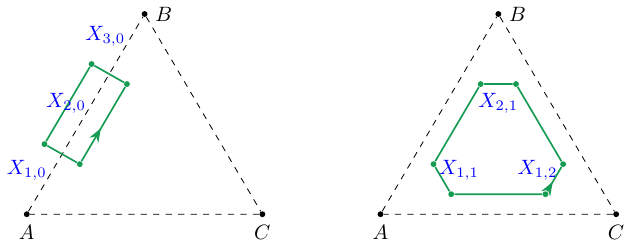}
	
	\caption{\label{Fig:faces} Rectangular and hexagonal faces. The cluster variables are given for $N=4$}
\end{figure}

It is sufficient to show this property for the faces of the graph, see Fig.~\ref{Fig:faces}.
For the rectangular face, we have
\begin{multline}
	S\; \prod\nolimits_{i=1}^{N-1} H_i(X_{i,0})\; S\; \prod\nolimits_{i=1}^{N-1} H_i(X_{N-i,0})
	\\ 
	= (-1)^{N-1}\prod\nolimits_{i=1}^{N-1} H_{N-i}(X_{i,0})^{-1}  \prod\nolimits_{i=1}^{N-1} H_i(X_{N-i,0})=(-1)^{N-1},
\end{multline}
where we used the third relation in~\eqref{eq:S conjugation}.

We omit the proof for the hexagonal face but illustrate the fact by computations for \(N=2\) and \(N=3\):
\begin{align}
	&PGL_2\colon& \qquad &S E_1 S E_1 S E_1= 
	\bigg(\begin{pmatrix}
		0 & 1 \\ -1 & 0
	\end{pmatrix} 
	\begin{pmatrix}
	1 & 1 \\ 0 & 1
	\end{pmatrix} \bigg)^3=-1,
	\\	
	&PGL_3\colon& \qquad &(S \; E_2 E_1 H_2(X_{1,1})E_1)^3=1.
\end{align}
We see from these computations that monodromy is trivial in \(PGL_N\) but not necessarily in \(SL_N\).
In order to have the moduli space of \(SL_N\)-local systems, one has to consider a certain cover of the moduli space of \(PGL_N\)-local systems.
In each cluster chart, this is an extension by certain roots of Laurent monomials.

To summarize, for any path \(\gamma\) in a rectangular-hexagonal graph, we assign a parallel transport matrix \(T_\gamma \in PGL_N\).
It follows from Lemma~\ref{Lem:faces trivial} that \(T_\gamma\) depends on the homotopy class of \(\gamma\).
Hence, we obtain a \(G\)-local system.

Moreover, for any puncture or marked point on the boundary \(p\), there is a path in the rectangular-hexagonal graph \(\gamma_p\) close to this point.
It consists of edges of types \(e_1\) and \(e_2\).
Since matrices \(T_1,T_2\) in formulas~\eqref{eq:T1 T2 T3} are upper triangular, the corresponding parallel transport \(T_{\gamma_p}\) is upper triangular.
This determines the choice of framing, see Definition~\ref{Def:X G,S}.
Finally, for any boundary segment, we can define parallel transport starting from it, hence, to any boundary segment we assign a pinning.
Therefore, we have defined a map from the cluster chart \(\mathcal{X}_{\mathcal{T}}\) to \(\mathcal{P}_{G,\Sigma}\).

Theorem~\ref{Th:P G,S} states that matrix elements of parallel transports between boundary segments are global functions on \(\mathcal{P}_{G, \Sigma}\).
Monodromies \(M_\gamma\) (parallel transports over closed loops \(\gamma\)) depend on the initial point, i.e., are defined up to conjugation.
Therefore the functions \(\operatorname{Tr}M_\gamma^k\) are well defined on \(\mathcal{P}_{G,\Sigma}\), i.e., are global functions.

\begin{Example}\label{Ex:sphere 4 puntures}
	Let \(\Sigma\) be a sphere with 4 punctures.
	Let us compare dimensions of \(\mathcal{P}_{G,\Sigma}\) and \(\mathcal{X}_{\mathcal{T}}\).
	Since we have no boundary components, we have
	\begin{multline}
		\dim \mathcal{P}_{G,\Sigma}= \dim \mathcal{X}_{G,\Sigma}=\dim \operatorname{Loc.\ Sys.}_{G,\Sigma}
		\\ 
		=\dim \Big\{M_1,M_2,M_3,M_4\in G\mid \prod M_i=1\Big\}\Big/G 	=2 \dim G=2(N^2-1).
	\end{multline}
	Here \(\operatorname{Loc.\ Sys.}_{G,\Sigma}\) denotes the moduli space of \(G\)-local systems on \(\Sigma\) and \(M_i\) denotes monodromy around a path that encircles \(p_i\) starting at some base point \(p\).
	
	On the other hand, a triangulation \(\mathcal{T}\) of \(\Sigma\) consists of \(4\) triangles and has \(6\) edges.
	This counting follows from the fact that the number of vertices is 4 and the computation of the Euler characteristic.
	For instance, one can take a triangulation that is topologically given by faces and edges of a tetrahedron.
	It follows from the description in Fig.~\ref{Fig:triangle} that the quiver has \(N-1\) vertices on each edge and \((N-1)(N-2)/2\) vertices inside each triangle.
	Therefore
	\begin{equation}
		\dim \mathcal{X}_{\mathcal{T}}=4\frac{(N-1)(N-2)}{2}+6(N-1)=2(N^2-1).
	\end{equation}
\end{Example}

\begin{Example}
	Consider \(\Sigma\) to be a rectangle \(ABCD\).
	Consider a moduli space intermediate between \(\mathcal{P}_{G,\Sigma}\) and \(\mathcal{X}_{G,\Sigma}\) with pinning data for sides \(AB\) and \(CD\) but not for sides \(BC\) and \(DA\).
	The corresponding plabic graph and cluster variables (for \(G=PGL_4\)) are shown in Fig.~\ref{Fig:square}.
	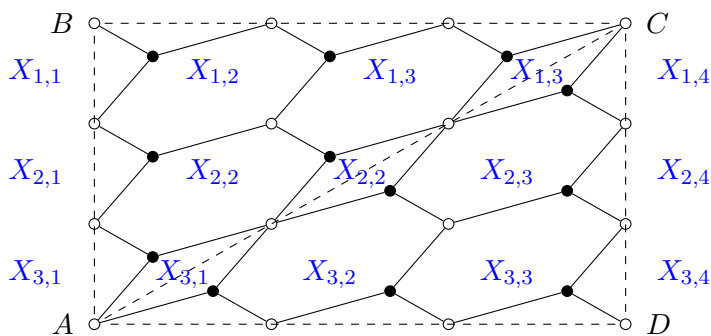
\begin{figure}[h]
		\centering
		\begin{tikzpicture}
			\def\xs{2.33}
			\def\ys{1.33}
			\def\NN{3}
			\pgfmathtruncatemacro{\Nminus}{\NN-1}
				
			\begin{scope}[shift={(0,0)}]
				\node[ label=left:{$A$}] (A) at (0,0) {};
				\node[ label=left:{$B$}] (B) at (0,\NN*\ys) {};
				\node[ label=right:{$C$}] (C) at (\NN*\xs,\NN*\ys) {};
				\node[ label=right:{$D$}] (D) at (\NN*\xs,0) {};
				
				\draw[dashed] (A) to (B) to (C) to (D) to (A) to (C); 	
				
				\foreach \i in {0,...,\NN}
				{
					\pgfmathtruncatemacro{\Ni}{\NN-\i}
					\foreach \j in {0,...,\Ni}
					{
						\node[circle, draw, inner sep=0pt, minimum size=4pt] (w\i\j) at (\j*\xs,\NN*\ys-\i*\ys) {};	
						
					}
					
				}
				\foreach \i in {1,...,\NN}
				{
					\pgfmathtruncatemacro{\Ni}{\NN-\i}
					\pgfmathtruncatemacro{\Nip}{\NN-\i+1}
					\pgfmathtruncatemacro{\im}{\i-1}
					\foreach \j in {\Nip,...,\NN}
					{
						\pgfmathtruncatemacro{\Nj}{\NN-\j}
						\pgfmathtruncatemacro{\Njp}{\NN-\j+1}
						\pgfmathtruncatemacro{\jm}{\j-1}
						\pgfmathtruncatemacro{\jp}{\j+1}
						
						\node[circle, draw, inner sep=0pt, minimum size=4pt] (w\i\j) at (\j*\xs,\NN*\ys-\i*\ys) {};	
						\node[circle, draw, fill, inner sep=0pt,  minimum size=4pt] (b1\Nip\Njp) at (-0.67*\xs+\Njp*\xs,0.67*\ys+\NN*\ys-\Nip*\ys) {};
						\node[circle, draw, fill, inner sep=0pt,  minimum size=4pt] (b2\i\j) at (-0.33*\xs+\j*\xs,0.33*\ys+\NN*\ys-\i*\ys) {};
						\draw (b1\Nip\Njp) to (w\Ni\Njp);
						\draw (b1\Nip\Njp) to (w\Ni\Nj);
						\draw (b1\Nip\Njp) to (w\Nip\Nj);
						
						\draw (b2\i\j) to (w\i\j);
						\draw (b2\i\j) to (w\im\j);
						\draw (b2\i\j) to (w\i\jm);

						\node[blue] at (0.33*\xs+\j*\xs,0.5*\ys+\NN*\ys-\i*\ys) {$X_{\i,\jp}$};						
						\node[blue] at (-1.33*\xs+\Njp*\xs,0.5*\ys+\NN*\ys-\Nip*\ys) {$X_{\Nip,\Njp}$};
					}
					\node[blue] at (-0.5*\xs+\Nip*\xs,0.5*\ys+\NN*\ys-\i*\ys) {$X_{\i,\Nip}$};
				}
			\end{scope}
		\end{tikzpicture}
		\caption{\label{Fig:square} Plabic graph and cluster variables for the rectangle, \(N=4\)}
	\end{figure}
	
	In this case, the only nontrivial parallel transport goes from left to right: \(T_{AB, DC}\).
	Therefore, the corresponding moduli space \(\mathcal{P}_{G,\Sigma}\) should be (birationally equivalent to) the group \(G\) itself.
	On the other hand, it is easy to see that the plabic graph in Fig.~\ref{Fig:square} coincides with the one for the open double Bruhat cell \(G^{w_0,w_0}\subset G\).
	
	One can identify sides \(AB\) and \(DC\), getting a cylinder from the rectangle.
	In terms of \(\mathcal{P}_{G,\Sigma}\) this would correspond to the replacement \(G^{w_0,w_0}\) by \(G^{w_0,w_0}/\operatorname{Ad}H\).
	Such a quotient by the adjoint action of the Cartan subgroup was used in Sec.~\ref{Sec:Rel Toda}.
\end{Example}

\subsection{Change of triangulation} 
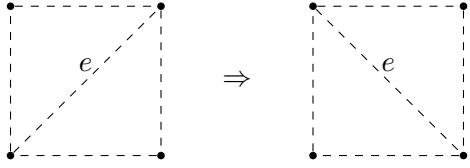
\begin{figure}[h]
	\centering
	\begin{tikzpicture}
		\def\xs{0.66}
		\def\ys{0.66}
		\def\NN{3}
		\begin{scope}[shift={(0,0)}]

			\node[circle, fill, inner sep=1pt, label=left:{}] (A) at (0,0) {};
			\node[circle, fill, inner sep=1pt, label=left:{}] (B) at (0,\NN*\ys) {};
			\node[circle, fill, inner sep=1pt, label=right:{}] (C) at (\NN*\xs,\NN*\ys) {};
			\node[circle, fill, inner sep=1pt, label=right:{}] (D) at (\NN*\xs,0) {};
			
			\draw[dashed] (A) to (B) to (C) to (D) to (A) to node[above]{$e$} (C);

		\end{scope}

		\node at (\NN*\xs+1,0.5*\NN*\ys) {$\Rightarrow$};
		
		\begin{scope}[shift={(\NN*\xs+2,0)}]

			\node[circle, fill, inner sep=1pt, label=left:{}] (A) at (0,0) {};
			\node[circle, fill, inner sep=1pt, label=left:{}] (B) at (0,\NN*\ys) {};
			\node[circle, fill, inner sep=1pt, label=right:{}] (C) at (\NN*\xs,\NN*\ys) {};
			\node[circle, fill, inner sep=1pt, label=right:{}] (D) at (\NN*\xs,0) {};
			
			\draw[dashed] (B) to (C) to (D) to (A) to (B) to node[above]{$e$} (D);

		\end{scope}

	\end{tikzpicture}
	\caption{\label{Fig:flip} Flip of triangulation at edge \(e\)}
\end{figure}

In the discussion above we worked with the seed \({\mathsf{s}_\mathcal{T}}\) constructed for a given triangulation \(\mathcal{T}\).
If the triangulation \(\mathcal{T}\) possesses a non-trivial automorphism, then it naturally induces a permutation of the seed~\(\mathsf{s}_\mathcal{T}\).
More interesting cluster transformations come from flips of the triangulation depicted in Fig.~\ref{Fig:flip}.
Here and below, we label the edge before and after the flip by the same letter.

\begin{Theorem}\label{Th:flip cluster}
	Let \(\mathcal{T},\mathcal{T}'\) be two triangulations connected by a flip at edge \(e\).
	Then there exists a sequence of mutations \(\boldsymbol{\mu}_e\) such that
	\begin{enumerate}[label=(\alph*)]
	
		\item The transformation \(\boldsymbol{\mu}_e\) transforms the quiver corresponding to \(\mathcal{T}\) to a quiver corresponding to \(\mathcal{T}'\).
		
		\item The transformation \(\boldsymbol{\mu}_e\colon \mathcal{X}_{\mathcal{T}} \to \mathcal{X}_{\mathcal{T}'}\) intertwines the maps \(\mathcal{X}_{\mathcal{T}},\mathcal{X}_{\mathcal{T}'} \to \mathcal{P}_{G,\Sigma}\).
		\label{it:flip monodromy}
		
		\item The composition satisfies \(\boldsymbol{\mu}_e\circ \boldsymbol{\mu}_e = \operatorname{id}\).
		\label{it:bigon}
		
	\end{enumerate}	
	Let \(\mathcal{T}''\) be a triangulation obtained from \(\mathcal{T}\) by a flip at another edge \(\tilde{e}\).
	Then
	\begin{enumerate}[label=(\alph*)]
		\setcounter{enumi}{3}
		
		\item If the edges \(e,\tilde{e}\) do not share the same triangle, then the corresponding cluster transformations commute: \(\boldsymbol{\mu}_e\circ \boldsymbol{\mu}_{\tilde{e}} =\boldsymbol{\mu}_{\tilde{e}}\circ \boldsymbol{\mu}_e\).
		\label{it:square}
		
		\item If the edges \(e,\tilde{e}\) belong to the same triangle, then the corresponding cluster transformations satisfy the pentagon relation  \(\boldsymbol{\mu}_e\circ \boldsymbol{\mu}_{\tilde{e}} =(e,\tilde{e})\boldsymbol{\mu}_e\circ \boldsymbol{\mu}_{\tilde{e}}\circ \boldsymbol{\mu}_e\).
		\label{it:pentagon}
	\end{enumerate}	
\end{Theorem}

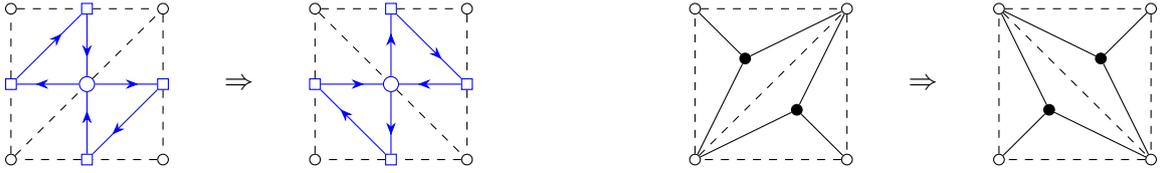
\begin{figure}[h]
	\centering
	\begin{tikzpicture}
		\def\xs{2}
		\def\ys{2}
		\def\NN{1}
		
%		mutation of quiver
		\begin{scope}[shift={(0,0)}]

			\node[circle, draw, inner sep=0pt, minimum size=4pt] (w00) at (0,0) {};
			\node[circle, draw, inner sep=0pt, minimum size=4pt] (w10) at (0,\NN*\ys) {};
			\node[circle, draw, inner sep=0pt, minimum size=4pt] (w11) at (\NN*\xs,\NN*\ys) {};
			\node[circle, draw, inner sep=0pt, minimum size=4pt] (w01) at (\NN*\xs,0) {};
			
			\node[circle,draw,inner sep=2, blue] (x) at (0.5*\xs,0.5*\ys) {};
			\node[rectangle,draw,inner sep=2, blue] (x1) at (0*\xs,0.5*\ys) {};
			\node[rectangle,draw,inner sep=2, blue] (x2) at (0.5*\xs,1*\ys) {};
			\node[rectangle,draw,inner sep=2, blue] (x3) at (1*\xs,0.5*\ys) {};
			\node[rectangle,draw,inner sep=2, blue] (x4) at (0.5*\xs,0*\ys) {};

			\draw[blue,special arrow=0.7] (x) to (x1);
			\draw[blue,special arrow=0.7] (x2) to (x);
			\draw[blue,special arrow=0.7] (x) to (x3);
			\draw[blue,special arrow=0.7] (x4) to (x);
			\draw[blue,special arrow=0.7] (x1) to (x2);
			\draw[blue,special arrow=0.7] (x3) to (x4);
			
			\draw[dashed] (w00) to (x1) to  (w10) to (x2) to (w11) to (x3) to (w01) to (x4) to (w00) to (x) to (w11); 	
		\end{scope}
		
		\node at (\NN*\xs+1,0.5*\NN*\ys) {$\Rightarrow$};
		
		\begin{scope}[shift={(\NN*\xs+2,0)}]
		
			\node[circle, draw, inner sep=0pt, minimum size=4pt] (w00) at (0,0) {};
			\node[circle, draw, inner sep=0pt, minimum size=4pt] (w10) at (0,\NN*\ys) {};
			\node[circle, draw, inner sep=0pt, minimum size=4pt] (w11) at (\NN*\xs,\NN*\ys) {};
			\node[circle, draw, inner sep=0pt, minimum size=4pt] (w01) at (\NN*\xs,0) {};

			\node[circle,draw,inner sep=2, blue] (x) at (0.5*\xs,0.5*\ys) {};
			\node[rectangle,draw,inner sep=2, blue] (x1) at (0*\xs,0.5*\ys) {};
			\node[rectangle,draw,inner sep=2, blue] (x2) at (0.5*\xs,1*\ys) {};
			\node[rectangle,draw,inner sep=2, blue] (x3) at (1*\xs,0.5*\ys) {};
			\node[rectangle,draw,inner sep=2, blue] (x4) at (0.5*\xs,0*\ys) {};

			\draw[blue,special arrow=0.7] (x) to (x2);
			\draw[blue,special arrow=0.7] (x3) to (x);
			\draw[blue,special arrow=0.7] (x) to (x4);
			\draw[blue,special arrow=0.7] (x1) to (x);
			\draw[blue,special arrow=0.7] (x4) to (x1);
			\draw[blue,special arrow=0.7] (x2) to (x3);
			
			\draw[dashed] (w10) to (x2) to (w11) to (x3) to (w01) to (x4) to (w00) to (x1) to (w10) to (x) to (w01); 	
		\end{scope}
		
%		spider for a plabic graph
		\begin{scope}[shift={(2*\NN*\xs+5,0)}]

			\node[circle, draw, inner sep=0pt, minimum size=4pt] (w00) at (0,0) {};
			\node[circle, draw, inner sep=0pt, minimum size=4pt] (w10) at (0,\NN*\ys) {};
			\node[circle, draw, inner sep=0pt, minimum size=4pt] (w11) at (\NN*\xs,\NN*\ys) {};
			\node[circle, draw, inner sep=0pt, minimum size=4pt] (w01) at (\NN*\xs,0) {};

			\node[circle, draw, fill, inner sep=0pt, minimum size=4pt] (b1) at (0.33*\xs,0.67*\ys) {};
			
			\node[circle, draw, fill, inner sep=0pt, minimum size=4pt] (b2) at (0.67*\xs,0.33*\ys) {};
			
			\draw (b1) to (w00);
			\draw (b1) to (w11);
			\draw (b1) to (w10);
			
			\draw (b2) to (w00);
			\draw (b2) to (w01);
			\draw (b2) to (w11);

			\draw[dashed] (w00) to (w10) to (w11) to (w01) to (w00) to  (w11);

		\end{scope}
		
		\node at (3*\NN*\xs+6,0.5*\NN*\ys) {$\Rightarrow$};
		
		\begin{scope}[shift={(3*\NN*\xs+7,0)}]
			
			\node[circle, draw, inner sep=0pt, minimum size=4pt] (w00) at (0,0) {};
			\node[circle, draw, inner sep=0pt, minimum size=4pt] (w10) at (0,\NN*\ys) {};
			\node[circle, draw, inner sep=0pt, minimum size=4pt] (w11) at (\NN*\xs,\NN*\ys) {};
			\node[circle, draw, inner sep=0pt, minimum size=4pt] (w01) at (\NN*\xs,0) {};

			\node[circle, draw, fill, inner sep=0pt, minimum size=4pt] (b1) at (0.33*\xs,0.33*\ys) {};
			
			\node[circle, draw, fill, inner sep=0pt, minimum size=4pt] (b2) at (0.67*\xs,0.67*\ys) {};
			
			\draw (b1) to (w00);
			\draw (b1) to (w01);
			\draw (b1) to (w10);
			
			\draw (b2) to (w01);
			\draw (b2) to (w10);
			\draw (b2) to (w11);

			\draw[dashed] (w10) to (w11) to (w01) to (w00) to  (w10) to (w01);

		\end{scope}
	\end{tikzpicture}
	\caption{\label{Fig:flip N=2} Flip of triangulation as a mutation and spider move for \(N=2\)}
\end{figure}

Let us first illustrate this theorem with the simplest non-trivial example, \(N=2\).
In this case, quiver vertices are in one-to-one correspondence with the edges of the triangulation.
The transformation \(\mu_e\) in this case is given by one cluster mutation or one spider move in terms of plabic graphs; see Fig.~\ref{Fig:flip N=2}.

\begin{figure}[h]
	\begin{center}
		\begin{tikzpicture} 
			\def\xs{1.2}
			\def\ys{1.2}

			\begin{scope}
				\node[circle, fill, inner sep=1pt, label=left:{}] (A) at (-0.59*\xs,-0.8*\ys) {};				
				\node[circle, fill, inner sep=1pt, label=left:{}] (B) at (-0.95*\xs,0.3*\ys) {};
				\node[circle, fill, inner sep=1pt, label=left:{}] (C) at (0,\ys) {};
				\node[circle, fill, inner sep=1pt, label=left:{}] (D) at (0.95*\xs,0.3*\ys) {};				
				\node[circle, fill, inner sep=1pt, label=left:{}] (E) at (0.59*\xs,-0.8*\ys) {};
				
				\draw[dashed] (A) to (B) to (C) to (D) to (E) to (A);
				
				\draw[dashed] (A) to node[left]{$e_1$} (C); 	
				\draw[dashed] (E) to node[right]{$e_2$} (C); 	
 			\end{scope}
			
			\node at (2*\xs,0) {$\xrightarrow{\mu_1}$};
			
			\begin{scope}[shift={(4*\xs,0)}]	
				\node[circle, fill, inner sep=1pt, label=left:{}] (A) at (-0.59*\xs,-0.8*\ys) {};				
				\node[circle, fill, inner sep=1pt, label=left:{}] (B) at (-0.95*\xs,0.3*\ys) {};
				\node[circle, fill, inner sep=1pt, label=left:{}] (C) at (0,\ys) {};
				\node[circle, fill, inner sep=1pt, label=left:{}] (D) at (0.95*\xs,0.3*\ys) {};				
				\node[circle, fill, inner sep=1pt, label=left:{}] (E) at (0.59*\xs,-0.8*\ys) {};
				
				\draw[dashed] (A) to (B) to (C) to (D) to (E) to (A);
				
				\draw[dashed] (B) to node[below]{$e_1$} (E); 	
				\draw[dashed] (E) to node[right]{$e_2$} (C); 	
			\end{scope}
			
			\node at (6*\xs,0) {$\xrightarrow{\mu_2}$};
			
			\begin{scope}[shift={(8*\xs,0)}]	
				\node[circle, fill, inner sep=1pt, label=left:{}] (A) at (-0.59*\xs,-0.8*\ys) {};				
				\node[circle, fill, inner sep=1pt, label=left:{}] (B) at (-0.95*\xs,0.3*\ys) {};
				\node[circle, fill, inner sep=1pt, label=left:{}] (C) at (0,\ys) {};
				\node[circle, fill, inner sep=1pt, label=left:{}] (D) at (0.95*\xs,0.3*\ys) {};				
				\node[circle, fill, inner sep=1pt, label=left:{}] (E) at (0.59*\xs,-0.8*\ys) {};
				
				\draw[dashed] (A) to (B) to (C) to (D) to (E) to (A);
				
				\draw[dashed] (B) to node[below]{$e_1$} (E); 	
				\draw[dashed] (B) to node[above]{$e_2$} (D); 	
			\end{scope}

			\node at (0,-1.4*\ys) {$\downarrow_{\mu_2}$};
			
			\begin{scope}[shift={(0,-3*\ys)}]
				\node[circle, fill, inner sep=1pt, label=left:{}] (A) at (-0.59*\xs,-0.8*\ys) {};				
				\node[circle, fill, inner sep=1pt, label=left:{}] (B) at (-0.95*\xs,0.3*\ys) {};
				\node[circle, fill, inner sep=1pt, label=left:{}] (C) at (0,\ys) {};
				\node[circle, fill, inner sep=1pt, label=left:{}] (D) at (0.95*\xs,0.3*\ys) {};				
				\node[circle, fill, inner sep=1pt, label=left:{}] (E) at (0.59*\xs,-0.8*\ys) {};
				
				\draw[dashed] (A) to (B) to (C) to (D) to (E) to (A);
				
				\draw[dashed] (A) to node[left]{$e_1$} (C); 	
				\draw[dashed] (A) to node[below]{$e_2$} (D); 	
			\end{scope}
			
			\node at (2*\xs,-3*\ys) {$\xrightarrow{\mu_1}$};
			
			\begin{scope}[shift={(4*\xs,-3*\ys)}]	
				\node[circle, fill, inner sep=1pt, label=left:{}] (A) at (-0.59*\xs,-0.8*\ys) {};				
				\node[circle, fill, inner sep=1pt, label=left:{}] (B) at (-0.95*\xs,0.3*\ys) {};
				\node[circle, fill, inner sep=1pt, label=left:{}] (C) at (0,\ys) {};
				\node[circle, fill, inner sep=1pt, label=left:{}] (D) at (0.95*\xs,0.3*\ys) {};				
				\node[circle, fill, inner sep=1pt, label=left:{}] (E) at (0.59*\xs,-0.8*\ys) {};
				
				\draw[dashed] (A) to (B) to (C) to (D) to (E) to (A);
				
				\draw[dashed] (B) to node[above]{$e_1$} (D); 	
				\draw[dashed] (A) to node[below]{$e_2$} (D); 	
			\end{scope}
			
			\node at (6*\xs,-3*\ys) {$\xrightarrow{\mu_2}$};
			
			\node at (8*\xs,-1.4*\ys) {$\downarrow_{(1,2)}$};

			\begin{scope}[shift={(8*\xs,-3*\ys)}]	
				\node[circle, fill, inner sep=1pt, label=left:{}] (A) at (-0.59*\xs,-0.8*\ys) {};				
				\node[circle, fill, inner sep=1pt, label=left:{}] (B) at (-0.95*\xs,0.3*\ys) {};
				\node[circle, fill, inner sep=1pt, label=left:{}] (C) at (0,\ys) {};
				\node[circle, fill, inner sep=1pt, label=left:{}] (D) at (0.95*\xs,0.3*\ys) {};				
				\node[circle, fill, inner sep=1pt, label=left:{}] (E) at (0.59*\xs,-0.8*\ys) {};
				
				\draw[dashed] (A) to (B) to (C) to (D) to (E) to (A);
				
				\draw[dashed] (B) to node[below]{$e_2$} (E); 	
				\draw[dashed] (B) to node[above]{$e_1$} (D); 	
			\end{scope}

		\end{tikzpicture}
	\end{center}
	\caption{\label{Fig:pentagon triang} Pentagon relation in terms of flips of triangulations}	
\end{figure}
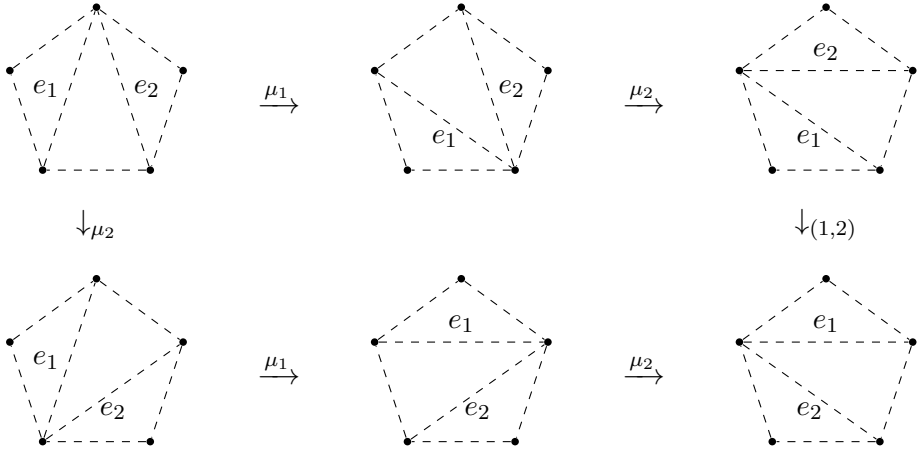

The properties \ref{it:bigon}, \ref{it:square}, \ref{it:pentagon} of Theorem~\ref{Th:flip cluster} in this case are equivalent to the relations among mutations given in Proposition~\ref{Prop:mutation relations}.
Furthermore, we see that the pentagon relation among mutations geometrically corresponds to the transformation of five triangulations of a pentagon, see Fig.~\ref{Fig:pentagon triang}.

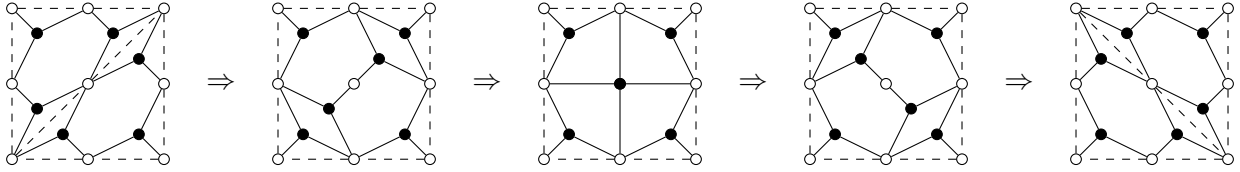
\begin{figure}[h]
	\centering
	\begin{tikzpicture}
		\def\xs{1}
		\def\ys{1}
		\def\NN{2}

%		spider for a plabic graph
		\begin{scope}[shift={(0,0)}]

			\node[circle, draw, inner sep=0pt, minimum size=4pt] (w00) at (0,0) {};
			\node[circle, draw, inner sep=0pt, minimum size=4pt] (w10) at (0,\ys) {};
			\node[circle, draw, inner sep=0pt, minimum size=4pt] (w20) at (0,\NN*\ys) {};
			\node[circle, draw, inner sep=0pt, minimum size=4pt] (w01) at (\xs,0) {};
			\node[circle, draw, inner sep=0pt, minimum size=4pt] (w11) at (\xs,\ys) {};
			\node[circle, draw, inner sep=0pt, minimum size=4pt] (w21) at (\xs,\NN*\ys) {};			
			\node[circle, draw, inner sep=0pt, minimum size=4pt] (w02) at (\NN*\xs,0) {};
			\node[circle, draw, inner sep=0pt, minimum size=4pt] (w12) at (2*\xs,\ys) {};
			\node[circle, draw, inner sep=0pt, minimum size=4pt] (w22) at (\NN*\xs,\NN*\ys) {};

			\node[circle, draw, fill, inner sep=0pt, minimum size=4pt] (b1) at (0.33*\xs,0.67*\ys) {};
			\node[circle, draw, fill, inner sep=0pt, minimum size=4pt] (b2) at (0.33*\xs,1.67*\ys) {};
			\node[circle, draw, fill, inner sep=0pt, minimum size=4pt] (b3) at (1.33*\xs,1.67*\ys) {};
			
			\node[circle, draw, fill, inner sep=0pt, minimum size=4pt] (b4) at (0.67*\xs,0.33*\ys) {};
			\node[circle, draw, fill, inner sep=0pt, minimum size=4pt] (b5) at (1.67*\xs,0.33*\ys) {};
			\node[circle, draw, fill, inner sep=0pt, minimum size=4pt] (b6) at (1.67*\xs,1.33*\ys) {};
			
			\draw (b1) to (w00);
			\draw (b1) to (w10);
			\draw (b1) to (w11);
			\draw (b2) to (w20);
			\draw (b2) to (w21);
			\draw (b2) to (w10);
			\draw (b3) to (w21);
			\draw (b3) to (w22);
			\draw (b3) to (w11);
			
			\draw (b4) to (w00);
			\draw (b4) to (w01);
			\draw (b4) to (w11);
			\draw (b5) to (w01);
			\draw (b5) to (w02);
			\draw (b5) to (w12);
			\draw (b6) to (w11);
			\draw (b6) to (w12);
			\draw (b6) to (w22);
			
			\draw[dashed] (w00) to (w10) to  (w20) to (w21) to (w22) to(w12) to (w02) to (w01) to (w00) to (w11) to (w22);

		\end{scope}
		
		\node at (\NN*\xs+0.75,0.5*\NN*\ys) {$\Rightarrow$};
		
		\begin{scope}[shift={(1*\NN*\xs+1.5,0)}]
			
			\node[circle, draw, inner sep=0pt, minimum size=4pt] (w00) at (0,0) {};
			\node[circle, draw, inner sep=0pt, minimum size=4pt] (w10) at (0,\ys) {};
			\node[circle, draw, inner sep=0pt, minimum size=4pt] (w20) at (0,\NN*\ys) {};
			\node[circle, draw, inner sep=0pt, minimum size=4pt] (w01) at (\xs,0) {};
			\node[circle, draw, inner sep=0pt, minimum size=4pt] (w11) at (\xs,\ys) {};
			\node[circle, draw, inner sep=0pt, minimum size=4pt] (w21) at (\xs,\NN*\ys) {};			
			\node[circle, draw, inner sep=0pt, minimum size=4pt] (w02) at (\NN*\xs,0) {};
			\node[circle, draw, inner sep=0pt, minimum size=4pt] (w12) at (2*\xs,\ys) {};
			\node[circle, draw, inner sep=0pt, minimum size=4pt] (w22) at (\NN*\xs,\NN*\ys) {};

			\node[circle, draw, fill, inner sep=0pt, minimum size=4pt] (b1) at (0.33*\xs,0.33*\ys) {};
			\node[circle, draw, fill, inner sep=0pt, minimum size=4pt] (b2) at (0.33*\xs,1.67*\ys) {};
			\node[circle, draw, fill, inner sep=0pt, minimum size=4pt] (b3) at (1.33*\xs,1.33*\ys) {};
			
			\node[circle, draw, fill, inner sep=0pt, minimum size=4pt] (b4) at (0.67*\xs,0.67*\ys) {};
			\node[circle, draw, fill, inner sep=0pt, minimum size=4pt] (b5) at (1.67*\xs,0.33*\ys) {};
			\node[circle, draw, fill, inner sep=0pt, minimum size=4pt] (b6) at (1.67*\xs,1.67*\ys) {};

			\draw (b1) to (w00);
			\draw (b1) to (w01);
			\draw (b1) to (w10);
			\draw (b2) to (w20);
			\draw (b2) to (w21);
			\draw (b2) to (w10);
			\draw (b3) to (w11);
			\draw (b3) to (w21);
			\draw (b3) to (w12);
			
			\draw (b4) to (w11);
			\draw (b4) to (w01);
			\draw (b4) to (w10);
			\draw (b5) to (w01);
			\draw (b5) to (w02);
			\draw (b5) to (w12);
			\draw (b6) to (w22);
			\draw (b6) to (w21);
			\draw (b6) to (w12);			
			
			\draw[dashed] (w00) to (w10) to  (w20) to (w21) to (w22) to(w12) to (w02) to (w01) to (w00);

		\end{scope}
		\node at (2*\NN*\xs+2.25,0.5*\NN*\ys) {$\Rightarrow$};
		
		\begin{scope}[shift={(2*\NN*\xs+3,0)}]
			
			\node[circle, draw, inner sep=0pt, minimum size=4pt] (w00) at (0,0) {};
			\node[circle, draw, inner sep=0pt, minimum size=4pt] (w10) at (0,\ys) {};
			\node[circle, draw, inner sep=0pt, minimum size=4pt] (w20) at (0,\NN*\ys) {};
			\node[circle, draw, inner sep=0pt, minimum size=4pt] (w01) at (\xs,0) {};
			\node[circle, draw, inner sep=0pt, minimum size=4pt] (w11) at (\xs,\ys) {};
			\node[circle, draw, inner sep=0pt, minimum size=4pt] (w21) at (\xs,\NN*\ys) {};			
			\node[circle, draw, inner sep=0pt, minimum size=4pt] (w02) at (\NN*\xs,0) {};
			\node[circle, draw, inner sep=0pt, minimum size=4pt] (w12) at (2*\xs,\ys) {};
			\node[circle, draw, inner sep=0pt, minimum size=4pt] (w22) at (\NN*\xs,\NN*\ys) {};

			\node[circle, draw, fill, inner sep=0pt, minimum size=4pt] (b1) at (0.33*\xs,0.33*\ys) {};
			\node[circle, draw, fill, inner sep=0pt, minimum size=4pt] (b2) at (0.33*\xs,1.67*\ys) {};

			\node[circle, draw, fill, inner sep=0pt, minimum size=4pt] (b) at (1*\xs,1*\ys) {};
			
			\node[circle, draw, fill, inner sep=0pt, minimum size=4pt] (b5) at (1.67*\xs,0.33*\ys) {};
			\node[circle, draw, fill, inner sep=0pt, minimum size=4pt] (b6) at (1.67*\xs,1.67*\ys) {};

			\draw (b1) to (w00);
			\draw (b1) to (w01);
			\draw (b1) to (w10);
			\draw (b2) to (w20);
			\draw (b2) to (w21);
			\draw (b2) to (w10);
			
			\draw (b) to (w21);
			\draw (b) to (w12);
			
			\draw (b) to (w01);
			\draw (b) to (w10);
	
			\draw (b5) to (w01);
			\draw (b5) to (w02);
			\draw (b5) to (w12);
			\draw (b6) to (w22);
			\draw (b6) to (w21);
			\draw (b6) to (w12);			
			
			\draw[dashed] (w00) to (w10) to  (w20) to (w21) to (w22) to(w12) to (w02) to (w01) to (w00);

		\end{scope}

		\node at (3*\NN*\xs+3.75,0.5*\NN*\ys) {$\Rightarrow$};
		
		\begin{scope}[shift={(3*\NN*\xs+4.5,0)}]
			
			\node[circle, draw, inner sep=0pt, minimum size=4pt] (w00) at (0,0) {};
			\node[circle, draw, inner sep=0pt, minimum size=4pt] (w10) at (0,\ys) {};
			\node[circle, draw, inner sep=0pt, minimum size=4pt] (w20) at (0,\NN*\ys) {};
			\node[circle, draw, inner sep=0pt, minimum size=4pt] (w01) at (\xs,0) {};
			\node[circle, draw, inner sep=0pt, minimum size=4pt] (w11) at (\xs,\ys) {};
			\node[circle, draw, inner sep=0pt, minimum size=4pt] (w21) at (\xs,\NN*\ys) {};			
			\node[circle, draw, inner sep=0pt, minimum size=4pt] (w02) at (\NN*\xs,0) {};
			\node[circle, draw, inner sep=0pt, minimum size=4pt] (w12) at (2*\xs,\ys) {};
			\node[circle, draw, inner sep=0pt, minimum size=4pt] (w22) at (\NN*\xs,\NN*\ys) {};

			\node[circle, draw, fill, inner sep=0pt, minimum size=4pt] (b1) at (0.33*\xs,0.33*\ys) {};
			\node[circle, draw, fill, inner sep=0pt, minimum size=4pt] (b2) at (0.33*\xs,1.67*\ys) {};
			\node[circle, draw, fill, inner sep=0pt, minimum size=4pt] (b3) at (1.33*\xs,0.67*\ys) {};
			
			\node[circle, draw, fill, inner sep=0pt, minimum size=4pt] (b4) at (0.67*\xs,1.33*\ys) {};
			\node[circle, draw, fill, inner sep=0pt, minimum size=4pt] (b5) at (1.67*\xs,0.33*\ys) {};
			\node[circle, draw, fill, inner sep=0pt, minimum size=4pt] (b6) at (1.67*\xs,1.67*\ys) {};

			\draw (b1) to (w00);
			\draw (b1) to (w01);
			\draw (b1) to (w10);
			\draw (b2) to (w20);
			\draw (b2) to (w21);
			\draw (b2) to (w10);
			\draw (b3) to (w11);
			\draw (b3) to (w01);
			\draw (b3) to (w12);
			
			\draw (b4) to (w11);
			\draw (b4) to (w21);
			\draw (b4) to (w10);
			\draw (b5) to (w01);
			\draw (b5) to (w02);
			\draw (b5) to (w12);
			\draw (b6) to (w22);
			\draw (b6) to (w21);
			\draw (b6) to (w12);			
			
			\draw[dashed] (w00) to (w10) to  (w20) to (w21) to (w22) to(w12) to (w02) to (w01) to (w00);

		\end{scope}

		\node at (4*\NN*\xs+5.25,0.5*\NN*\ys) {$\Rightarrow$};
		
		\begin{scope}[shift={(4*\NN*\xs+6,0)}]

			\node[circle, draw, inner sep=0pt, minimum size=4pt] (w00) at (0,0) {};
			\node[circle, draw, inner sep=0pt, minimum size=4pt] (w10) at (0,\ys) {};
			\node[circle, draw, inner sep=0pt, minimum size=4pt] (w20) at (0,\NN*\ys) {};
			\node[circle, draw, inner sep=0pt, minimum size=4pt] (w01) at (\xs,0) {};
			\node[circle, draw, inner sep=0pt, minimum size=4pt] (w11) at (\xs,\ys) {};
			\node[circle, draw, inner sep=0pt, minimum size=4pt] (w21) at (\xs,\NN*\ys) {};			
			\node[circle, draw, inner sep=0pt, minimum size=4pt] (w02) at (\NN*\xs,0) {};
			\node[circle, draw, inner sep=0pt, minimum size=4pt] (w12) at (2*\xs,\ys) {};
			\node[circle, draw, inner sep=0pt, minimum size=4pt] (w22) at (\NN*\xs,\NN*\ys) {};

			\node[circle, draw, fill, inner sep=0pt, minimum size=4pt] (b1) at (0.33*\xs,0.33*\ys) {};
			\node[circle, draw, fill, inner sep=0pt, minimum size=4pt] (b2) at (1.33*\xs,0.33*\ys) {};
			\node[circle, draw, fill, inner sep=0pt, minimum size=4pt] (b3) at (0.67*\xs,1.67*\ys) {};
			
			\node[circle, draw, fill, inner sep=0pt, minimum size=4pt] (b4) at (0.33*\xs,1.33*\ys) {};
			\node[circle, draw, fill, inner sep=0pt, minimum size=4pt] (b5) at (1.67*\xs,0.67*\ys) {};
			\node[circle, draw, fill, inner sep=0pt, minimum size=4pt] (b6) at (1.67*\xs,1.67*\ys) {};

			\draw (b1) to (w00);
			\draw (b1) to (w01);
			\draw (b1) to (w10);
			\draw (b2) to (w01);
			\draw (b2) to (w02);
			\draw (b2) to (w11);
			\draw (b3) to (w11);
			\draw (b3) to (w20);
			\draw (b3) to (w21);
			
			\draw (b4) to (w11);
			\draw (b4) to (w20);
			\draw (b4) to (w10);
			\draw (b5) to (w11);
			\draw (b5) to (w02);
			\draw (b5) to (w12);
			\draw (b6) to (w22);
			\draw (b6) to (w21);
			\draw (b6) to (w12);			
			
			\draw[dashed]   (w20) to (w21) to (w22) to(w12) to (w02) to (w01) to (w00)  to (w10) to (w20) to (w11) to (w02); 	
			
		\end{scope}
		
	\end{tikzpicture}
	\caption{\label{Fig:flip N=3} Flip of triangulation as a sequence of spider moves, \(N=3\)}
\end{figure}

In Fig.~\ref{Fig:flip N=3} we presented a sequence of spider moves, concatenations, and unconcatenations that gives \(\boldsymbol{\mu}_e\) for \(N=3\).
It is not difficult to guess its generalization for higher \(N\).

Recall that the \emph{mapping class group} of $\Sigma$ is the group of homotopy classes of diffeomorphisms of the surface $\Sigma$.
It was shown in \cite[Prop 0.1]{Fock:1999} that the mapping class groupoid is generated in some sense by flips of triangulations and automorphisms of triangulations.
The relations between such morphisms are as follows: a flip is an involution, flips of edges that do not share the same triangle commute, and flips of edges that belong to the same triangle satisfy pentagon relations.
Therefore, it follows from properties \ref{it:bigon}, \ref{it:square}, \ref{it:pentagon} of Theorem~\ref{Th:flip cluster} that the cluster structure of \(\mathcal{P}_{G, \Sigma}\) is invariant under the action of the mapping class group.
In other words, the elements of the mapping class group can be realized by sequences of mutations and permutations (i.e., belong to the group \(G_\mathcal{Q}\)).

\begin{Example}
	Consider \(\Sigma\) to be an annulus with one marked point on each boundary.
	The mapping class group of \(\Sigma\) is generated by a \emph{Dehn twist} that rotates the internal circle by \(360^\circ\), preserving the external circle.
%	This transformation can be realized as follows: cut the annulus into two annuli by an auxiliary medium circle, rotate the internal one annulus by \(360^\circ\), and then glue back.
	
	In terms of triangulations, the Dehn twist in this case can be realized via a composition of a flip and a permutation.
	We depicted this in Fig.~\ref{Fig:Dehn}.
	We colored two edges orange and brown; in Fig.~\ref{Fig:Dehn} we first perform a flip at the brown edge and then swap the brown and orange colors.
	It is easy to see that the result is equivalent to the Dehn twist.
	
	\begin{figure}[h]
		\centering
		\begin{tikzpicture}
			\def\ys{0.5}
			\begin{scope}
				\node[circle, fill, inner sep=1pt, label=left:{}] (A) at (0,3*\ys) {};				
				\node[circle, fill, inner sep=1pt, label=left:{}] (B) at (0,\ys) {};
				
			    \draw[dashed] (0,0) circle[radius=3*\ys]; 
			    \draw[dashed] (0,0) circle[radius=\ys]; 
			    
			    \draw[thick,orange] (A) to (B);
			    \draw[thick,Brown] (A) to[out=225,in=180] (0,-2.5*\ys)  to[out=0, in=270]  (1.5*\ys,0) to[out=90,in=45] (B);
			\end{scope}
			
			\node at (3*\ys+0.75,0) {$\Rightarrow$};
			
			\begin{scope}[shift={(6*\ys+1.5,0)}]
				
				\node[circle, fill, inner sep=1pt, label=left:{}] (A) at (0,3*\ys) {};				
				\node[circle, fill, inner sep=1pt, label=left:{}] (B) at (0,\ys) {};
				
				\draw[dashed] (0,0) circle[radius=3*\ys]; 
				\draw[dashed] (0,0) circle[radius=\ys]; 
				
				\draw[thick,orange] (A) to (B);
				\draw[thick,Brown] (A) to[out=315,in=0] (0,-2.5*\ys)  to[out=180, in=270]  (-1.5*\ys,0) to[out=90,in=135] (B);
			\end{scope}

			\node at (9*\ys+2.25,0) {$\Rightarrow$};

			\begin{scope}[shift={(12*\ys+3,0)}]
				
				\node[circle, fill, inner sep=1pt, label=left:{}] (A) at (0,3*\ys) {};				
				\node[circle, fill, inner sep=1pt, label=left:{}] (B) at (0,\ys) {};
				
				\draw[dashed] (0,0) circle[radius=3*\ys]; 
				\draw[dashed] (0,0) circle[radius=\ys]; 
				
				\draw[thick,Brown] (A) to (B);
				\draw[thick,orange] (A) to[out=315,in=0] (0,-2.5*\ys)  to[out=180, in=270]  (-1.5*\ys,0) to[out=90,in=135] (B);
			\end{scope}

		\end{tikzpicture}
		\caption{\label{Fig:Dehn} Dehn twist as a composition of flip and permutation}
	\end{figure}
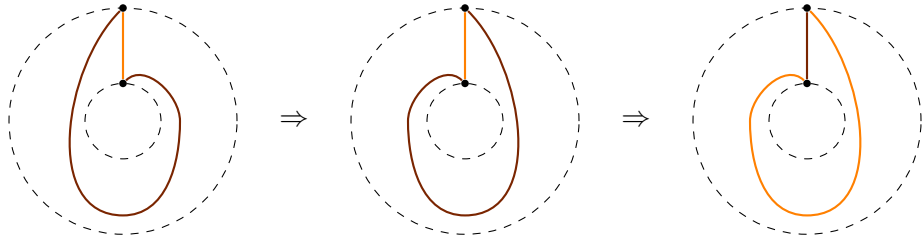	
\end{Example}

\begin{Example}
	Consider \(\Sigma\) to be a torus with one puncture.
	To any element \(g\) of the mapping class group, we can assign an element of \(SL(2,\mathbb{Z})\) by considering the action of \(g\) on \(H_1(\Sigma)\).
	In our (torus with one puncture) case, this is an isomorphism.
	The group \(SL(2,\mathbb{Z})\) is generated by the elements
	\begin{equation}
		S=\begin{pmatrix}
			0 & -1 \\ 1 & 0
		\end{pmatrix},\quad
		T=\begin{pmatrix}
			1 & 1 \\ 0 & 1
		\end{pmatrix}
	\end{equation}
	with relations \(S^4=1\), \((S T)^3=S^2\).
	The transformation \(T\) is a Dehn twist and \(S\) is a \(90^\circ\) rotation.
	
	Let us take \(N=2\).
	We have a triangulation of \(\Sigma\) that is obtained from the triangulation of a square by gluing two pairs of edges.
	Then the corresponding quiver has 3 vertices and double arrows between them, see Fig.~\ref{Fig:torus N=2}.
	This quiver is called the Markov quiver.
	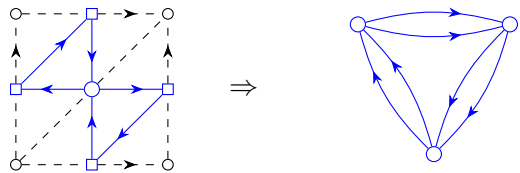
\begin{figure}[h]
		\centering
		\begin{tikzpicture}
			\def\xs{2}
			\def\ys{2}
						
			\begin{scope}[shift={(0,0)}]

				\node[circle, draw, inner sep=0pt, minimum size=4pt] (w00) at (0,0)  {};
				\node[circle, draw, inner sep=0pt, minimum size=4pt] (w10) at (0,\ys) {};
				\node[circle, draw, inner sep=0pt, minimum size=4pt] (w11) at (\xs,\ys) {};
				\node[circle, draw, inner sep=0pt, minimum size=4pt] (w01) at (\xs,0) {};

				\node[circle,draw,inner sep=2, blue] (x) at (0.5*\xs,0.5*\ys) {};
				\node[rectangle,draw,inner sep=2, blue] (x1) at (0*\xs,0.5*\ys) {};
				\node[rectangle,draw,inner sep=2, blue] (x2) at (0.5*\xs,1*\ys) {};
				\node[rectangle,draw,inner sep=2, blue] (x3) at (1*\xs,0.5*\ys) {};
				\node[rectangle,draw,inner sep=2, blue] (x4) at (0.5*\xs,0*\ys) {};
				
				\draw[dashed, special arrow=0.8] (w00) to (x1) to (w10); 
				\draw[dashed, special arrow=0.8] (w10) to (x2) to (w11); 
				\draw[dashed, special arrow=0.8] (w01) to (x3) to (w11); 
				\draw[dashed, special arrow=0.8] (w00) to (x4) to (w01); 
				\draw[dashed] (w00) to (x) to (w11);

				\draw[blue,special arrow=0.7] (x) to (x1);
				\draw[blue,special arrow=0.7] (x2) to (x);
				\draw[blue,special arrow=0.7] (x) to (x3);
				\draw[blue,special arrow=0.7] (x4) to (x);
				\draw[blue,special arrow=0.7] (x1) to (x2);
				\draw[blue,special arrow=0.7] (x3) to (x4);
			\end{scope}
			
			\node at (\xs+1,0.5*\ys) {$\Rightarrow$};
			
			\begin{scope}[shift={(2*\xs+1.5,0.5*\ys)}]
				\node[circle,draw,inner sep=2, blue] (x) at (0*\xs,-0.43*\ys) {};
				\node[circle,draw,inner sep=2, blue] (x1) at (-0.5*\xs,0.43*\ys) {};
				\node[circle,draw,inner sep=2, blue] (x2) at (0.5*\xs,0.43*\ys) {};

				\draw[blue,special arrow=0.7] (x) to[bend left=15] (x1);
				\draw[blue,special arrow=0.7] (x) to[bend right=15] (x1);
				\draw[blue,special arrow=0.7] (x1) to[bend left=15] (x2);
				\draw[blue,special arrow=0.7] (x1) to[bend right=15] (x2);
				\draw[blue,special arrow=0.7] (x2) to[bend left=15] (x);
				\draw[blue,special arrow=0.7] (x2) to[bend right=15] (x);

			\end{scope}	
		\end{tikzpicture}
		\caption{\label{Fig:torus N=2} Triangulation of the torus and corresponding quiver for \(N=2\)}
	\end{figure}
	
	The natural automorphisms of the Markov quiver are given by \(120^\circ\) rotations and mutation at one vertex composed with transposition.
	These cluster transformations generate the group \(PSL(2,\mathbb{Z})\), which has the relations \(S^2=(ST)^3=1\).
	The latter cluster transformation can be identified with the Dehn twist \(T\), while the former has order \(3\) and corresponds to the generator \(ST\).
\end{Example}

\subsection{Poisson structure} The cluster structure ensures a Poisson structure on \(\mathcal{P}_{G,\Sigma}\).
Since the seed is obtained as an amalgamation of the triangle seeds in Fig.~\ref{Fig:triangle}, it is natural to compute the Poisson brackets in that case first.

\begin{figure}[h]
	\centering
	\includegraphics[]{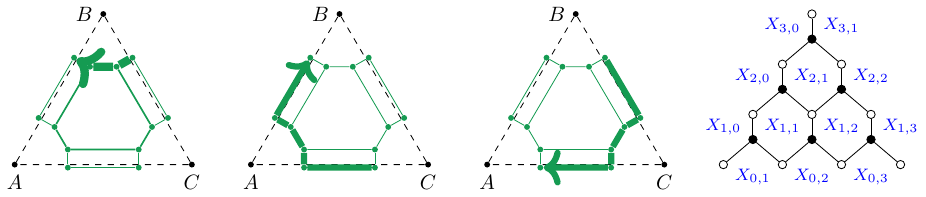}
	\caption{\label{Fig:monodromy triangle} Paths for \(T_{BC,BA}\), \(T_{CA,BA}\), \(T_{BC,AC}\) and plabic graphs for \(N=4\)}
\end{figure}

Consider paths in the rectangular-hexagonal graph depicted in Fig.~\ref{Fig:monodromy triangle}.
Let us denote the corresponding parallel transport matrices by \(T_{BC,BA}\), \(T_{CA,BA}\) and \(T_{BC,AC}\), respectively.

For clarity, let us also write formulas for these matrices in the case \(N=4\), using cluster variables as in Fig.~\ref{Fig:monodromy triangle}:
\begin{subequations}
	\begin{multline}
		T_{BC,BA}= \mathbb{L}_{w_0}(\mathbf{X})=H_3(X_{1,0})H_2(X_{2,0})H_1(X_{3,0})
		\\ 
		E_3E_2E_1  H_3(X_{1,2}) H_2(X_{1,1}) E_3E_2  H_3(X_{2,1}) E_3\; H_3(X_{1,3})H_2(X_{2,2})H_1(X_{3,1});
	\end{multline}
	\begin{multline}
		T_{CA,BA}= S^{-1}\mathbb{L}_{w_0}(\mathbf{X})^{-1}S=\mathbb{L}_{\overline{w_0}}(\mathbf{X})=H_3(X_{1,0})H_2(X_{2,0})H_1(X_{3,0})
		\\ 
		F_1F_2F_3  H_1(X_{2,1}) H_2(X_{1,1}) F_1F_2  H_1(X_{1,2}) F_1\; H_3(X_{0,1})H_2(X_{0,2})H_1(X_{0,3});
	\end{multline}
	\begin{multline}
		T_{BC,AC}= S^{-1}\mathbb{L}_{w_0}(\mathbf{X})^{-1}S=\mathbb{L}_{\overline{w_0}}(\mathbf{X})=H_3(X_{0,3})H_2(X_{0,2})H_1(X_{0,1})
		\\ 
		F_1F_2F_3  H_1(X_{1,1}) H_2(X_{1,2}) F_1F_2  H_1(X_{2,1}) F_1\; H_3(X_{1,3})H_2(X_{2,2})H_1(X_{3,1}).
	\end{multline}
\end{subequations}
Lemma~\ref{Lem:faces trivial} ensures that a certain product of these parallel transports is trivial
\begin{equation}
	T_{BC,AC}^{-1}\, S\, T_{CA,BA}^{-1}\,T_{BC,BA}=1.
\end{equation}
Note that the transport $T_{BC, BA}$ is upper triangular, while $T_{CA, BA}$ and $T_{BC, AC}$ are lower triangular.

In order to write Poisson brackets for these transport matrices, we need a modification of the \(r\)-matrix introduced in formula~\eqref{eq:r matrix}:
	\begin{equation}
	r^+
	=r+\frac12 \operatorname{P}_{\mathbb{C}^N\otimes\mathbb{C}^N}
	=\sum\nolimits_{a<b} E_{a,b}\otimes E_{b,a}  + \frac12 \sum\nolimits_{a} E_{a,a}\otimes E_{a,a},
\end{equation}
where \(\operatorname{P}_{\mathbb{C}^N\otimes\mathbb{C}^N}\) is an operator that permutes factors.
The matrix \(r^+\) is not antisymmetric; on the other hand, it satisfies the ordinary, not modified, classical Yang--Baxter equation, that is, equation~\eqref{eq:modified classical YB} with \(c=0\).
It follows from the definition that
\begin{equation}
	[r^+,L_1+L_2]=[r,L_1+L_2], \qquad [r^+,L_1L_2]=[r,L_1L_2],
\end{equation}
i.e., in all computations above one can replace \(r\) by \(r^+\).
Let us denote
\begin{equation}
	\tilde{r}^+=r^+-\frac{1}{2N}\operatorname{Id}_{\mathbb{C}^N\otimes \mathbb{C}^N},
\end{equation}
where $\operatorname{Id}_{\mathbb{C}^N\otimes \mathbb{C}^N}$ denotes the identity operator acting on \(\mathbb{C}^N\otimes \mathbb{C}^N\).
The coefficient \(\frac{1}{2N}\) can be motivated by the property \(\operatorname{Tr}\tilde{r}^+=0\).

\begin{Theorem}\label{Th:transport triangle}
	\begin{enumerate}[label=(\alph*)]
		\item Each of the transport matrices \(T_{BC,BA}\), \(T_{CA,BA}\), \(T_{BC,AC}\) satisfies 
		\begin{equation} \label{eq:parallel transport 6 term}
			\{T_1,T_2\}=[r^+,T_1T_2]. 
		\end{equation}
		\item The Poisson bracket of the transport matrices with the same target has the form
		\begin{equation}\label{eq:parallel transport 5 term target}
			\{T_{BC,BA,1},T_{CA,BA,2}\}=\tilde{r}^+T_{BC,BA,1}T_{CA,BA,2}.
		\end{equation}
		\item The Poisson bracket of the transport matrices with the same source has the form
		\begin{equation}\label{eq:parallel transport 5 term source}
			\{T_{BC,BA,1},T_{BC,AC,2}\}=-T_{BC,BA,1}T_{BC,AC,2}\tilde{r}^+.
	\end{equation}
	\end{enumerate}
\end{Theorem}
The tensor notations here are \(T_{BC,BA,1}=T_{BC,BA}\otimes 1\), \(T_{CA,BA,2}=1\otimes T_{CA,BA}\), etc.

The formula~\eqref{eq:parallel transport 6 term} follows from Theorem \ref{Th:Double Bruhat cluster} and the observation that \(T_{BC,BA}\), \(T_{CA,BA}\), \(T_{BC,AC}\) have the form \(\mathbb{L}_w\) for \(w=w_0\) or \(w=\overline{w_0}\).
The formulas~\eqref{eq:parallel transport 5 term target}, \eqref{eq:parallel transport 5 term source} were proven in \cite[Sec. 2]{Chekhov:2020darboux}.

\begin{figure}[h]
	\centering
	\includegraphics[]{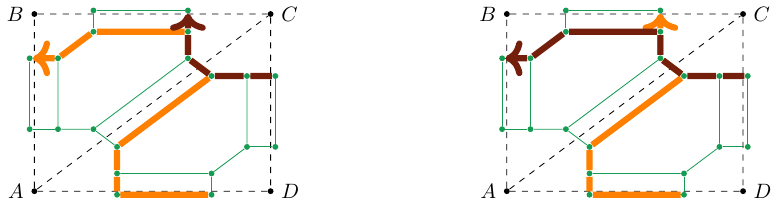}
	\caption{\label{Fig:square transport} On the left paths for \(T_{DA,BA}, T_{CD,CB}\), on the right paths for \(T_{CD,BA}, T_{DA,CB}\)}
\end{figure}

Using these formulas, we can compute the Poisson bracket for any two parallel transports.
Consider, for example, the rectangle $ABCD$.
In Fig.~\ref{Fig:square transport} we depicted paths corresponding to transports \(T_{DA,BA}, T_{CD,CB}\), \(T_{CD,BA}, T_{DA,CB}\).

Each of the matrices \(T_{DA, BA}, T_{CD, CB}\) can be written as a product of two elementary ones corresponding to simple triangles.
Cluster variables for the rectangle are obtained via the amalgamation of coordinates corresponding to triangles.
Cluster variables corresponding to different triangles Poisson commute.
Using formulas~\eqref{eq:parallel transport 5 term target} and~\eqref{eq:parallel transport 5 term source} we obtain (following \cite[Sec. 3]{Chekhov:2020darboux}):
\begin{multline}\label{eq: Poisson bracket T 1}
	\{T_{CD,CB,1},T_{DA,BA,2}\}=\{T_{CA,CB,1}T_{CD,CA,1},T_{CA,BA,2}T_{DA,CA,2}\}
	\\=-T_{CA,CB,1}T_{CA,BA,2} \tilde{r}^+T_{CD,CA,1}T_{DA,CA,2}+T_{CA,CB,1}T_{CA,BA,2} \tilde{r}^+T_{CD,CA,1}T_{DA,CA,2}= 0;
\end{multline}
The resulting vanishing agrees with the fact that paths $CD \to CB$ and $DA \to BA$ have no transverse intersection.
Another way to obtain this vanishing is to make a flip of the triangulation; then the transports \(T_{DA, BA}, T_{CD, CB}\) would depend on variables in different triangles ($ABD$ and $CDB$, respectively) and hence Poisson commute.
Here we used Theorem~\ref{Th:flip cluster} \ref{it:flip monodromy} that ensures that the cluster Poisson bracket between transports can be computed using any triangulation.

On the other hand, the paths $CD \to BA$ and $DA \to CB$ have a transverse intersection.
Using~\eqref{eq:parallel transport 5 term source}, \eqref{eq:parallel transport 5 term target} we obtain
\begin{multline}\label{eq: Poisson bracket T 2}
	\{T_{CD,BA,1},T_{DA,CB,2}\}=\{T_{CA,BA,1}T_{CD,CA,1},T_{CA,CB,2}T_{DA,CA,2}\}
	\\=T_{CA,BA,1}T_{CA,CB,2} (\tilde{r}^+)^tT_{CD,CA,1}T_{DA,CA,2}+T_{CA,BA,1}T_{CA,CB,2} \tilde{r}^+T_{CD,CA,1}T_{DA,CA,2}
	\\= T_{CA,BA,1}T_{CA,CB,2} \left(\operatorname{P}_{\mathbb{C}^N\otimes\mathbb{C}^N} -\frac{1}{N}\operatorname{Id}_{\mathbb{C}^N\otimes\mathbb{C}^N} \right)T_{CD,CA,1}T_{DA,CA,2}
	\\ 
	=T_{DA,BA,1}T_{CD,CB,2}  \operatorname{P}_{\mathbb{C}^N\otimes\mathbb{C}^N} -\frac1N T_{CD,BA,1}T_{DA,CB,2}.
\end{multline}
Here we used transposition of formula~\eqref{eq:parallel transport 5 term target}:
\begin{equation}
	\{T_{BC,AC,1},T_{BC,BA,2}\}=T_{BC,AC,1}T_{BC,BA,2}(\tilde{r}^+)^t.
\end{equation}

Recall that \(M_\gamma\) denotes monodromy along a closed path \(\gamma\) on a hexagonal-rectangular graph.
The trace of \(M_{\gamma}\) is a well-defined function on \(\mathcal{P}_{G,\Sigma}\) (for \(G=SL_N\)).
Using formulas~\eqref{eq: Poisson bracket T 1} and~\eqref{eq: Poisson bracket T 2}, we can compute Poisson brackets between such functions.
Namely, for any two paths \(\alpha,\beta\) we have
\begin{equation}\label{eq:P.b. trace}
	\{\operatorname{Tr}M_{\alpha},\operatorname{Tr}M_{\beta}\}=\sum_{p \in \alpha\cap \beta } \epsilon_{p;\alpha,\beta} \Big( \operatorname{Tr} M_{\alpha_p \beta_p } - \frac1N  \operatorname{Tr}M_{\alpha} \operatorname{Tr}M_{\beta}\Big)
\end{equation}
Here \(\epsilon_{p;\alpha,\beta} \) denotes the sign of intersection of paths \(\alpha\) and \(\beta\) at the point \(p\).
By \(\alpha_p \beta_p\) we denote a path that starts at the point \(p\), first goes along \(\alpha\) and then along \(\beta\).
We obtained the celebrated Goldman Poisson bracket on the moduli space of local systems \cite{Goldman:1986invariant}.
Moreover, the  Poisson bracket between transport matrices above gives (a version of) Fock--Rosly Poisson brackets \cite{Fock:1998poisson}.

For any puncture \(p\), let \(\gamma_p\) be a path that encircles \(p\) and no other punctures.
Let \(M_p\) be a monodromy along this path.
It follows from the formulas above that \(\operatorname{Tr}M_p^k\) Poisson commutes with any parallel transport, hence it is a Casimir function.
In other words, symmetric functions of the eigenvalues of \(M_p\) are Casimir functions.
Moreover, since our local systems are framed, the eigenvalues of \(M_p\) (or rather ratios of eigenvalues for \(PGL_N\)) are well defined on \(\mathcal{P}_{G, \Sigma}\) and are Casimir functions too.

This can also be seen using cluster coordinates.
Namely, for any puncture there are \(N-1\) cycles \(\mathsf{C}_1,\dots, \mathsf{C}_{N-1}\) around it, such that, for any vertex, the numbers of incoming and outgoing edges to each of these cycles are equal.
For each \(j\), the product of variables corresponding to vertices in \(\mathsf{C}_j\) is a Casimir function.

\begin{figure}[h]
	\centering
	\includegraphics[]{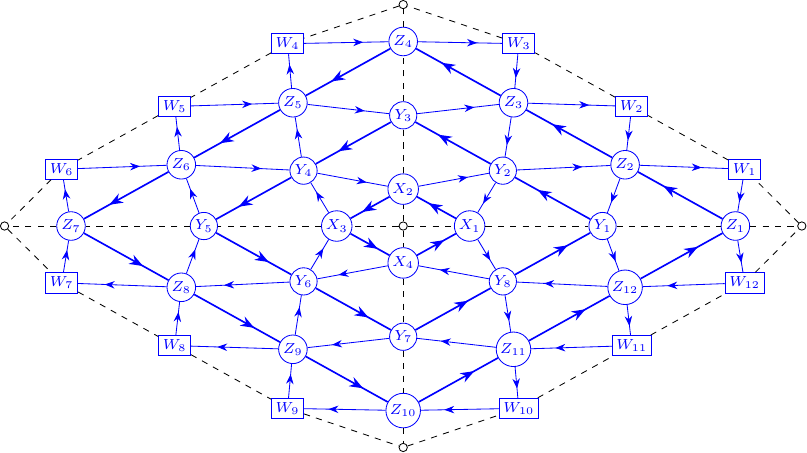}
	\caption{\label{Fig:Casimirs} Quiver for a puncture with four adjacent triangles, \(N=4\).}
\end{figure}

For example, take a puncture \(p\) adjacent to 4 triangles in the triangulation \(\mathcal{T}\).
The corresponding quiver is depicted in Fig.~\ref{Fig:Casimirs} for \(N=4\).
The cycles are
\begin{multline}
	\mathsf{C}_1=(X_1,X_2,X_3,X_4),\quad \mathsf{C}_2=(Y_1,Y_2,Y_3,Y_4,Y_5,Y_6,Y_7,Y_8),
	\\
	\mathsf{C}_3=(Z_1,Z_2,Z_3,Z_4,Z_5,Z_6,Z_7,Z_8,Z_9,Z_{10},Z_{11},Z_{12}).
\end{multline}
The corresponding Casimir functions are given by 
\begin{equation}
	C_1=\prod_{i=1}^4 X_i, \quad C_2=\prod_{i=1}^8 Y_i, \quad C_3=\prod_{i=1}^{12} Z_i.
\end{equation}
The monodromy around the path close to \(p\) is upper-triangular and has the form
\begin{equation}
	M_p=\begin{pmatrix}
		 C_1^{3/4} C_2^{1/2} C_3^{1/4}& * & * & *
		 \\
		 0 & C_1^{-1/4}C_2^{1/2}C_3^{1/4} & * &*
		 \\ 
		 0 & 0 & C_1^{-1/4} C_2^{-1/2} C_3^{1/4}& *
		 \\
		 0& 0& 0&  C_1^{-1/4} C_2^{-1/2} C_3^{-3/4}
	\end{pmatrix}.
\end{equation}
Hence the Casimir functions \(C_1, C_2, C_3\) are ratios of eigenvalues of the monodromy matrix \(M_p\).

%\bigskip 

\begin{figure}[h]
	\centering
	\includegraphics[]{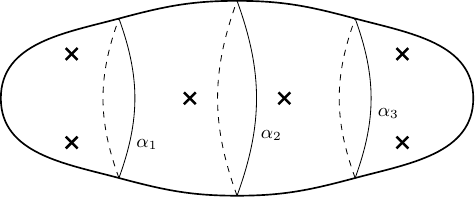}
	\caption{\label{Fig:sph 6 punt} Sphere with 6 punctures}
\end{figure}

Finally, let us return to the integrable systems.
It follows from formula~\eqref{eq:P.b. trace} that traces of monodromies along non-intersecting closed paths Poisson commute.
Consider the case where \(\Sigma\) is a sphere with \(m\) punctures.
In this case we can draw \(m-3\) closed curves \(\alpha_1,\dots,\alpha_{m-3}\) on \(\Sigma\) that are non-intersecting, represent different homology classes and do not encircle zero or one puncture (since in these cases \(\operatorname{Tr}M_{\alpha_j}\) is either trivial or a Casimir function).
If we cut \(\Sigma\) along these curves, we obtain a decomposition of \(\Sigma\) into \(m-2\) pairs of pants.
See Fig.~\ref{Fig:sph 6 punt} for the example with \(m=6\) punctures.

Let us assume that \(N=2\).
In this case, the cluster variables are assigned to the triangulation edges.
It is easy to see that a triangulation consists of \(2m-4\) triangles and has \(3m-6\) edges.
Therefore, \(\dim \mathcal{P}_{G,\Sigma}=3m-6\).
This can be also seen directly, similarly to the \(m=4\) case discussed in Example~\ref{Ex:sphere 4 puntures}.

On this \((3m-6)\)-dimensional variety, we have \(m\) Casimir functions assigned to the punctures.
Therefore, the dimension of a generic symplectic leaf is (not greater than) \(2m-6\).
The functions \(\operatorname{Tr}M_{\alpha_1},\dots, \operatorname{Tr}M_{\alpha_{m-3}}\) give a system of Poisson commuting functions.

Furthermore, one can find a seed in which the cluster quiver has the form shown in Fig.~\ref{Fig:quiver sphere m=6} (again for \(m=6\)):
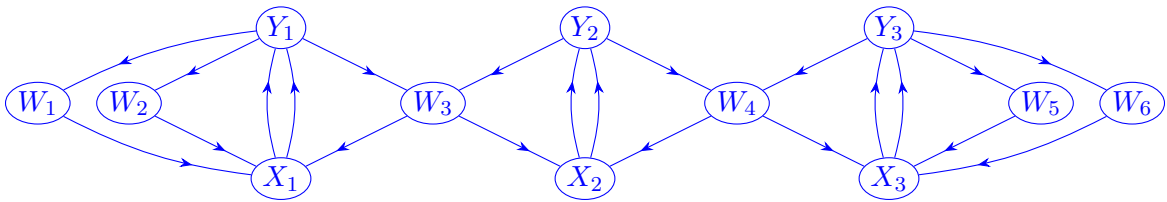
\begin{figure}[h]
	\centering 
	\begin{tikzpicture}
		\def\xs{2} 		
		\def\ys{1}	
		
		\foreach \i in {1,...,3}
		{
			\node[styleNode,blue] (X\i) at (2*\i*\xs,-1*\ys) {$X_\i$};
			\node[styleNode, blue] (Y\i) at (2*\i*\xs,1*\ys) {$Y_\i$};
			\draw[blue,special arrow=0.7] (X\i) to[bend right=15] (Y\i);
			\draw[blue,special arrow=0.7] (X\i) to[bend left=15] (Y\i);			
		}
		\foreach \i in {3,...,4}
		{
			\node[styleNode,blue] (W\i) at (2*\i*\xs-3*\xs,0*\ys) {$W_\i$};
			\pgfmathtruncatemacro{\iminustwo}{\i-2}
			\pgfmathtruncatemacro{\iminusone}{\i-1}
			\draw[blue,special arrow=0.7] (Y\iminustwo) to (W\i);
			\draw[blue,special arrow=0.7] (Y\iminusone) to (W\i);
			\draw[blue,special arrow=0.7] (W\i) to (X\iminustwo);			
			\draw[blue,special arrow=0.7] (W\i) to (X\iminusone);
		}
		\node[styleNode,blue] (W1) at (0.4*\xs,0*\ys) {$W_1$};
		\node[styleNode,blue] (W2) at (1*\xs,0*\ys) {$W_2$};
		\draw[blue,special arrow=0.7] (W1) to[bend right=10] (X1);			
		\draw[blue,special arrow=0.7] (W2) to (X1);			
		\draw[blue,special arrow=0.7] (Y1) to[bend right=10] (W1);			
		\draw[blue,special arrow=0.7] (Y1) to (W2);			
		
		\node[styleNode,blue] (W6) at (7.6*\xs,0*\ys) {$W_6$};
		\node[styleNode,blue] (W5) at (7*\xs,0*\ys) {$W_5$};
		\draw[blue,special arrow=0.7] (W6) to[bend left=10] (X3);			
		\draw[blue,special arrow=0.7] (W5) to (X3);			
		\draw[blue,special arrow=0.7] (Y3) to[bend left=10] (W6);			
		\draw[blue,special arrow=0.7] (Y3) to (W5);			
		
	\end{tikzpicture}
	\caption{\label{Fig:quiver sphere m=6} Quiver for sphere with 6 punctures, \(N=2\)}
\end{figure}
and \(\operatorname{Tr}M_{\alpha_i}=H_{\text{Toda}}(X_i,Y_i)\), where \(H_{\text{Toda}}\) is the \(SL_2\) Toda Hamiltonian.
Therefore, the functions \(\operatorname{Tr}M_{\alpha_i}\), \(i=1,\dots,m-3\), define an integrable system on \(\mathcal{P}_{G,\Sigma}\) where \(G=SL_2\) and \(\Sigma\) is a sphere with \(m\) punctures.

For higher rank \(N>2\), this construction gives a so-called superintegrable system.

\begin{Remark} \label{Rem:GMN}
	There is a close connection between cluster varieties \(\mathcal{P}_{G,\Sigma}, \mathcal{X}_{G,\Sigma}\) and 4d supersymmetric theories, namely so-called class \(S\) theories \cite{Gaiotto:2013wall}.
	Cf.~Remark~\ref{Rem:4d Coulomb} above.
\end{Remark}

\section{Loop groups and Goncharov--Kenyon integrable systems}\label{Sec:GK}
We follow \cite{Fock:2016},\cite{Goncharov:2013} in this section (see also the exposition in \cite[Sec. 1,2]{Bocklandt:2016dimer} and \cite[Sec. 2]{Bershtein:2024cluster}).
Our goal is to replace the constructions in Sec.~\ref{Sec:PL groups} and Sec.~\ref{Sec:Rel Toda} by their affine analogues.

\subsection{Bruhat cells in loop group} 
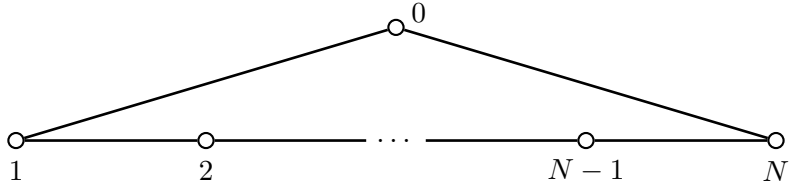
\begin{figure}[h]
	\centering
	\begin{tikzpicture}[elt/.style={circle,draw=black!100,thick, inner sep=0pt,minimum size=2mm}]
			\def\xs{2.5} 		
			\def\ys{1.5}

			\path (0,\ys) 	node 	(a0) [elt] {}
			(-2*\xs,0) 	node 	(a1) [elt] {}
			(-\xs,0) 	node 	(a2) [elt] {}
			(0,0)	node 	(a3)  {$\dots$}
			(\xs,0) 	node 	(a4) [elt] {}
			(2*\xs,0) 	node 	(a5) [elt] {};
			
			\draw [black,line width=1pt] (a0) -- (a1) -- (a2) -- (a3); 
			\draw [black,line width=1pt] (a3) -- (a4) -- (a5) -- (a0);
			
			\node at (0.3,\ys+0.2) 	{$0$};
			\node at  (-2*\xs,-0.4) {$1$};
			\node at  (-\xs,-0.4)  {$2$};
			\node at (\xs,-0.4)  {$N-2$};
			\node at (2*\xs,-0.4)   {$N-1$};	
			
	\end{tikzpicture}
	\caption{\label{Fig:Dynkin} Dynkin diagram for \(\widehat{PGL}^{\sharp}_N\)}
\end{figure}

Consider a coextended loop group \(\widehat{PGL}^{\sharp}_N\).
This group can be realized as a group of expressions of the form \(A(\lambda)T_X\), where \(A(\lambda)\) is a Laurent polynomial with values in \(PGL_N\) (for \(\lambda\neq 0\)) and \(T_X\) is a multiplicative shift by \(X\)
\begin{equation}
	T_X=\exp\Big( \lambda \partial_\lambda  \log(X)\Big)=
	X^{\lambda \partial_\lambda}.
\end{equation}
The multiplication between such expressions has the form 
\begin{equation}
	A_1(\lambda)T_{X_1}A_2(\lambda)T_{X_2}=A_1(\lambda)A_2(X_1\lambda)T_{X_1X_2}.
\end{equation}
The Dynkin diagram for \(\widehat{PGL}^{\sharp}_N\) has the form of a cycle with \(N\) vertices, see Fig.~\ref{Fig:Dynkin}.

Let us now define analogs of the elementary matrices~\eqref{eq:EHF} in the affine setting:
\begin{subequations}\label{eq:EHF affine}
	\begin{align}
		&E_i=1+E_{i,i+1}=\exp(E_{i,i+1}),\; i=1,\dots,N-1, \qquad E_0= 1+\lambda E_{N,1}=\exp(\lambda E_{N,1}),
		\\ 
		& E_{\bar{i}}=1+E_{i+1,i}=\exp(E_{i+1,i}),\; i=1,\dots,N-1, \qquad E_{\bar{0}}= 1+\lambda^{-1} E_{1,N}=\exp(\lambda^{-1} E_{1,N}),
		\\
		&H_i(X) =\operatorname{diag}(\underbrace{X^{\frac{N-i}{N}},\dots, X^{\frac{N-i}{N}}}_{i}, \underbrace{X^{-\frac{i}{N}},\dots ,X^{-\frac{i}{N}}}_{N-i} ) T_X.
	\end{align}	
\end{subequations}
We can also use the notation \(F_i=E_{\bar{i}}\).
Note that \(H_0(X)=T_X\).
One can recognize the formulas for the affine root generators in the expressions for \(E_0\) and \(F_0\).
These generators satisfy the analog of relations~\eqref{eq:rel EHF} where now the indices are considered modulo \(N\).
In particular, the shift part in the definition of \(H_i\) was introduced in order to have
\begin{equation}
	H_i(X) E_0=E_0 H_i(X),\quad H_i(X) F_0=F_0 H_i(X),\qquad  i\neq 0.
\end{equation}

The matrix \(A(\lambda)\) with values in \(GL_N\) for \(\lambda \neq 0\) should have determinant of the form \(c \lambda^k\), where \(c \in \mathbb{C}^\times\) and \(k \in \mathbb{Z}\).
Since we are working with the group \(\widehat{PGL}^{\sharp}_N\), we can multiply \(A\) by scalar matrices.
Hence, only the residue \(k\) modulo \(N\) is invariant.
In other words, \(\widehat{PGL}^{\sharp}_N\) has \(N\) connected components parametrized by \(k \pmod N\).
In order to parameterize elements in all components, we introduce a matrix
\begin{equation}
	\Lambda=\sum_{i=1}^{N-1} E_{i,i+1}+\lambda E_{N,1}.
\end{equation}
It has the properties 
\begin{equation}\label{eq:Lambda_matrix_definition}
	\Lambda E_i \Lambda^{-1}=E_{i+1},\quad \Lambda F_i \Lambda^{-1}=F_{i+1},\quad \Lambda H_i(X) \Lambda^{-1}=H_{i+1}(X),\qquad i\in \mathbb{Z}/N\mathbb{Z}.
\end{equation}

The affine Weyl group \(W^{a}(A_{N-1})\) for \(\widehat{SL}_N\) is generated by \(s_0,\dots,s_{N-1}\) subject to braid relations.
Correspondingly, the double affine Weyl group \(W^{a}(A_{N-1}+A_{N-1})\) for \(\widehat{SL}_N\) is generated by \(s_0,\dots,s_{N-1}, \bar{s}_0, \dots, \bar{s}_{N-1}\) subject to braid relations~\eqref{eq:braid} where indices are considered in \(\mathbb{Z}/N\mathbb{Z}\).\footnote{Note that for \(N=2\) there is no relation between \(s_0\) and \(s_1\).
This is due to the fact that the affine Dynkin diagram for \(\widehat{SL}^{\sharp}_2\) is not simply laced.}
We also extend the affine Weyl group to \(W^{ae}(A_{N-1}+A_{N-1})\) by adding the generator \(\Lambda\) with relations
\begin{equation}
	\Lambda s_i \Lambda^{-1}=s_{i+1},\quad \Lambda \bar{s}_i \Lambda^{-1}=\bar{s}_{i+1}, \qquad i\in \mathbb{Z}/N\mathbb{Z}.
\end{equation}
In terms of the Dynkin diagram (see Fig.~\ref{Fig:Dynkin}), \(\Lambda\) corresponds to the automorphism given by rotation by \(2\pi/N\).
It follows from these relations that any element \(w \in W^{ae}(A_{N-1}+A_{N-1})\) can be written in the form \(w=s_{i_1}s_{i_2}\cdot\dots \cdot s_{i_l}\Lambda^k\).
Now we can give an analog of Definition~\ref{Def:factorization}.

\begin{Definition}
	For any reduced word \(w=s_{i_1}s_{i_2}\cdot\dots \cdot s_{i_l}\Lambda^k\), \(i_1,\dots, i_l \in \{0,\dots,N-1,\bar{0},\dots,\overline{N-1}\}\) we assign a product
	\begin{equation}\label{eq:factorization affine}
		\mathbb{L}_{\mathbf{s}}(\mathbf{X},\lambda)=H_1(X_1)\cdot \dots \cdot H_{N-1}(X_{N-1})H_0(X_N) E_{i_1} H_{i_1}(X_{N+1})E_{i_2} H_{i_2}(X_{N+2})\cdot \dots E_{i_l} H_{i_l}(X_{N+l})\Lambda^k.
	\end{equation}
\end{Definition}
Note that now the matrix \(\mathbb{L}\) depends on the spectral parameter. 

In order to construct an integrable system, we need an additional constraint on coordinates \(\mathbf{X}\), namely 
\begin{equation}\label{eq:q=1}
	\prod\nolimits_{i=1}^{N+l} X_i=1.
\end{equation}
This is a necessary and sufficient condition to cancel the multiplicative shift part in \(\mathbb{L}_{\mathbf{s}}(\mathbf{X},\lambda)\).
Under this condition we have \(\mathbb{L}_{\mathbf{s}}(\mathbf{X},\lambda)\in PGL_N[\lambda^{\pm 1}]\) and can define its characteristic polynomial
\begin{equation}\label{eq:Z=det}
	\mathcal{Z}(\mathbf{X}|\lambda,\mu)= \det (\mathbb{L}_{\mathbf{s}}(\mathbf{X},\lambda)+\mu).
\end{equation}
We will often suppress the dependence on variables \(\mathbf{X}\) and write simply $\mathcal{Z}(\lambda,\mu)$.
The coefficients of $\mathcal{Z}(\lambda,\mu)$ define the integrable system, see Theorem~\ref{Th:IS GK} below.

Geometrically, the polynomial \(\mathcal{Z}\) defines a complex curve \(\mathcal{C}=\{(\lambda,\mu)|\mathcal{Z}(\lambda,\mu)=0\}\subset \mathbb{C}^*\times \mathbb{C}^*\).
The compactification \(\bar{\mathcal C}\), obtained by filling in the punctures of \(\mathcal C\), is called the \emph{spectral curve}. The matrix \(\mathbb{L}(\lambda)\) is called the \emph{Lax matrix}.

Note that the definition of \(\mathcal{Z}\) as a characteristic polynomial of \(\mathbb{L}(\lambda)\) was in fact a bit vague, since \(\mathbb{L}(\lambda) \in PGL_N[\lambda^{\pm 1}]\).
The most essential freedom is given by conjugation by shift operators \(T_X\) that corresponds to rescaling the spectral parameter \(\lambda\mapsto \lambda X\).
We illustrate this issue in Examples~\ref{Ex:Toda affine SL2}, \ref{Ex:Toda affine SL3} below.
Before discussing them, we would like to construct plabic graphs and a cluster structure similarly to the non-affine case above.

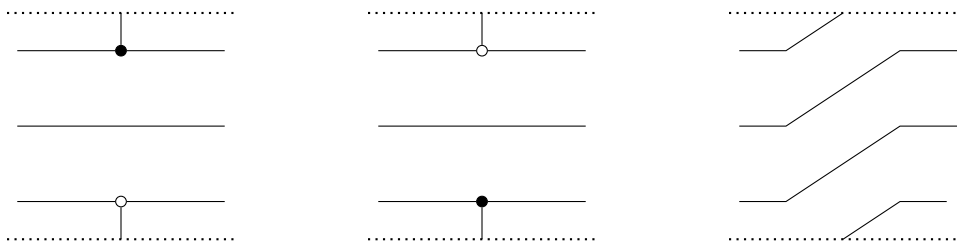
\begin{figure}[h]
	\centering
	\begin{tikzpicture}
		\def\xs{1.5}
		\def\ys{1}

		\begin{scope}
			\node[circle, draw, fill=black, inner sep=0pt, minimum size=4pt] (d1) at (0,-1*\ys) {};
			
			\node[circle, draw,  inner sep=0pt, minimum size=4pt] (u1) at (0,-3*\ys) {};

			\draw (d1) -- + (0,0.5*\ys);
			\draw (u1) -- + (0,-0.5*\ys);
			
			%			sources and targets			
			\node (tau3) at (-1*\xs,-3*\ys) {};	
			\node (tau2) at (-1*\xs,-2*\ys) {};			
			\node (tau1) at (-1*\xs,-\ys) {};			
			\node (sigma3) at (1*\xs,-3*\ys) {};	
			\node (sigma2) at (1*\xs,-2*\ys) {};	
			\node (sigma1) at (1*\xs,-\ys) {};
			
			\draw (sigma1) to (d1) to (tau1);
			\draw (sigma2)  to (tau2);
			\draw (sigma3) to (u1) to (tau3);
			
			%			boundary
			\draw[thick, dotted] (-1*\xs,-0.5*\ys) -- (1*\xs,-0.5*\ys);
			\draw[thick, dotted] (-1*\xs,-3.5*\ys) -- (1*\xs,-3.5*\ys);
		\end{scope}
		
		\begin{scope}[shift={(2.5*\xs+1,0)}]
			\node[circle, draw,  inner sep=0pt, minimum size=4pt] (d1) at (0,-1*\ys) {};
			
			\node[circle, draw, fill=black, inner sep=0pt, minimum size=4pt] (u1) at (0,-3*\ys) {};

			\draw (d1) -- + (0,0.5*\ys);
			\draw (u1) -- + (0,-0.5*\ys);
			
%			sources and targets			
			\node (tau3) at (-1*\xs,-3*\ys) {};	
			\node (tau2) at (-1*\xs,-2*\ys) {};			
			\node (tau1) at (-1*\xs,-\ys) {};			
			\node (sigma3) at (1*\xs,-3*\ys) {};	
			\node (sigma2) at (1*\xs,-2*\ys) {};	
			\node (sigma1) at (1*\xs,-\ys) {};
			
			\draw (sigma1) to (d1) to (tau1);
			\draw (sigma2)  to (tau2);
			\draw (sigma3) to (u1) to (tau3);
			
%			boundary
			\draw[thick, dotted] (-1*\xs,-0.5*\ys) -- (1*\xs,-0.5*\ys);
			\draw[thick, dotted] (-1*\xs,-3.5*\ys) -- (1*\xs,-3.5*\ys);
		\end{scope}
		
		\begin{scope}[shift={(5.0*\xs+2,0)}]
			
			%			sources and targets			
			\node (tau3) at (-1*\xs,-3*\ys) {};	
			\node (tau2) at (-1*\xs,-2*\ys) {};			
			\node (tau1) at (-1*\xs,-\ys) {};			
			\node (sigma3) at (1*\xs,-3*\ys) {};	
			\node (sigma2) at (1*\xs,-2*\ys) {};	
			\node (sigma1) at (1*\xs,-\ys) {};
			
			\draw (tau1) to ++ (0.5*\xs,0) to ++ (0.5*\xs,0.5*\ys);
			\draw (tau2) to ++ (0.5*\xs,0) to ++ (1*\xs,1*\ys) to ++ (0.5*\xs,0);
			\draw (tau3) to ++ (0.5*\xs,0) to ++ (1*\xs,1*\ys) to ++ (0.5*\xs,0);
			\draw (sigma3) to ++ (-0.5*\xs,0) to ++ (-0.5*\xs,-0.5*\ys);
			
			%			boundary
			\draw[thick, dotted] (-1*\xs,-0.5*\ys) -- (1*\xs,-0.5*\ys);
			\draw[thick, dotted] (-1*\xs,-3.5*\ys) -- (1*\xs,-3.5*\ys);
		\end{scope}

	\end{tikzpicture}		
	\caption{\label{Fig:s0 Lambda} Left to right: plabic graphs corresponding to \(s_0, \bar{s}_0, \Lambda\) for \(N=3\)}
\end{figure}

The plabic graph is defined similarly to Definition~\ref{Def:Gamma i}.
The vertices are drawn on \(N\) parallel horizontal lines that are now considered to belong to a cylinder.
This allows us to draw vertical edges connecting the top and bottom horizontal lines that correspond to \(s_0\) and \(\bar{s}_0\).
The generator \(\Lambda\) corresponds to the cyclic shift of the horizontal lines, see Fig.~\ref{Fig:s0 Lambda}.

The quiver is defined by a plabic graph via Definition~\ref{Def:quiver from plabic}.
Finally, the plabic graph and quiver corresponding to \(\widehat{PGL}^{\sharp}_N/\operatorname{Ad}H\) are obtained by amalgamation of the left and right boundaries.
As a result, we will get a plabic graph and a quiver drawn on a \emph{torus} \(\Sigma=\mathbb{T}^2\).

Recall Definition~\ref{Def:Coxeter} of the Coxeter element.
Let us consider the case of the Coxeter cell, i.e., the cell corresponding to \(w=c\bar{c}\), where now \(c\) is the Coxeter element of the affine Weyl group.
\begin{Example}	\label{Ex:Toda affine SL2}
	Let us take \(G=\widehat{PGL}^{\sharp}_2\) and \(w=\bar{s}_1s_1\bar{s}_0s_0\). 
	The Lax matrix is given by the formula
	\begin{equation}
		\mathbb{L}(\lambda)=H_1(X_1)H_0(X_0)F_1H_1(Y_1)E_1H_1(Z_1)F_0H_0(Y_0)E_0H_0(Z_0).
	\end{equation}
	The corresponding plabic graph \(\Gamma_{\mathbf{i}}\) and quiver are depicted on the left in Fig.~\ref{Fig:closed Toda N=2}.
	
	\begin{figure}[h]
		\centering
		\includegraphics[]{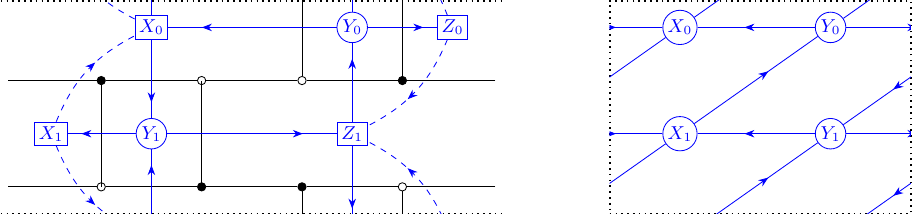}
		\caption{\label{Fig:closed Toda N=2} On the left: plabic graph and quiver on a cylinder corresponding to \(w=\bar{s}_1s_1\bar{s}_0s_0\); on the right: the quiver obtained by amalgamation, drawn on a torus.}
	\end{figure}
	
	The polynomial~\eqref{eq:Z=det} determining the spectral curve depends on \(\mathbb{L}(\lambda)\) up to conjugation.
	As in the finite case, we want to conjugate by \(\prod H_i(Z_i)^{-1}\) and then amalgamate \(X_i\) and \(Z_i\). The problem is that now it also contains the shift operator. Hence 
	\begin{equation}
		\det(\mathbb{L}(\lambda)+\mu) =\det(\mathbb{L}(\lambda)+\mu) |_{X_i\to X_i Z_i, Z_i\to 1, \lambda \to \lambda \prod_i Z_i^{-1}}.
	\end{equation}
	Note that now the spectral parameter \(\lambda\) depends on the dynamical variables. If we want to make it a Casimir, we need to rescale it further to compensate for the \(\prod_i Z_i\) term. 

	In our example, for the computation of the polynomial~\eqref{eq:Z=det}, we assume that \(Z_0=Z_1=1\) and also impose the integrability condition~\eqref{eq:q=1}. 
	Then we have a 3-dimensional phase space with local coordinates \(X_1, Y_1, Y_0\). 
	They can be considered as cluster coordinates for the quiver obtained by amalgamating the left and right boundaries, depicted in Fig.~\ref{Fig:closed Toda N=2} right. 
	Here the vertices corresponding to the products \(X_iZ_i\) are labeled simply by \(X_i\).
	
	We have
	\begin{equation}\label{eq:Z SL2 Toda}
		\mu^{-1}\det (\mathbb{L}_{\mathbf{s}}(\mathbf{X},\lambda)+\mu)|_{\lambda \mapsto \lambda X_0^{1/2}Y_0^{1/2}}=\mathcal{Z}(\lambda,\mu)=\lambda+\mu^{-1}+\mu+\lambda^{-1} C+H,
	\end{equation}
	where \(C=Y_0Y_1\) and
	\begin{equation}
		H=X_1^{-1/2}Y_1^{-1/2}+X_1^{-1/2}Y_1^{1/2}+X_1^{1/2}Y_1^{1/2}+X_1^{1/2}Y_1^{-1/2}Y_0Y_1.
	\end{equation}
	Note the rescaling of the spectral parameter in formula~\eqref{eq:Z SL2 Toda}; it compensates the \(\prod_i Z_i\) term. 
	Generally speaking, it is not unique: it can be shifted by any Casimir on cluster coordinates.

	Considering \(\mathcal{Z}(\lambda,\mu)\) as a polynomial in \(\lambda\) and \(\mu\), we see that its coefficients Poisson commute.
	More precisely, three of the coefficients are just equal to 1, one is equal to \(C\) and is a Casimir function (see the quiver on the right in Fig.~\ref{Fig:closed Toda N=2}).
	The most non-trivial coefficient \(H\) can be identified with the Hamiltonian of the closed relativistic \(SL_2\) Toda system~\cite{Marshakov:2013}.
\end{Example}

The following example will be our running example throughout this section.

\begin{Example}	\label{Ex:Toda affine SL3} 
	Let us take \(G=\widehat{PGL}^{\sharp}_3\) and \(w=\bar{s}_1s_1\bar{s}_2s_2\bar{s}_0s_0\). 
	The Lax matrix is given by the formula
	\begin{equation}\label{eq:L(lambda) Sl3}
		\mathbb{L}(\lambda)=H_1(X_1)H_2(X_2)H_0(X_0)F_1H_1(Y_1)E_1H_1(Z_1)F_2H_2(Y_2)E_2H_2(Z_2)F_0H_0(Y_0)E_0H_0(Z_0).
	\end{equation}
	The corresponding plabic graph \(\Gamma_{\mathbf{i}}\) and quiver are depicted on the left in Fig.~\ref{Fig:closed Toda N=3}.
	The amalgamated quiver is depicted on the right in Fig.~\ref{Fig:closed Toda N=3}.
	
	\begin{figure}[h]
		\centering
		\includegraphics[]{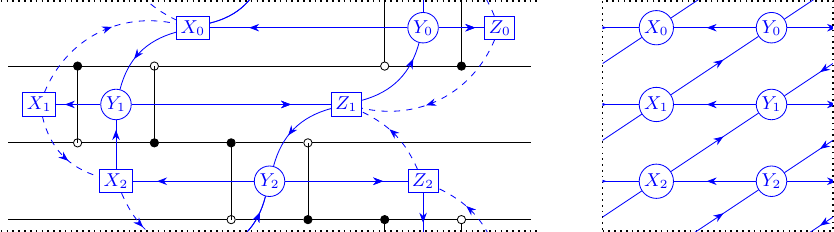}
		\caption{\label{Fig:closed Toda N=3} On the left: plabic graph and quiver on a cylinder corresponding to \(w=\bar{s}_1s_1\bar{s}_2s_2\bar{s}_0s_0\); on the right: the quiver obtained by amalgamation, drawn on a torus.}
	\end{figure}
	
	The phase space of the integrable system has coordinates \(X_0,X_1,X_2, Y_0,Y_1,Y_2\) subject to the constraint~\eqref{eq:q=1} given by \(\prod_{i=0}^2 X_iY_i=1\).
	The spectral curve equation has the form
		\begin{equation}\label{eq:Z SL3 Toda}
		\mu^{-1}\det (\mathbb{L}_{\mathbf{s}}(\mathbf{X},\lambda)+\mu)|_{\lambda \mapsto \lambda X_0^{2/3}X_2^{1/3}Y_0^{2/3}Y_2^{1/3}}=\mathcal{Z}(\lambda,\mu)=
		\lambda\mu+\mu^{-1}+\mu^2-\lambda^{-1} C+H_1+H_2\mu,
	\end{equation}
	where \(C=Y_0Y_1Y_2\) and
	\begin{subequations}
%		\begin{equation}
%			C=Y_1Y_2Y_3,
%		\end{equation}
		\begin{multline}
			H_1=
			C \left(X_1^2 X_2 Y_1^{-1} Y_2^{-2}\right)^{1/3} 
			+ \left(X_1^2 X_2 Y_1^2 Y_2\right)^{1/3} 
			+ \left(X_2 Y_1^2 Y_2 X_1^{-1}\right)^{1/3}
			\\ + \left(X_2 Y_2 X_1^{-1} Y_1^{-1}\right)^{1/3} 
			+ \left(Y_2 X_1^{-1} X_2^{-2} Y_1^{-1}\right)^{1/3} 
			+ \left(X_1^{-1} X_2^{-2} Y_1^{-1} Y_2^{-2}\right)^{1/3},
		\end{multline}
		\begin{multline}	
			H_2=	
			C \left(X_2^2 X_1 Y_1^{-2} Y_2^{-1}\right)^{1/3} 
			+ \left(X_2^2 X_1 Y_1 Y_2^2\right)^{1/3} 
			+ \left(X_1 Y_1 Y_2^2 X_2^{-1}\right)^{1/3}
			\\ + \left(X_1 Y_1 X_2^{-1} Y_2^{-1}\right)^{1/3} 
			+ \left(Y_1 X_1^{-2} X_2^{-1} Y_2^{-1}\right)^{1/3} 
			+ \left(X_1^{-2} X_2^{-1} Y_1^{-2} Y_2^{-1}\right)^{1/3}.
		\end{multline}
	\end{subequations}
	It is straightforward to check that the coefficients of \(\mathcal{Z}\) define an integrable system.
	Namely, \(C\) is a Casimir function, and \(H_1\) and \(H_2\) Poisson commute and serve as Hamiltonians.
	This is the closed \(SL_3\) Toda system.

	Note the rescaling of the spectral parameter \(\lambda\) in formula~\eqref{eq:Z SL3 Toda}. 
	As in the previous case, it compensates the \(\prod_i Z_i\) rescaling of \(\lambda\) arising from conjugation.
\end{Example}

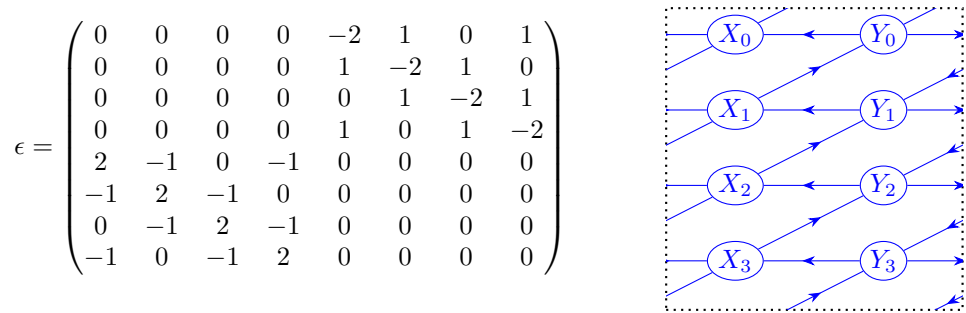
\begin{figure}[h]
	\begin{center}
		\begin{tikzpicture}[font=\small]
			\def\xs{1.3} 		
			\def\ys{1}	
			\def\NN{4} 
			\pgfmathtruncatemacro{\Nplusone}{\NN+1}
			\pgfmathtruncatemacro{\Nminusone}{\NN-1}
			
			\node at ($(-\Nplusone*\xs/2,-\Nplusone*\ys/2)$) {	\(\epsilon=\begin{pmatrix} 
					0 & 0 & 0 & 0 & -2 & 1 &0 &1
					\\
					0 & 0 & 0 & 0 & 1 & -2 &1 &0
					\\
					0 & 0 & 0 & 0 & 0 & 1 & -2 &1
					\\
					0 & 0 & 0 & 0 & 1 & 0 &1 &-2
					\\
					2 & -1 &0 &-1 & 0 & 0 & 0 & 0
					\\
					-1 & 2 &-1 &0 & 0 & 0 & 0 & 0
					\\
					0 & -1 & 2 &-1 & 0 & 0 & 0 & 0
					\\
					-1 & 0 &-1 &2 & 0 & 0 & 0 & 0					
				\end{pmatrix}\)	};
			\begin{scope}[shift={(2*\xs,-0.5*\NN*\ys+1.5*\ys)}]
				\draw[thick, dotted] (-0.7*\xs,-0.15*\ys-\NN*\ys) to ++ (3*\xs,0) to ++ (0,\NN*\ys) to ++ (-3*\xs,0) to ++ (0,-\NN*\ys);
				
				\begin{scope}
					\clip (-0.7*\xs,-0.15*\ys-\NN*\ys) to ++ (3*\xs,0) to ++ (0,\NN*\ys) to ++ (-3*\xs,0) to ++ (0,-\NN*\ys);
					
					\foreach \i in {-1,...,\Nplusone}
					{
						\node[styleNode,blue] (a\i) at (-1.5*\xs,-\i*\ys-0.5*\ys) {$A_{\i}$};
						\node[styleNode,blue] (x\i) at (0*\xs,-\i*\ys-0.5*\ys) {$X_{\i}$};
						\node[styleNode,blue] (y\i) at (1.5*\xs,-\i*\ys-0.5*\ys) {$Y_{\i}$};
						\node[styleNode,blue] (z\i) at (3*\xs,-\i*\ys-0.5*\ys) {$Z_{\i}$};
					}
					
					\foreach \i in {0,...,\Nplusone}
					{
						\pgfmathtruncatemacro{\iminusone}{\i-1}
						\draw[special arrow=0.6, blue] (y\i) to (x\i); 
						\draw[special arrow=0.6, blue] (y\i) to (z\i);
						\draw[special arrow=0.6, blue] (a\i) to (x\i);
						\draw[special arrow=0.6, blue] (x\iminusone) to (a\i);
						\draw[special arrow=0.6, blue] (z\iminusone) to (y\i);
						\draw[special arrow=0.6, blue] (x\i) to (y\iminusone);
					}
				\end{scope}			
				
			\end{scope}
			
			%vertices
%			\foreach \j in {1,...,\Nminusone}
%			{
%				\node[styleNode] (X\j) at(\xs,-\j*\ys){$X_{\j}$};
%				\node[styleNode] (Y\j) at(2*\xs,-\j*\ys){$Y_{\j}$};
%			}
%			\node[styleNode] (X\NN) at(\xs,-\NN*\ys){$X_{0}$};
%			\node[styleNode] (Y\NN) at(2*\xs,-\NN*\ys){$Y_{0}$};
%			
%			%arrows
%			\foreach \j in {2,...,\NN}
%			{
%				\pgfmathtruncatemacro{\jminusone}{\j-1}				\draw[styleArrow] (Y\j) to (X\j);
%				\draw[styleArrow] (X\jminusone) to (Y\j);
%				\draw[styleArrow] ($(X\j)-(0.5*\xs,0)$) to (X\j);
%				\draw[styleArrow](Y\j) to ($(Y\j)+(0.5*\xs,0)$) ;
%				\draw[styleArrow] ($(Y\jminusone)+(0.5*\xs,-0.5*\ys)$) to (Y\jminusone);
%				\draw[styleArrow] (X\j) to ($(X\j)+(-0.5*\xs,0.5*\ys)$);
%				
%			}
%			%arrows near boundary
%			\draw[styleArrow](Y1) to (X1);
%			\draw[styleArrow] ($(X1)-(0.5*\xs,0)$) to (X1);
%			\draw[styleArrow] (Y1) to ($(Y1)+(0.5*\xs,0)$);
%			\draw[styleArrow] (X1) to ($(X1)-(0.5*\xs,-0.5*\ys)$);
%			\draw[styleArrow] (X\NN) to ($(X\NN)+(0.5*\xs,-0.5*\ys)$);
%			\draw[styleArrow] ($(Y1)-(0.5*\xs,-0.5*\ys)$) to (Y1) ;
%			\draw[styleArrow] ($(Y\NN)+(0.5*\xs,-0.5*\ys)$) to (Y\NN);
%			
%			%boundary of torus
%			\draw[thick, dotted] (0.5*\xs,-0.5*\ys) -- ($(0.5*\xs,-\NN*\ys)+(0,-0.5*\ys)$)  -- ($(2.5*\xs,-\NN*\ys)+(0,-0.5*\ys)$) --
%			(2.5*\xs,-0.5*\ys) -- (0.5*\xs,-0.5*\ys);
		\end{tikzpicture}
		\caption{\label{Fig:Toda closed} On the left: the matrix \(\epsilon\); on the right: the quiver for the closed (periodic) Toda system for $\widehat{SL}_4$. The~quiver is drawn on a torus.}
	\end{center}
\end{figure}

The results of these two examples can be generalized to arbitrary \(N\).
The resulting quiver and its adjacency matrix are depicted in Fig.~\ref{Fig:Toda closed}.
Note that, similarly to the open case, the matrix has the form  \(\epsilon=\begin{pmatrix}
	0 & -C \\ C &0 
\end{pmatrix}\), where \(C\) is the Cartan matrix of the \(A_{N-1}^{(1)}\) root system. The corresponding integrable system is a closed relativistic Toda system. See \cite{Fock:2016} for more details.

The other examples of integrable systems that can be constructed this way include the XXZ chain (see \cite{Marshakov:2019cluster}) and the pentagram map (see, e.g., \cite{Gekhtman:2016integrable} and references therein).

See also \cite{Williams:2013double}, \cite{Williams:2013cluster} for a more general discussion of the Poisson and cluster structures on the Kac--Moody groups.

\subsection{Path interpretation} Now we will reformulate the constructions above in more combinatorial terms (essentially in the framework of \cite{Gekhtman:2012Poisson}).
Let us start with an analog of Lemma~\ref{Lem:L=T}.
As before, we can introduce a perfect orientation (such that all horizontal lines go from right to left and all vertical edges go from black to white vertices) and (infinitely remote) boundary source vertices \(\sigma_i=(+\infty,-i)\) and target vertices \(\tau_i=(-\infty,-i)\), \(1\le i \le N\).
By definition, this graph is drawn on the cylinder.

It is convenient to promote it to a graph on the plane, which is a universal cover of the cylinder.
On the plane we label the target vertices by  \(\tau_k=(-\infty,-k)\), \( k \in\mathbb{Z}\).
Let \(p_0\) be the horizontal path from \(\sigma_N\) to \(\tau_N\).
Then, for any path
\(p\) from \(\sigma_j\), \(1\le j \le N\), to \(\tau_k\), \(k\in \mathbb{Z}\), we define
\begin{equation}\label{eq:weight path affine}
	\wt(p)=\lambda^{(i-k)/N}\overline{\wt}(p)=\lambda^{(i-k)/N}\; \prod_{f \text{ below } p \text{ and above } p_0}X_f\; \prod_{f \text{ above } p \text{ and below } p_0}X_f^{-1},
\end{equation}
where \(1\le i\le N\), \(k\equiv i \pmod N\), and \(X_f\) is a variable corresponding to the face \(f\).
Note that if \(1 \le k \le N\) and the path \(p\) is inside the strip \(-N\le y \le -1\), this definition agrees with the definition above: product of variables assigned to faces below the path.

Then the transfer matrix has the form 
\begin{equation}
	\tilde{T}_{i,j}=\sum_{k \equiv i \!\!\!\!\pmod N}\; \sum_{p \colon \sigma_j \to \tau_k} \wt(p)
\end{equation}
One can view \(\tilde{T}\) as an element of \(\widehat{PGL}^{\sharp}_N\), i.e., it is defined up to a scalar factor.
Or, one can multiply \(\tilde{T}\) by a monomial in face variables \(X_f\) and get a normalized transfer matrix \(T\) such that \(\det T =\lambda^k\) for some \(k\in \mathbb{Z}\).
Then we have \(T=\mathbb{L}_{\mathbf{s}}(\mathbf{X};\lambda)\).

The path \(p_0\) can be viewed as some normalization according to formula~\eqref{eq:weight path affine}.
If we change the path \(p_0\) to another path \(p_0'\), then all weights (and the whole matrix \(\tilde{T}\)) will be multiplied by some monomial.
Such transformations do not change \(\tilde{T}\) as an element of \(\widehat{PGL}^{\sharp}_N\) or the normalized transfer matrix \(T\).
On the other hand, one can consider the renormalization of \(\tilde{T}\) to \(T\) as a redefinition of weights: by replacing \(p_0\) by another path (or rather a linear combination of paths with rational coefficients), one can make the transfer matrix equal to \(T\).

\begin{Example}
	Let us take \(G=\widehat{PGL}^{\sharp}_3\) and \(w=\bar{s}_1s_1\bar{s}_2s_2\bar{s}_0s_0\), see Example~\ref{Ex:Toda affine SL3} above.
	The oriented plabic graph on the plane with face variables is depicted in Fig.~\ref{Fig:network affine}.
		
	\begin{figure}[h]
		\centering
		\includegraphics[]{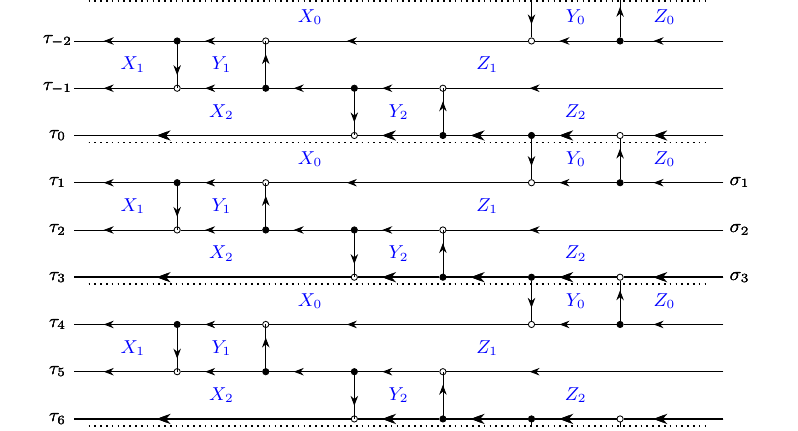}
		\caption{\label{Fig:network affine} Network on the universal cover corresponding to  \(G=\widehat{PGL}^{\sharp}_3\), \(w=\bar{s}_1s_1\bar{s}_2s_2\bar{s}_0s_0\). }		
	\end{figure}
	
	The corresponding transfer matrix is equal to 
	\begin{equation}
		\tilde{T}=\begin{pmatrix}
			X_{12}Y_{12}Z_{12} (1+Y_0+\lambda X_{123}Y_{123})& X_{12}Y_{12}Z_2 & X_{12}Y_{12}+\lambda^{-1}X_0
			\\ 
			X_2Y_{12}Z_{12}(1+Y_0+\lambda X_{012}Y_{02}(1+Y_1)) & X_2Y_2Z_2(1+Y_1) & X_2Y_2 (1+Y_1)+\lambda^{-1}X_{01}^{-1}
			\\
			\lambda X_{012}Y_{012}Z_{12}(1+Y_2)& Y_2Z_2 & 1+Y_2
		\end{pmatrix}.
	\end{equation}
	Here we used the shorthand notations \(X_{i_1\cdots i_k}=X_{i_1}\cdot\ldots\cdot X_{i_k}\) and similarly for \(Y\) and \(Z\).
	It is straightforward to check that \(T\) is equal to \(\mathbb{L}(\lambda)\) given by formula~\eqref{eq:L(lambda) Sl3}.
\end{Example}

Let us now interpret the spectral curve equation~\eqref{eq:Z=det}.
By definition we have
\begin{equation}\label{eq:Z=sum Lambda^l}
	\mathcal{Z}(\mathbf{X}|\lambda,\mu)= \det (\mathbb{L}(\lambda)+\mu)=\sum_{l=0}^N\mu^{N-l}\operatorname{Tr}\bigwedge\nolimits^l \mathbb{L}(\lambda).
\end{equation}
As was explained above, we can assume that \(\mathbb{L}(\lambda)\) is a transfer matrix.
Note also that the renormalization of \(\tilde{T}\) is equivalent to rescaling of \(\mu\) in formula~\eqref{eq:Z=sum Lambda^l}.
So we can write each term in~\eqref{eq:Z=sum Lambda^l} as a sum:
\begin{equation}
	\operatorname{Tr}\bigwedge\nolimits^l \mathbb{L}(\lambda)=\sum _{1 \le i_1<\dots<i_l\le N}\; \sum_{\alpha \in S_l}(-1)^{l(\alpha)}  \prod_{r=1}^l\; \sum_{k_r \in\mathbb{Z},\; k_r \equiv i_{\alpha(r)} \!\!\!\! \pmod N}\; \sum_{p_r\colon \sigma_{i_r}\to \tau_{k_r}} \wt(p_r).
\end{equation}
In other words, the summation goes over the sets of paths \(P(\{\sigma_{i_r}\}\to \{\tau_{k_r}\})\).
Using the Lindström--Ges\-sel--Vien\-not lemma for the paths on the acyclic graph, we can rewrite this formula as a sum of non-intersecting tuples of paths \(P_{\mathrm{nc}}(\{\sigma_{i_r}\}\to \{\tau_{k_r}\})\) between the same sources and targets.
Moreover, we can also assume that these paths do not intersect on the cylinder.
Such a tuple of paths exists only if \(k_1<k_2<\dots <k_l<k_1+N\), hence we have
\begin{equation}
	\operatorname{Tr}\bigwedge\nolimits^l \mathbb{L}(\lambda)=\sum_{1 \le i_1<\dots<i_l\le N}\; \sum_{k_1<\dots<k_l<k_1+N}  \;\sum_{\mathbf{p} \in P_{\mathrm{nc}}(\{\sigma_{i_r}\}\to \{\tau_{k_r}\})} (\pm) \wt(\mathbf{p}).
\end{equation}
Note that here the integers \(k_r\) should satisfy \( k_r \equiv i_{\alpha(r)} \!\!
\pmod N\) for some permutation \(\alpha\in S_l\) and the sign \(\pm\) is equal to \((-1)^{l(\alpha)}\).
Moreover, it follows from inequalities among \(k_1,\dots k_l\) that the permutation has the form \((1,2,\dots,l)^{a}\), where \(a=(\sum_r i_r-\sum_r k_r)/N\).
Therefore we get
\begin{equation}\label{eq:Z=sum paths}
	\mathcal{Z}(\mathbf{X}|\lambda,\mu)= \sum_{l=0}^N \mu^{N-l} \sum_{\mathbf{i},\mathbf{k}}\; \sum_{\mathbf{p} \in P_{\mathrm{nc}}(\{\sigma_{i_r}\}\to \{\tau_{k_r}\})} (-1)^{a(l-1)} \lambda^{a} \overline{\wt}(\mathbf{p}) 
%	\\ 
	=\sum_{a,b} %(-1)^{ab +a(N+1)}
	\lambda^a\mu^b \mathcal{Z}_{a,b}(\mathbf{X}).
\end{equation}
Here in the sum, we assume the same conditions as above, i.e. \(1 \le i_1<\dots<i_l\le N\), 
\(k_1<\dots<k_l<k_1+N\), \(a=(\sum_r i_r-\sum_r k_r)/N\) and \( k_r \equiv i_{r+a \!
\pmod l} \!\!
\pmod N\).
The quantity \(\overline{\wt}\) is the \(\lambda\)-independent part of the weight, see formula~\eqref{eq:weight path affine}, and we used the variable \(b=N-l\).

Functions \(\mathcal{Z}_{a,b}(\mathbf{X})\) defined in formula~\eqref{eq:Z=sum paths} are Laurent polynomials in face variables \(\mathbf{X}\) (in general with fractional powers) with positive integer coefficients.
These functions form an integrable system, see Theorem~\ref{Th:IS GK} below.
%Changing the sign of \(\lambda\) one can exclude the linear sign \((-1)^{a(N+1)}\), but quadratic term \((-1)^{ab}\) remains and is essential (geometrically it corresponds to the spinor structure or quadratic form on a torus \(\Sigma=\mathbb{T}^2\)). 

\subsection{Dimer models} 
\begin{figure}[h]
	\centering
	\begin{tikzpicture}[font=\small]
		\def\xs{1.3}
		\def\ys{1.3}
		
		\begin{scope}[shift={(0*\xs,0*\ys)}]
			
			\begin{scope}
				\clip (-0*\xs,-0.25*\ys) to ++ (7*\xs,0) to ++  (0*\xs,-3*\ys) to ++ (-7*\xs,0) to ++ (0,3*\ys);
				
				\node[circle, draw, fill=black, inner sep=0pt, minimum size=4pt] (b2) at (0.5*\xs,-2*\ys) {};				
				\node[circle, draw, fill=black, inner sep=0pt, minimum size=4pt] (b3) at (2.5*\xs,-3*\ys) {};
				
				\node[circle, draw, inner sep=0pt, minimum size=4pt] (d1) at (\xs,-2*\ys) {};
				\node[circle, draw, fill=black, inner sep=0pt, minimum size=4pt] (u1) at (\xs,-\ys) {};		
				\node[circle, draw, fill=black, inner sep=0pt, minimum size=4pt] (d2) at (2*\xs,-2*\ys) {};
				\node[circle, draw,   inner sep=0pt, minimum size=4pt] (u2) at (2*\xs,-1*\ys) {};
				
				\node[circle, draw, inner sep=0pt, minimum size=4pt] (m2) at (2.5*\xs,-2*\ys) {};
				
				\node[circle, draw, inner sep=0pt, minimum size=4pt] (d3) at (3*\xs,-3*\ys) {};
				\node[circle, draw, fill=black, inner sep=0pt, minimum size=4pt] (u3) at (3*\xs,-2*\ys) {};		
				\node[circle, draw, fill=black, inner sep=0pt, minimum size=4pt] (d4) at (4*\xs,-3*\ys) {};
				\node[circle, draw,   inner sep=0pt, minimum size=4pt] (u4) at (4*\xs,-2*\ys) {};
				
				\node[circle, draw, inner sep=0pt, minimum size=4pt] (m3) at (4.5*\xs,-3*\ys) {};
				\node[circle, draw, fill=black, inner sep=0pt, minimum size=4pt] (m1) at (3.5*\xs,-1*\ys) {};
				
				\node[circle, draw, inner sep=0pt, minimum size=4pt] (d5) at (5*\xs,-1*\ys) {};
				\node[circle, draw, fill=black, inner sep=0pt, minimum size=4pt] (u5) at (5*\xs,-3*\ys) {};
				\node[circle, draw, fill=black, inner sep=0pt, minimum size=4pt] (d6) at (6*\xs,-1*\ys) {};
				\node[circle, draw,   inner sep=0pt, minimum size=4pt] (u6) at (6*\xs,-3*\ys) {};
				
				\node[circle, draw, inner sep=0pt, minimum size=4pt] (b1) at (6.5*\xs,-1*\ys) {};

				\foreach \i in {1,3,5}
				{
					\draw (d\i) -- ++ (0,\ys);
					\draw[special arrow=0.7] (u\i) -- ++ (0,-\ys);
				}
				\draw[special arrow=0.7]  (5*\xs,0*\ys) to (d5);	
				\foreach \i in {2,4,6}
				{
					\draw[special arrow=0.7] (d\i) -- ++ (0,\ys);
					\draw[] (u\i) -- ++ (0,-\ys);
				}	
				%			sources and targets			
				
				\node (tau3) at (0*\xs,-3*\ys) {};	
				\node (tau2) at (0*\xs,-2*\ys) {};			
				\node (tau1) at (0*\xs,-\ys) {};			
				
				\node (sigma3) at (7*\xs,-3*\ys) {};	
				\node (sigma2) at (7*\xs,-2*\ys) {};	
				\node (sigma1) at (7*\xs,-\ys) {};

				\draw[special arrow=0.7] (sigma1) to (b1);
				\draw[yellow, line width=2pt] (b1) to (d6);
				\draw[special arrow=0.7] (b1) to (d6);
				\draw[special arrow=0.7] (d6) to (d5);
				\draw[yellow, line width=2pt] (d5) to (m1);
				\draw[special arrow=0.7] (d5) to (m1);
				\draw[special arrow=0.7] (m1) to (u2);
				\draw[yellow, line width=2pt] (u2) to (u1);
				\draw[special arrow=0.7] (u2) to (u1);
				\draw[special arrow=0.7] (u1) to (tau1);
				\draw[special arrow=0.7] (sigma2) to (u4);
				\draw[yellow, line width=2pt] (u4) to (u3);
				\draw[special arrow=0.7] (u4) to (u3);
				\draw[special arrow=0.7] (u3) to (m2); 
				\draw[yellow, line width=2pt] (m2) to (d2);
				\draw[special arrow=0.7] (m2) to (d2);
				\draw[special arrow=0.7] (d2) to (d1);
				\draw[yellow, line width=2pt] (d1) to (b2);
				\draw[special arrow=0.7] (d1) to (b2);
				\draw[special arrow=0.7] (b2) to (tau2);
				\draw[special arrow=0.7] (sigma3) to (u6);
				\draw[yellow, line width=2pt](u6) to (u5);
				\draw[special arrow=0.7](u6) to (u5);				
				\draw[special arrow=0.7] (u5) to (m3);
				\draw[yellow, line width=2pt] (m3) to (d4);
				\draw[special arrow=0.7] (m3) to (d4);
				\draw[special arrow=0.7] (d4) to (d3);				
				\draw[yellow, line width=2pt] (d3) to (b3);
				\draw[special arrow=0.7] (d3) to (b3);
				\draw[special arrow=0.7] (b3) to (tau3);
				
				\node[blue] (x2) at (1.5*\xs,-2.5*\ys) {$X_2$};
				\node[blue] (x1) at (0.5*\xs,-1.5*\ys) {$X_1$};
				\node[blue] (x0) at (2.5*\xs,-0.5*\ys) {$X_0$};
				\node[blue] (y2) at (3.5*\xs,-2.5*\ys) {$Y_2$};
				\node[blue] (y1) at 
				(1.5*\xs,-1.5*\ys) {$Y_1$};
				\node[blue] (y0) at (5.5*\xs,-0.5*\ys) {$Y_0$};
				\node[blue] (z2) at (5.5*\xs,-2.5*\ys) {$Z_2$};
				\node[blue] (z1) at (4.5*\xs,-1.5*\ys) {$Z_1$};
				\node[blue] (z0) at (6.5*\xs,-0.5*\ys) {$Z_0$};

			\end{scope}
			
			%				boundary
			\draw[thick, dotted] (0*\xs,-0.25*\ys) -- ++ (7*\xs,0);
			\draw[thick, dotted] (0*\xs,-3.25*\ys) -- ++ (7*\xs,0);
		\end{scope}

	\end{tikzpicture}
	\caption{\label{Fig:dimers closed Toda N=3} Bipartite graph equivalent to the one in Fig.~\ref{Fig:closed Toda N=3} with dimer configuration \(D_0\)}		
\end{figure}
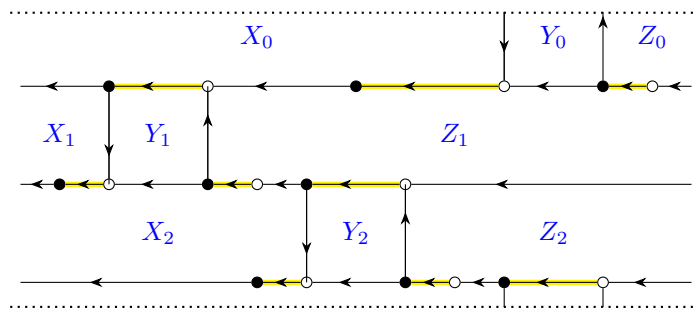

Assume that the graph is bipartite, and each horizontal line starts with a white vertex and ends with a black vertex (going from right to left).
One can easily transform the graph to this form using the insertion of 2-valent vertices in Fig.~\ref{Fig:contr 2 alent}.
For instance, the graph drawn on the left in Fig.~\ref{Fig:closed Toda N=3} can be transformed to the graph in Fig.~\ref{Fig:dimers closed Toda N=3}.

Note that our orientation was taken such that there is one outgoing edge from each white vertex.
Let \(D_0\) denote the set of such edges.
Since the graph is bipartite and there is only one incoming edge at each black vertex, the set \(D_0\) is a dimer cover.

\begin{Definition}
	A \emph{dimer cover} (an equivalent notion is a \emph{perfect matching}) of a graph \(\Gamma\) is a subset \(D \subset E(\Gamma)\) such that for any vertex \(v \in V(\Gamma)\) there exists a unique edge \(e \in D\) incident to \(v\).
\end{Definition}

In Fig.~\ref{Fig:dimers closed Toda N=3} we highlighted the dimer cover \(D_0\) in yellow.
Note also that while the graph in Fig.~\ref{Fig:dimers closed Toda N=3} is depicted on an infinite cylinder, from now on we glue its ends and consider the graph on a torus \(\Sigma=\mathbb{T}^2\).

Let \(D\) be another dimer configuration.
Then, the difference \(D-D_0\) is a union of cycles.
Since the edges in \(D_0\) go from white to black vertices, and the edges in \(D\setminus D_0\) go from black to white vertices, we see that \(D-D_0=\mathbf{p}=\{p_1,\dots,p_r\}\) is a collection of oriented paths.
Since all horizontal edges go from right to left, any simple path $p_i$ must go through the source and target vertices on the cylinder.
Hence, we got a collection of paths on the right side of formula~\eqref{eq:Z=sum paths}.
On the other hand, for any collection of paths \(\mathbf{p}\), the sum \(D=D_0+\mathbf{p}\) is a dimer configuration.
Hence, formula~\eqref{eq:Z=sum paths} can be rewritten as
\begin{equation}\label{eq:Z=sum dimers}
	\mathcal{Z}(\mathbf{X}|\lambda,\mu)= \sum_{a,b} (-1)^{ab +a(N+1)}\lambda^a\mu^b \sum_{D, [D-D_0]=a \mathsf{A}+b \mathsf{B}}\wt(D-D_0).
\end{equation}
Here \([D-D_0]\) is considered as an element of \(H_1(\Sigma)\) and \(\mathsf{A}, \mathsf{B}\) are generators corresponding to vertical and horizontal cycles in our figures.
The formula for the sign also follows from~\eqref{eq:Z=sum paths}.
Changing the sign of \(\lambda\), one can exclude the linear sign \((-1)^{a(N+1)}\), but the quadratic term \((-1)^{ab}\) remains and is essential.
Geometrically, it corresponds to a spin structure or a quadratic form on a torus \(\Sigma=\mathbb{T}^2\), while combinatorially it corresponds to the choice of Kasteleyn sign.

\begin{figure}[h]
	\centering
	\includegraphics[]{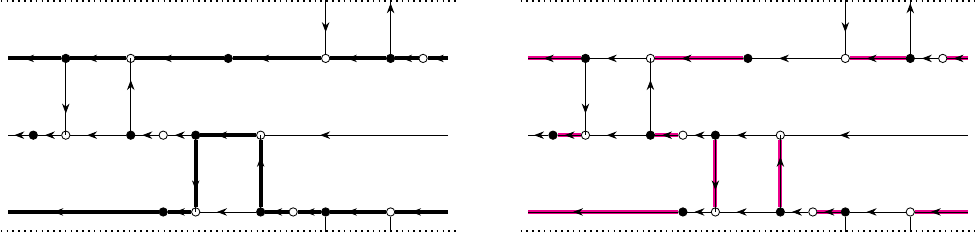}
	\caption{\label{Fig:paths and dimer configuration} On the left, the tuple of paths, on the right, the dimer configuration}		
\end{figure}

In Fig.~\ref{Fig:paths and dimer configuration} we presented an example of a collection of paths \(\mathbf{p}\) and a dimer configuration \(D\) such that \(D-D_0=\mathbf{p}\), where \(D_0\) is given in Fig.~\ref{Fig:dimers closed Toda N=3}.
In this case \([D-D_0]=2\mathsf{B} \in H_1(\Sigma)\).

The definition of \(\wt(D-D_0)\) in formula~\eqref{eq:Z=sum dimers} is a bit implicit; it reduces to the definition of the weight of a path in formula~\eqref{eq:weight path affine}.
In the setting of the dimer model, it is more convenient to define everything in terms of an edge weight function \(\wt \colon E(\Gamma) \to \mathbb{C}^*\).
This function is defined up to gauge transformations \(g \colon V(\Gamma) \rightarrow \mathbb{C}^*\), acting by \(\wt(e) \mapsto g(b)\wt(e)g(w)^{-1} \), where \(e=bw\), \(b\) is a black vertex and \(w\) is a white vertex.
Then for any path \(\gamma=(e_1,\dots, e_k)\), with \(e_j\in E(\Gamma)\) we define its weight to be \(\wt(\gamma)=\prod_{j=1}^k \wt(e_j)^{\pm 1}\), where the sign is plus if the path goes through an edge from a black to a white vertex, and minus in the opposite case.
Note that the orientation of edges from black to white used in the definition of \(\wt(\gamma)\) differs from the perfect orientation used above on exactly the edges in \(D_0\).

For any face \(f\), let us denote by the same letter the path which goes counterclockwise around it (equivalently, the path which turns left at any vertex).
The graph \(\Gamma\) is embedded in the torus, but as above, we usually draw it on the fundamental rectangle, where inside the strip \(-N-0.5<y<-0.5\), source and target vertices are on the left and right vertical boundaries (and are glued in the torus).

It is easy to see that there exists a unique up to gauge transformation edge weight function \(\wt\) such that
\begin{enumerate}
	\item For any face \(f\) we have \(X_f=\wt(f)\)
	\item The vertical edges intersecting the boundary have weight proportional to \(\lambda^{-1}\), and weights of all other edges are independent of \(\lambda\).
	\item The horizontal edges intersecting the boundary have weight proportional to \(\mu^{-1}\), and weights of all other edges are independent of \(\mu\).
\end{enumerate}
With this weight function we can compute \(\wt(D-D_0) =\prod_{e \in D}\wt(e) \prod_{e\in D_0}\wt(e)^{-1}\) and use formula~\eqref{eq:Z=sum dimers}.

Note the symmetry between \(\lambda\) and \(\mu\) in the definition of the weight function.
This is in contrast to the Lax matrix construction above, see formula~\eqref{eq:Z=det}.
One can also say that for any closed path \(p\) we have \(\wt(p)\sim \lambda^a\mu^b\), where \([p]=a\mathsf{A}+b\mathsf{B}\in H_1(\Sigma)\).
Here \(\mathsf{A}, \mathsf{B}\) are the generators corresponding to vertical and horizontal cycles, as above.

Furthermore, we can now reverse the logic, and just start from a bipartite graph \(\Gamma\) on a torus \(\Sigma\) (i.e., not from the word \(\mathbf{i}\) corresponding to the element \(w\in W^{ae}(A_{N-1}+A_{N-1})\)).
Let us fix some weight function \(\wt\).
It determines face variables by the formula \(X_f=\wt(f)\).
For a given symplectic basis \([\mathsf{A}],[\mathsf{B}]\in H_1(\Sigma)\) and given paths \(\mathsf{A}\) and \(\mathsf{B}\) which represent these cycles, one can define \emph{spectral parameters} \(\lambda = \wt(\mathsf{A})\) and \(\mu = \wt(\mathsf{B})\).
Fixing a particular dimer configuration \(D_0\), we can compute the partition function by formula~\eqref{eq:Z=sum dimers}.

Note that the definition of the spectral variables \((\lambda, \mu)\) depends on the choice of the paths \((\mathsf{A}, \mathsf{B})\).
For example, the rescaling of \(\lambda\) by \(\prod_i Z_i\) that we encountered in Toda Examples \ref{Ex:Toda affine SL2}, \ref{Ex:Toda affine SL3} corresponds to the path that separates the \(Z\)'s from the \(Y\)'s, i.e. the ultimate faces in each row from the penultimate ones. 
If one modifies the chosen representatives, the variables \((\lambda,\mu)\) are multiplied by monomials in the face variables~\(\mathbf{X}\).

Under the change of symplectic basis \([\mathsf{A}],[\mathsf{B}]\in H_1(\Sigma)\), the spectral parameters are transformed as \(\lambda \mapsto \lambda^a\mu^b, \mu \mapsto \lambda^c\mu^d\), where \(\begin{pmatrix}
	a & b \\ c &d
\end{pmatrix}\in SL(2,\mathbb{Z})\). Here we do not assume that a bipartite graph is presented in a form \(\Gamma_{\mathbf{i}}\) for some rank \(N\) and word \(\mathbf{i}\). It turns out that for consistent graphs (which we will define below), such a presentation always exists. Moreover, there are infinitely many such presentations, with \(\lambda,\mu\) related by \(SL(2,\mathbb{Z})\)-transformations.

Note also that a change of the reference dimer configuration \(D_0 \mapsto D_0'\) results in multiplication of \(\mathcal{Z}\) by a monomial factor (and, possibly, a change of sign of spectral variables \(\lambda,\mu\)).
This transformation does not change the spectral curve \(\mathcal{C}\).

\subsection{Zigzag paths} The Poisson bracket on face variables above is a cluster one~\eqref{eq:PB cluster} for the quiver defined in Definition~\ref{Def:quiver from plabic}.
It also has a more geometric definition using the dual surfaces.

One can thicken the graph \(\Gamma\) to make a ribbon graph.
Topologically this ribbon graph is a surface \(\Sigma=\mathbb{T}^2\) with \(F(\Gamma)\) holes.
For a ribbon graph, the cyclic order of the edges at each vertex is fixed.
Let us define a \emph{dual bipartite ribbon graph} \(\Gamma^D\) by reversing the cyclic order at all black vertices.
This dual bipartite ribbon graph is topologically a \emph{dual surface} \(\Sigma^D\) with holes.

\begin{figure}[h]
	\centering
	\includegraphics[]{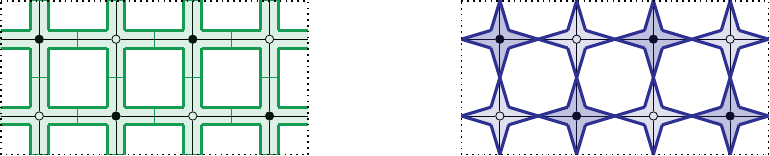}
	\caption{\label{Fig:dual graph} On the left: ribbon graph; on the right: dual ribbon graph}		
\end{figure}

Equivalently, one can say that we replace the edges of the graph \(\Gamma\) by thin ribbons and twist each of them once.
See an example in Fig.~\ref{Fig:dual graph}.

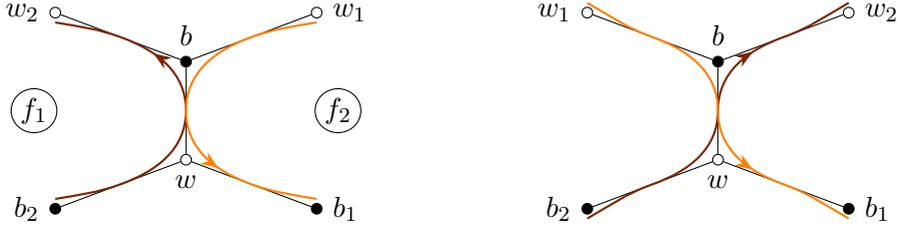
\begin{figure}[h]
	\centering
\begin{tikzpicture}
	\def\xs{2}
	\def\ys{1.3}

	\begin{scope}
		
		% Define nodes with labels
		\node[circle, draw, inner sep=0pt, fill=black, minimum size=4pt, label=above:{$b$}] (b) at (0,\ys) {};
		\node[circle, draw, inner sep=0pt, minimum size=4pt,
		label=below:{$w$}]	(w) at (0,0) {};
		\node[circle, draw, inner sep=0pt, minimum size=4pt,
		label=right:{$w_1$}] (w1) at (0.86*\xs,1.5*\ys) {};
		\node[circle, draw, inner sep=0pt, minimum size=4pt,
		label=left:{$w_2$}]	(w2) at (-0.86*\xs,1.5*\ys) {};
		
		\node[circle, draw, inner sep=0pt, fill=black, minimum size=4pt,label=right:{$b_1$}] (b1) at (0.86*\xs,-0.5*\ys) {};
		
		\node[circle, draw, inner sep=0pt, fill=black, minimum size=4pt, label=left:{$b_2$}] (b2) at (-0.86*\xs,-0.5*\ys) {};
		
		% Draw arrows
		\draw (b) -- (w);
		\foreach \i in {1,2} {
			\draw (b) -- (w\i);
			\draw (b\i) -- (w);
		}
		
%		face
		\draw[special arrow=0.7, Brown, thick] (-0.86*\xs,-0.4*\ys) to[out=10,in=200] (-0.43*\xs,-0.25*\ys) to[out=20,in=270] (0*\xs,0.5*\ys) to[out=90,in=340] (-0.43*\xs,1.25*\ys) to[out=160,in=350] (-0.86*\xs,1.4*\ys);
		\draw[special arrow=0.7, orange, thick] (0.86*\xs,1.4*\ys) to[out=190,in=20] (0.43*\xs,1.25*\ys) to[out=200,in=90] (0*\xs,0.5*\ys) to[out=270,in=160] (0.43*\xs,-0.25*\ys) to[out=340,in=170] (0.86*\xs,-0.4*\ys);
				
		\node[styleNode] at (-\xs,0.5*\ys) {$f_1$};
		\node[styleNode] at (\xs,0.5*\ys) {$f_2$};
	\end{scope}
	
	\begin{scope}[shift={(2*\xs+3,0*\ys)}]
		\def\xs{2}
		\def\ys{1.3}
		
		% Define nodes with labels
		\node[circle, draw, inner sep=0pt, fill=black, minimum size=4pt, label=above:{$b$}] (b) at (0,\ys) {};
		\node[circle, draw, inner sep=0pt, minimum size=4pt,
		label=below:{$w$}]	(w) at (0,0) {};
		\node[circle, draw, inner sep=0pt, minimum size=4pt,
		label=right:{$w_2$}] (w1) at (0.86*\xs,1.5*\ys) {};
		\node[circle, draw, inner sep=0pt, minimum size=4pt,
		label=left:{$w_1$}]	(w2) at (-0.86*\xs,1.5*\ys) {};
		
		\node[circle, draw, inner sep=0pt, fill=black, minimum size=4pt,label=right:{$b_1$}] (b1) at (0.86*\xs,-0.5*\ys) {};
		
		\node[circle, draw, inner sep=0pt, fill=black, minimum size=4pt, label=left:{$b_2$}] (b2) at (-0.86*\xs,-0.5*\ys) {};
		
		% Draw arrows
		\draw (b) -- (w);
		\foreach \i in {1,2} {
			\draw (b) -- (w\i);
			\draw (b\i) -- (w);
		}
		
		%		face
		\draw[special arrow=0.7, Brown, thick] (-0.86*\xs,-0.6*\ys) to[out=30,in=200] (-0.43*\xs,-0.25*\ys) to[out=20,in=270] (0*\xs,0.5*\ys) to[out=90,in=200] (0.43*\xs,1.25*\ys) to[out=20,in=210] (0.86*\xs,1.6*\ys);
		
		\draw[special arrow=0.7, orange, thick] (-0.86*\xs,1.6*\ys) to[out=330,in=160] (-0.43*\xs,1.25*\ys) to[out=340,in=90] (0*\xs,0.5*\ys) to[out=270,in=160] (0.43*\xs,-0.25*\ys) to[out=340,in=150] (0.86*\xs,-0.6*\ys);
		
	\end{scope}
\end{tikzpicture}
	\caption{\label{Fig:dual faces} On the left: faces on graph \(\Gamma\); on the right: the corresponding paths in dual graph $\Gamma^D$}
\end{figure}

\begin{Lemma}\label{Lem:dual bracket}
	Let \(f_1, f_2\) be two faces of \(\Gamma\subset \Sigma\).
	Then, the number of arrows between the corresponding vertices in the quiver \(\mathcal{Q}\) from Definition~\ref{Def:quiver from plabic} is equal to the intersection number of the corresponding paths on the dual surface.
\end{Lemma}
The proof is explained in Fig.~\ref{Fig:dual faces}.
Note that the cyclic order for the black vertex \(b\) is reversed.
Paths $f_1,f_2$ on the original graph \(\Gamma\) and the dual graph \(\Gamma^D\) are drawn through the midpoints of the edges.

\begin{figure}[h]

	\centering
	\includegraphics[]{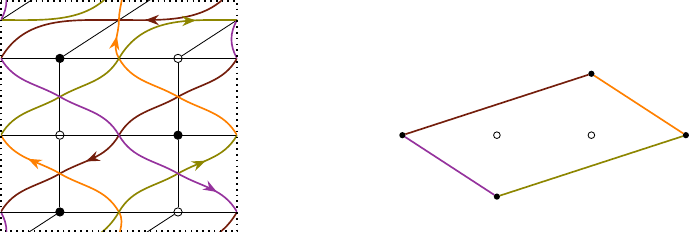}

	\caption{\label{Fig:dimers closed Toda zigzags} Left: bipartite graph with zigzags, right: Newton polygon}		
\end{figure}

It follows from Lemma~\ref{Lem:dual bracket} that Casimir functions for the cluster Poisson bracket correspond to weights of homologically trivial paths in \(\Gamma^D\subset \Sigma^D\).
Such paths are generated by faces of \(\Gamma^D\subset \Sigma^D\).
It follows from the definition that in terms of the original graph, they correspond to zigzag paths, i.e., paths which turn \emph{right at black vertices} and \emph{left at white vertices}.
We will draw zigzags as paths that go through the middles of the edges.
Locally, the correspondence between faces and zigzags can be seen in Fig.~\ref{Fig:dual faces}.
In Fig.~\ref{Fig:dimers closed Toda zigzags}, we depicted all zigzags in our running example.

We denote zigzag paths by \(\zeta_1,\dots,\zeta_B\).
For any \(j\) let us denote the zigzag variable by \(z_j=\pm \wt(\zeta_j)\).
Here the sign ``\(\pm\)'' depends on the length of \(\zeta_j\) and the Kasteleyn orientation, see the references.
The functions \(z_1,\dots,z_B\) are Casimir functions for the Poisson bracket given by intersection on the dual surface \(\Sigma^D\).
It is easy to see that any edge \(e \in E(\Gamma)\) belongs to exactly two zigzags that go through~\(e\) in the opposite directions on it.
Therefore, in the product of zigzag variables, the edge weights are canceled, and we have \(\prod_{j=1}^B z_j=1\).

From now on, we will always assume that the paths \(\mathsf{A}, \mathsf{B}\) are chosen to be zigzags or formal combinations of zigzags with rational coefficients.
Hence \(\lambda,\mu\) become Casimir functions.
Similarly to formula~\eqref{eq:weight path affine}, let \(\overline{\wt}\) denote the \(\lambda,\mu\)-independent part of the weight function and let \(\bar{z}_j=\pm \overline{\wt}(\zeta_j)\).
Then \(\bar{z}_1,\dots, \bar{z}_B\) are expressed in terms of face variables \(\mathbf{X}\) and are Casimir functions of the cluster variety \(\mathcal{X}\).\footnote{To be more precise, on its subvariety given by equation \(\prod_{f\in F(\Gamma)} X_f=1\).}
These Casimir functions are subject to 3 constraints: \(\prod \bar{z}_j=1\), \(\overline{\wt}(\mathsf{A})=1\), \(\overline{\wt}(\mathsf{B})=1\).
Hence, the total number of Casimir functions is equal to \(B-3\), where \(B\) is the number of zigzags.

It appears that the class of dimer models that lead to the construction of cluster integrable systems is also defined in terms of zigzag paths.
Let \(\widehat{\Gamma}\subset \mathbb{R}^2\) denote the preimage of the bipartite graph \(\Gamma\subset \Sigma\) on the universal cover.
One can similarly define zigzags \(\hat{\zeta}\in Z(\widehat{\Gamma})\) on the graph \(\widehat{\Gamma}\).

\begin{Definition}\label{def:consistancy}
	The bipartite graph on a torus is called \emph{consistent} if it satisfies the following conditions:
	\begin{enumerate}[label=(\alph*)]
		\item \label{it:1} Any zigzag \(\zeta\) represents a nontrivial homology class \([\zeta]\neq 0 \in H_1(\Sigma)\).
		
		\item \label{it:2} There are no parallel \emph{bigons} on the universal cover, namely any two zigzags \(\hat{\zeta}_1, \hat{\zeta}_2\) do not have a pair of intersections such that both paths go in the same direction from one intersection to the other.
		
		\item \label{it:3} Any zigzag \(\hat{\zeta}\) on the universal cover does not have self-intersections.
	\end{enumerate}
\end{Definition}

Such graphs are also called minimal in \cite{Goncharov:2013} \footnote{The notion of intersection of two zigzag paths is a bit subtle in the case of vertices of valency 2, see, e.g.,
\cite[Def. 3.4]{Ishii:2011note}.
However, it follows from consistency conditions that any 2-valent vertex is connected with two \emph{different} vertices and therefore can be contracted.}.
A consistent dimer model is a pair of a consistent bipartite graph \(\Gamma\) and a weight function.
Since the weight function is defined on the edges up to gauge transformations assigned to vertices, one can view the weight function as an element \([\wt]\in H^1(\Gamma,\mathbb{C}^*)\).
Equivalently, the weight function is determined by spectral variables \(\lambda,\mu\) and face variables \(X_1,\dots, X_F\) subject to \(\prod_{f\in F(\Gamma)} X_f=1\).

For a consistent dimer model, one can compute the dimer partition function by formula~\eqref{eq:Z=sum dimers}.
It can be proven \cite{Goncharov:2013}, \cite{Ishii:2009:2} that for any consistent dimer model, there exists a dimer configuration, so the sum is not empty.
The dimer partition function \(\mathcal{Z}(\lambda,\mu)\) depends on the choice of reference dimer configuration, i.e., is defined up to a monomial factor.

The following lemma is straightforward to check.

\begin{Lemma} 
	The 4-gon mutation shown in Fig.~\ref{fi:face4} preserves the dimer partition function.
\end{Lemma}

This lemma means that different dimer models could give the same integrable system.
In terms of quivers, this would also mean that quivers related by mutation correspond to the same integrable system.
The combinatorial invariant preserved under this mutation is the \emph{Newton polygon}~\(\Delta\) of \(\mathcal{Z}(\lambda,\mu)\).
By definition \(\Delta\) is the convex hull of \((a,b)\in \mathbb{Z}^2\) such that \(\mathcal{Z}_{a,b}\neq 0\), where we used the decomposition $\mathcal{Z}=\sum_{a,b\in \mathbb{Z}} \lambda^a\mu^b \mathcal{Z}_{a,b}$, see formula~\eqref{eq:Z=sum paths}.

Note that since the partition function $\mathcal{Z}$ is defined up to a monomial factor (e.g., hidden in the choice of \(D_0\)), the Newton polygon \(\Delta\) is defined up to translation.
Furthermore, a change of symplectic basis in \(H_1(\Sigma)\) leads to an \(SL(2,\mathbb{Z})\)-transformation of \(\Delta\), hence, overall \(\Delta\) is defined up to the action of the group of affine transformations \(SA(2,\mathbb{Z})=SL(2,\mathbb{Z})\ltimes \mathbb{Z}^2\).

Let \(I\) denote the number of integral points inside \(\Delta\).
Let \(B\) denote the number of integral points on the boundary of \(\Delta\) (i.e., the number of vertices and points in the interiors of the sides).
Clearly, the numbers \(I, B\) are invariant under the \(SA(2,\mathbb{Z})\) action.

It appears that for a consistent dimer model \(\Gamma\), the number of zigzags is equal to \(B\).
Moreover, the whole Newton polygon \(\Delta\) can also be reconstructed from the zigzag paths.
Since any edge \(e \in E(\Gamma)\) belongs to exactly two zigzags that go through \(e\) in opposite directions on it, we have \(\sum [\zeta_j]=0\), where \([\zeta]\in H_1(\Sigma)=\mathbb{Z}^2\) denotes the homology class of the zigzag.
It follows from the consistency conditions that any zigzag has no self-intersections on \(\Sigma\).
Ordering the primitive vectors \([\zeta_j]\) counterclockwise and concatenating them head-to-tail, we obtain a convex integral polygon. 
It is defined up to translation; by construction, its elementary segments are given by $[\zeta_j]$.
The following theorem states that this polygon coincides with \(\Delta\).

\begin{Theorem}[{\cite{George2022inverse}, \cite{Broomhead:2010}}]  \label{Th:zigzags boundary}
	For any zigzag \(\zeta\) there is a side~\(E\) such that~\(E\) is parallel to~\([\zeta]\).
	Furthermore, for any side \(E\) we have
	\begin{equation}\label{eq:Z|E prod}
		\mathcal{Z}(\mathbf{X}|\lambda,\mu)|_E \sim \prod_{ \zeta_j\in Z(\Gamma),\; [\zeta_j] \text{ parallel to } E} (1+z_j ).
	\end{equation}
\end{Theorem}
Here, the boundary of \(\Delta\) (namely, all vectors \(E\) corresponding to sides) is oriented counterclockwise.
We say that two vectors \(u_1,u_2 \in \mathbb{R}^2\) are parallel if \(u_1=ku_2\) with \( k > 0\).
We used the notation
\begin{equation}
	\mathcal{Z}(\mathbf{X}|\lambda,\mu)|_E=\sum\nolimits_{(a,b)\in E} \lambda^a \mu^b \mathcal{Z}_{a,b}(\mathbf{X}).
\end{equation}
Note that the sign in the definition of zigzag variables \(z_j\) above was chosen such that there are no signs in formula~\eqref{eq:Z|E prod}.

In particular, this theorem means that the number of zigzags parallel to the side \(E\) is equal to the number of primitive segments on this side.
Therefore, the total number of zigzags is equal to \(B\) (number of integral points on the boundary of \(\Delta\)).

We illustrate this theorem in Fig.~\ref{Fig:dimers closed Toda zigzags}.
The bipartite graph used there is equivalent to the one in Fig.~\ref{Fig:dimers closed Toda N=3} using contractions of 2-valent vertices.
The partition function is given by formula~\eqref{eq:Z SL3 Toda}.
Its Newton polygon coincides with the one on the right in Fig.~\ref{Fig:dimers closed Toda zigzags}.

\begin{proof}[Sketch of the proof of Theorem~\ref{Th:zigzags boundary}]
	Consider a zigzag \(\zeta=(e_1,\dots,e_{2l})\).
	The length of \(\zeta\) is even since the graph is bipartite.
	Half of the edges in \(\zeta\) go from black to white vertices.
	We will call such edges \emph{zigs} and assume (without loss of generality) that these edges are \(e_1,e_3, \dots, e_{2l-1}\).
	Similarly, we call the edges of \(\zeta\) \emph{zags} if they go from white to black vertices.
	Under our assumption they are \(e_2,e_4,\dots,e_{2l}\).
	
	Clearly for any dimer configuration \(D\) the number of edges in the intersection \(D\cap \zeta\) is not greater than \(l\).
	Moreover, if \(|D\cap \zeta|=l\), then either \(D\cap \zeta=\{\text{zigs}\}\) or \(D\cap \zeta=\{\text{zags}\}\).
	
	It can be proven that there exists a dimer configuration \(D_1\) such that \(D_1\cap \zeta=\{\text{zigs}\}\).
	Then we can define the dimer configuration \(D_2\) by swapping all zigs for zags in \(\zeta\).
	It follows from the definitions that \(\wt(D_1)/\wt(D_2)=\pm z\), where \(z\) is the zigzag variable corresponding to \(\zeta\).
	
	Let \(D\) be any other dimer configuration.
	Then \(D-D_1\) is a union of closed curves.
	It is easy to show that such curves have non-positive intersection with zigzag \(\zeta\), i.e., \((D-D_1,\zeta)\le 0\).
	See Fig.~\ref{Fig:zigzag} for an illustration.
	
	\begin{figure}[h]
		\centering
		\includegraphics[]{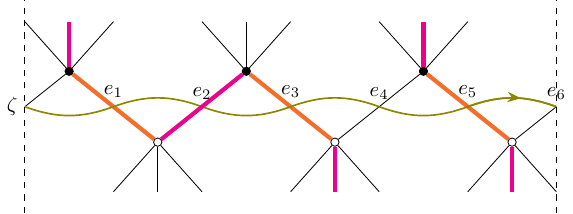}
		\caption{\label{Fig:zigzag} Zigzag \(\zeta\) in olive, edges of \(D_1\) are in orange, edges of \(D\) in magenta}
	\end{figure}
	
	Fix now some reference dimer configuration \(D_0\) and consider the dimer partition function~\eqref{eq:Z=sum dimers}.
	Each term \(\wt(D-D_0)=\lambda^a\mu^b\overline{\wt}(D-D_0)\) corresponds to the point \(P_D=(a,b)\in \mathbb{Z}^2\).
	The arguments above show that the points \(P_1=P_{D_1}\) and \(P_2=P_{D_2}\) differ by the vector \([\zeta]\), and, in particular, belong to the line parallel to \([\zeta]\).
	All other points \(P_D\) lie in one half-plane determined by the line \(P_1P_2\).
	Hence, this line contains the side of the Newton polygon \(\Delta\).
	We denote this side by \(E\).
	The first assertion of the theorem is proven.
	
	Let \(\zeta_1=\zeta, \zeta_2, \dots, \zeta_d\) be all zigzags parallel to \(\zeta\).
	It can be shown that for any \(1\le j\le d\) the intersection \(D_1\cap \zeta_j\) consists either of all zags or all zigs of \(\zeta_j\).
	Hence, swapping zigs and zags of \(\zeta_1,\dots,\zeta_d\) in \(D_1\), we can obtain \(2^d\) dimer configurations \(D_1,\dots, D_{2^d}\).
	Similarly to the argument above, points \(P_{D_k}\) corresponding to these configurations belong to \(E\).
	The sum of contributions of these configurations to the partition function is proportional to \(\prod_{j=1}^d (1+z_j)\).
	Furthermore, it can be shown that all dimer configurations \(D\) such that \(P_D\in E\) belong to the set \(D_1,\dots, D_{2^d}\) constructed above.
\end{proof}

Theorem~\ref{Th:zigzags boundary} has a clear corollary in terms of the Poisson structure.
Namely, since all zigzags' weights \(\{\bar{z}_i\}\) are Casimir functions, there exists a normalization of \(\mathcal{Z}(\mathbf{X}|\lambda,\mu)\) such that all \(\mathcal{Z}_{a,b}(\mathbf{X})\) for \((a,b) \in (\text{boundary of } \Delta)\) are Casimir functions.
The following theorem states that the functions \(\mathcal{Z}_{a,b}(\mathbf{X})\) for \((a,b) \in (\text{interior of } \Delta)\) can be taken as Hamiltonians of an integrable system.

\begin{Theorem}[\cite{Goncharov:2013}]\label{Th:IS GK}
	\begin{enumerate}[label=(\alph*)]
	
		\item \label{it:Th: IS GK a} For any convex integral polygon \(\Delta\) there exists a consistent dimer model \((\Gamma,\wt)\) with the Newton polygon of its partition function equal to \(\Delta\).
		
		\item \label{it:Th: IS GK b} The dimension of the \(\mathcal{X}\) cluster variety corresponding to \(\Gamma\) (i.e., the number of faces \(F(\Gamma)\)) is equal to \(2\operatorname{Area}\Delta\).
		
		\item \label{it:Th: IS GK c} The functions \(\mathcal{Z}_{a,b}(\mathbf{X})\) for \((a,b) \in (\text{interior of } \Delta)\) Poisson commute and are algebraically independent on the subvariety given by the equation \(\prod_{f\in F(\Gamma)} X_f=1\) in the cluster variety \(\mathcal{X}\).
	\end{enumerate}
\end{Theorem}
See also the earlier works \cite{Gulotta:2008properly}, \cite{Ishii:2009:2} for the combinatorial items \ref{it:Th: IS GK a}, \ref{it:Th: IS GK b}.

The claim that we obtained an integrable system requires an elementary, instructive counting argument.
Recall Pick's formula for the area of an integral polygon \(2\operatorname{Area}(\Delta)=2I+B-2\).
Hence, the dimension of the subvariety given by the equation \(\prod_{f\in F(\Gamma)} X_f=1\) is equal to \(2I+(B-3)\).
On the other hand, the number of Hamiltonians is equal to \(I\), and the number of Casimirs is equal to \(B-3\) (recall that there are three relations among \(\bar{z}_1,\dots,\bar{z}_B\)).

Theorem~\ref{Th:IS GK} is the main result of this section.
We conclude with several remarks.

\begin{Remark}
	It was proven in \cite{Fock:2016} that the class of integrable systems obtained by the construction with a Lax matrix (see formulas~\eqref{eq:factorization affine} and~\eqref{eq:Z=det} above) coincides with the class of integrable systems constructed from consistent bipartite graphs.
	
	See also \cite{Izosimov:2022dimers} about the relation between approaches in \cite{Gekhtman:2012Poisson} and in \cite{Goncharov:2013}.
\end{Remark}

\begin{Remark}
	Recall that the spectral curve \(\bar{\mathcal{C}}\) is by definition a compactification of the open curve  \(\mathcal{C}=\{(\lambda,\mu)|\mathcal{Z}(\lambda,\mu)=0\}\subset \mathbb{C}^*\times \mathbb{C}^*\).
	It appears that topologically \(\mathcal{C}\) is isomorphic to the dual ribbon graph  \(\Gamma^D\).
	Indeed, let us first compare the genera of these oriented surfaces.
	The standard result says that the genus of the spectral curve is equal to \(g(\overline{\mathcal{C}})=I\) (for generic values of the variables~\(\mathbf{X}\)), see, e.g.,
	\cite{Khovanskii:1978}.
	On the other hand, the genus of the dual surface can be computed via the Euler formula
	\begin{multline}
		2-2g(\Sigma^D)= |V(\Gamma^D)|-|E(\Gamma^D)|+|F(\Gamma^D)|
		=|V(\Gamma)|-|E(\Gamma)|+|Z(\Gamma)|+|F(\Gamma)|-2\operatorname{Area}(\Delta)
		\\=B-(2I+B-2)=2-2I.
	\end{multline}
	Here we used many results from above: correspondence between faces in \(\Gamma^D\) and zigzags in \(\Gamma\), equality \(|F(\Gamma)|=\dim \mathcal{X}=2\operatorname{Area}(\Delta)\) (see Theorem~\ref{Th:IS GK}), equality \(|Z(\Gamma)|=B\) (see Theorem~\ref{Th:zigzags boundary}), and Pick's formula.
	
	Second, let us compare the number of punctures.
	Namely, the punctures in $\mathcal{C}$ are points of \(\overline{\mathcal{C}}\setminus \mathcal{C}\) (points at infinity).
	They correspond to the roots of $\mathcal{Z}(\mathbf{X}|\lambda,\mu)|_E$, for all sides \(E\).
	The number of such roots is equal to \(B\).
	On the other hand, holes of \(\Gamma^D\subset\Sigma^D\) are faces of the embedded graph and we have \(|F(\Gamma^D)|=|Z(\Gamma)|=B\).
\end{Remark}

\begin{Remark} 
	There is a deep connection between such integrable systems and 5d supersymmetric theories (see, e.g., \cite{Bershtein:2018cluster} and references therein).
	This can be compared with relations of open Toda systems and moduli spaces of local systems to 4d theories, see Remarks~\ref{Rem:4d Coulomb} and \ref{Rem:GMN} above.
\end{Remark}

\begin{Remark}
	Due to the cluster structure of the phase space of the integrable system, the cluster modular group \(G_{\mathcal{Q}}\) gives a natural construction of discrete symmetries.
	In the previous section, the mapping class group of the surface was constructed this way.
	In the setting of Goncharov--Kenyon integrable systems the group \(G_{\mathcal{Q}}\) contains a lattice (related to the Picard group of \(\overline{\mathcal{C}}\)), see \cite{Fock:2016}, \cite{George:2022discrete}.
	Usually, these lattices can be extended to non-commutative groups like affine Weyl groups.
	For example, the symmetries of \(q\)-Painlev\'e equations can be constructed this way \cite{Bershtein:2018cluster}.
\end{Remark}

\section{Quantization of cluster varieties} \label{Sec:quantum}

In general, the quantization of Poisson manifolds is a very non-trivial problem.
But for a constant Poisson bracket, there is a standard solution.
Namely, for a pair of Darboux coordinates \(\mathsf{x},\mathsf{p}\) with Poisson bracket \(\{\mathsf{p},\mathsf{x}\}=1\), the natural quantization is \(\widehat{\mathsf{x}}=\mathsf{x}, \widehat{\mathsf{p}}=\hbar \partial_\mathsf{x}\) with commutation relations \([ \widehat{\mathsf{p}}, \widehat{\mathsf{x}}]=\hbar\).

This procedure can be applied to the cluster Poisson bracket~\eqref{eq:PB cluster}:
\begin{equation}
	\{X_i,X_j\}=\epsilon_{ij}X_iX_j,\qquad \{x_i,x_j\}=\epsilon_{ij},
\end{equation}
where \(x_i=\log X_i\).
We can quantize this bracket as
\begin{equation}
	\widehat{X}_i\widehat{X}_j=q^{2\epsilon_{ij}}\widehat{X}_j\widehat{X}_i,\qquad [\widehat{x}_i,\widehat{x}_j]=\epsilon_{ij} \hbar.
\end{equation}
where \(q=\exp(\hbar/2)\) and \(\widehat{X}_i=\exp(\widehat{x}_i)\).
The algebra  \(\mathbb{C}[\mathcal{X}_{\mathsf{s}}]_q= \mathbb{C}\langle \widehat{X}_i^{\pm 1} | i\in I\rangle / (\widehat{X}_i\widehat{X}_j-q^{2\epsilon_{ij}}\widehat{X}_j\widehat{X}_i) \) is called a quantum torus algebra.

Recall the definition of a seed as a quadruple \((\Lambda,\mathbf{e}, I, I_{\mathrm{f}})\), see Remark~\ref{Rem:seed lattice}.
For any vector \(\lambda=\sum n_i e_i \in \Lambda\), we assign an element \(\widehat{X}_\lambda\in\mathbb{C}\langle \widehat{X}_i^{\pm 1}\mid i\in I \rangle \) given by \(\widehat{X}_\lambda=\exp(\sum n_i \widehat{x}_i)\).
Then we have \(\widehat{X}_{e_i}=\widehat{X}_i\) and
\begin{equation}
	q^{-(\lambda,\mu)} \widehat{X}_\lambda \widehat{X}_\mu =\widehat{X}_{\lambda+\mu}= q^{-(\mu,\lambda)} \widehat{X}_\mu \widehat{X}_\lambda.
\end{equation}
In other words, the operator \(\widehat{X}_\lambda\) is an ordered product of \(\widehat{X}_1^{n_1},\widehat{X}_2^{n_2},\dots, \widehat{X}_{|I|}^{n_{|I|}} \).
Such a product rule is also called \emph{Weyl ordering}.

\medskip

Let us now define a quantum analog of mutation \(\mu_k \colon \mathsf{s} \to \mathsf{s}'\).
It is convenient (see \cite{Fock:2009cluster}) to decompose it into the composition \(\mu_{k}=  {\mu}_{k,+}^\sharp \circ \mu_{k,+}'\).
The first transformation is a monomial one and corresponds to the transformation of the basis given by formula~\eqref{eq:basis mutation}.
In terms of quantum cluster variables, it reads
\begin{equation}
	\mu_{k,+}' \colon \widehat{X}_i \mapsto
	\begin{cases}
		\widehat{X}_k^{-1} \qquad	 &\text{ if } i=k,
		\\
		\widehat{X}_{e_i+ \epsilon_{ik}e_k}= q^{-\epsilon_{ik}^2}\widehat{X}_i\widehat{X}_k^{\epsilon_{ik}}\quad	 &\text{ if } \epsilon_{ik}>0,
		\\
		\widehat{X}_i \qquad &\text{ if }i\neq k, \epsilon_{ik}\le 0,
	\end{cases}
\end{equation}	
The second transformation is conjugation by the function \(\varphi(\widehat{x}_k')^{-1}=\varphi(-\widehat{x}_k)^{-1}\), where
\begin{equation}\label{eq:varphi(b)}
	\varphi(x)=\prod\nolimits_{j=1}^\infty(1 + q^{2j-1}\re^{x}).
\end{equation} 
The function \(\varphi(x)\) essentially depends on \(x\) through \(\re^x\), so we will also use the notation \(\Phi\) for the function \(\Phi(\re^x)=\varphi(x)\).

Recall also the notation for the $q$-Pochhammer symbol $(y;p)_k=\prod_{j=1}^k(1-p^{j-1}y)$, where \(k\) is a positive integer or \(\infty\).
Then we can write \(\Phi(X)=(-q X;q^2)_\infty\).

We summarize some properties of the function \(\varphi\) in the following lemma.
Here and always below we assume that operators \(\widehat{\mathsf{p}}, \widehat{\mathsf{x}}\) satisfy \([ \widehat{\mathsf{p}}, \widehat{\mathsf{x}}]=\hbar\).
\begin{Lemma} 
	\begin{enumerate}[label=(\alph*)]
		\item The function \(\varphi\) satisfies the recurrence relation
			\begin{equation}\label{eq:recurrence phi}
				\varphi(x-\hbar)=\varphi(x)(1+q^{-1}\re^x),\qquad \Phi(q^{-2}X)=\Phi(X)(1+q^{-1}X).
			\end{equation}
			
		\item The function \(\varphi\) has the following expansions
		\begin{align}
			\label{eq:Phi expansion}
			\Phi(X)&=\sum_{k\ge 0}\frac{ q^{k^2}}{(q^2;q^2)_k} X^{k}
			\\
			\Phi(X)^{-1}&=\sum_{k\ge 0}\frac{(-1)^k q^{k}}{(q^2;q^2)_k} X^{k}
			\\
			\label{eq:log Phi expansion}
			\log \Phi(X)&=\sum_{k> 0}(-1)^{k-1}\frac{ q^{k}}{k(1-q^{2k})} X^k,
		\end{align}
		
		\item The function \(\varphi\) satisfies
		\begin{equation}
			\label{eq:Phi sum}
			\Phi(\re^{\widehat{\mathsf{p}}})\Phi(\re^{\widehat{\mathsf{x}}}) = \Phi(\re^{\widehat{\mathsf{x}}}+\re^{\widehat{\mathsf{p}}}).
		\end{equation}
		\item The function \(\varphi\) satisfies the pentagon identity
		\begin{equation}\label{eq:pentagon}
			\varphi(\widehat{\mathsf{x}})\varphi(\widehat{\mathsf{p}}) = \varphi(\widehat{\mathsf{p}}) \varphi(\widehat{\mathsf{x}}+\widehat{\mathsf{p}}) \varphi(\widehat{\mathsf{x}}).
		\end{equation}
	\end{enumerate}		
\end{Lemma}

\begin{Remark}
	Consider the limit \(\hbar \to 0\) assuming \(X\sim \hbar Y\).
	Then it follows from the expansion~\eqref{eq:Phi expansion} that \(\Phi(\widehat{X})\to \exp(Y)\).
	This also agrees with the property~\eqref{eq:Phi sum}.
	
	On the other hand, one can consider the limit \(\hbar \to 0\) assuming \(X\) is fixed.
	Then it follows from the expansion~\eqref{eq:log Phi expansion} that \(\log \Phi(X) \sim \dfrac{1}{\hbar} \operatorname{Li}_2(-X)\), where \(\operatorname{Li}_2(X)=\sum_{k>0} \dfrac{X^k}{k^2}\) is the dilogarithm function.
	In this limit, the pentagon relation~\eqref{eq:pentagon} goes to the pentagon relation for the dilogarithm.
		
	Therefore, the function \(\Phi\) can be called a \(q\)-analog of the exponential function.
	On the other hand, \(\Phi\) (or rather \(\log \Phi\)) can be called a \(q\) (or quantum) analog of the dilogarithm.
\end{Remark}
Using the recurrence relation~\eqref{eq:recurrence phi} we can calculate the action of the quantum mutation on the cluster variables.
Indeed, for \(\epsilon_{ik}=-n\le 0\), \(i\neq k\), we have
\begin{multline}
	\label{eq:q mut epsik<0}
	\mu_k(\widehat{X}_i)=\Phi^{-1}(\widehat{X}^{-1}_k)\left(\widehat{X}_{i}\right)\Phi(\widehat{X}^{-1}_k)=\widehat{X}_{i}\Phi^{-1}(\widehat{X}^{-1}_k q^{-2n})\Phi(\widehat{X}^{-1}_k)
	\\
	=\widehat{X}_{i}(1+q^{-1}\widehat{X}_k^{-1})^{-1}\cdot\ldots\cdot (1+q^{1-2n}\widehat{X}_k^{-1})^{-1}.
\end{multline}
In the second case, \(\epsilon_{ik}=n> 0\), we have
\begin{multline}
	\label{eq:q mut epsik>0}
	\mu_k(\widehat{X}_i)=\Phi^{-1}(\widehat{X}^{-1}_k)\left(q^{-n^2}\widehat{X}_i\widehat{X}_k^{n}\right)\Phi(\widehat{X}^{-1}_k)=q^{-n^2}\widehat{X}_i\widehat{X}_k^{n}\Phi^{-1}(\widehat{X}^{-1}_k q^{2n})\Phi(\widehat{X}^{-1}_k)
	\\
	=q^{-n^2}\widehat{X}_i\,\widehat{X}_k^{n}\,
	(1+q\widehat{X}_k^{-1})\cdot\ldots\cdot (1+q^{2n-1}\widehat{X}_k^{-1})
	=\widehat{X}_{i}(1+q^{-1}\widehat{X}_k)\cdot\ldots\cdot (1+q^{1-2n}\widehat{X}_k).
\end{multline}
Finally, since \(\Phi(\widehat{X}^{-1}_k)\) commutes with \(\widehat{X}_k\), we have \(\mu_k(\widehat{X}_k)=\widehat{X}_k^{-1}\).

We used the index \(+\) in the notation for \(\mu_k={\mu}_{k,+}^\sharp \circ \mu_{k,+}'\) because there is also another factorization \(\mu_k={\mu}_{k,-}^\sharp \circ \mu_{k,-}'\). We will not use the second factorization in the text, so from now on \(\mu^\sharp_{k}=\mu^\sharp_{k,+}\) and \(\mu'_{k}=\mu_{k,+}'\).

Now, after the definition of the mutation, several remarks are in order.

\begin{Remark}

		For \(q=1\), the formulas for mutation~\eqref{eq:q mut epsik<0}, \eqref{eq:q mut epsik>0} reduce to the classical expression~\eqref{eq:mutation rule}.
\end{Remark}
\begin{Remark}
	Mutation is an algebra homomorphism (between appropriate localizations of \(\mathbb{C}[\mathcal{X}_{\mathsf{s}}]_q\) and \(\mathbb{C}[\mathcal{X}_{\mathsf{s}'}]_q\)).
	Indeed, the monomial part \(\mu_{k}'\) is an automorphism since the transformation of the matrix \(\epsilon\) follows from the basis transformation~\eqref{eq:basis mutation}, while the second part \(\mu^{\sharp}_{k}\) is a conjugation.
\end{Remark}
\begin{Remark}
	\begin{enumerate}[label=(\alph*)]
		\item It is easy to check that \(\mu_k\) is an involution \(\mu_k\mu_k=\operatorname{id}\), cf.~Prop.~\ref{Prop:mutation relations}\ref{item:a}.
		
				\item Quantum mutations also satisfy the commutation relations given in Prop.
				\ref{Prop:mutation relations}\ref{item:b},\ref{item:c}.
				The commutativity of mutations at two non-connected vertices is obvious; let us comment on the pentagon relation.
				Let us assume that \(\epsilon_{ij}=1\) and denote \(\widehat{x}_i=\widehat{\mathsf{p}}\), \(\widehat{x}_j=\widehat{\mathsf{x}}\).
				Then one can easily see that the monomial parts of the mutations (equivalently, the basis transformations~\eqref{eq:basis mutation}) satisfy
		\(\mu'_{j}\mu'_{i} \mu'_{j}=(i,j)\mu'_{j}\mu'_{i}\).
		In particular, the action on \(\widehat{x}_i,\widehat{x}_j\) has the form
		\begin{align}
			(i,j)\mu'_j\mu'_i&\colon (\widehat{\mathsf{p}},\widehat{\mathsf{x}}) \xrightarrow{\mu'_i} (-\widehat{\mathsf{p}},\widehat{\mathsf{x}})
			\xrightarrow{\mu'_j} (-\widehat{\mathsf{p}},-\widehat{\mathsf{x}})
			\xrightarrow{(i,j)} (-\widehat{\mathsf{x}},-\widehat{\mathsf{p}}),
			\\
			\mu'_j\mu'_i \mu'_j&\colon (\widehat{\mathsf{p}},\widehat{\mathsf{x}}) \xrightarrow{\mu'_j} (\widehat{\mathsf{p}}+\widehat{\mathsf{x}},-\widehat{\mathsf{x}})
			\xrightarrow{\mu'_i} (-\widehat{\mathsf{p}}-\widehat{\mathsf{x}},\widehat{\mathsf{p}})
			\xrightarrow{\mu'_j} (-\widehat{\mathsf{x}},-\widehat{\mathsf{p}}).
		\end{align}
		The agreement of the conjugation parts of mutations exactly boils down to the pentagon relation for the quantum dilogarithm~\eqref{eq:pentagon}:
		\begin{multline}
			\operatorname{Ad}_{\varphi(-\mu_i\widehat{x}_j)^{-1}} \operatorname{Ad}_{\varphi(-\widehat{x}_i)^{-1}}=\operatorname{Ad}_{\varphi(-\widehat{\mathsf{p}})^{-1}}\operatorname{Ad}_{\varphi(-\widehat{\mathsf{x}})^{-1}}=
			\\ \operatorname{Ad}_{\varphi(-\widehat{\mathsf{x}})^{-1}}\operatorname{Ad}_{\varphi(-\widehat{\mathsf{x}}-\widehat{\mathsf{p}})^{-1}}\operatorname{Ad}_{\varphi(-\widehat{\mathsf{p}})^{-1}}=\operatorname{Ad}_{\varphi(-\mu_i\mu_j\widehat{x}_j)^{-1}} \operatorname{Ad}_{\varphi(-\mu_j\widehat{x}_i)^{-1}} \operatorname{Ad}_{\varphi(-\widehat{x}_j)^{-1}}.
		\end{multline}
	\end{enumerate}
\end{Remark}		
\begin{Remark}
	If \(\epsilon_{ik}=1\) then the formula \eqref{eq:q mut epsik>0} simply means that \(\mu_{k}(\widehat{X}_{e_i})=\widehat{X}_{e_i}+\widehat{X}_{e_i+e_k}\), i.e., a sum of elements \(\widehat{X}_\lambda\) (Weyl ordered products).
	In order to explain formula \eqref{eq:q mut epsik>0} for \(\epsilon_{ik}=n>1\), we can assume that \(\widehat{X}_i\) appears as an amalgamation of \(n\) simple vertices.
	In terms of the variables this means that \(\widehat{X}_i=\widehat{X}_i^{(1)}\cdot \dots \cdot \widehat{X}_i^{(n)}\) such that \(\widehat{X}_i^{(a)}\widehat{X}_k=q^2\widehat{X}_k\widehat{X}_i^{(a)}\), \(\forall a\).
	Hence
	\begin{multline}
		\mu_{k}(\widehat{X}_{i})= \mu_{k}(\widehat{X}_{i}^{(1)})\cdot \ldots \cdot \mu_{k}(\widehat{X}_{i}^{(n)} )
		\\
		=\widehat{X}_{i}^{(1)}(1+q^{-1}\widehat{X}_k)\cdot \ldots \cdot \widehat{X}_{i}^{(n)}(1+q^{-1}\widehat{X}_k)=
		\widehat{X}_{i}(1+q^{-1}\widehat{X}_k)\cdot \ldots \cdot (1+q^{1-2n}\widehat{X}_k).
	\end{multline}
	In other words, the factor \((1+q^{-1}\widehat{X}_k)\cdot \ldots \cdot (1+q^{1-2n}\widehat{X}_k)\) actually emerges from a non-commutative power.
\end{Remark}
\begin{Remark}
	Let \(\mathbb{L}[\mathcal{X}_{|\mathsf{s}|}]_q\) be the algebra of elements of \(\mathbb{C}[\mathcal{X}_{\mathsf{s}}]_q\) that are Laurent polynomials after any sequence of mutations.
	It can be proven (see \cite{Davison:2021strong}) that \(\mathbb{L}[\mathcal{X}_{|\mathsf{s}|}]_q\) is a flat deformation of the algebra of global functions \(\mathbb{L}[\mathcal{X}_{|\mathsf{s}|}]\).
\end{Remark}

In principle, for any statement above, one can ask for its quantum analogue.
Moreover, sometimes the quantum story is cleaner and makes some additional features more transparent.
We restrict ourselves to a couple of examples.

\begin{Example}
	One can define quantum transfer matrices for a network similarly to formula~\eqref{eq:T=Transfer matrix}, see \cite{Schrader:2017continuous} for some details.
	Namely, for each path \(p\), one can assign an element \(\lambda_p\in \Lambda\), that is, the sum of \(e_i\) corresponding to the faces \emph{below} the path with an overall shift corresponding to \(\tilde{T} \mapsto T\) above.
	Then the quantum transfer matrix is defined by \(\widehat{T}_{i,j}=\sum_{p \colon \sigma_j \to \tau_i} \widehat{X}_{\lambda_p}\).
	
	In particular, one can define parallel transport matrices assigned to the triangle, \(\widehat{T}_{BC, BA}\), \(\widehat{T}_{CA, BA}\), \(\widehat{T}_{BC, AC}\), see Fig.~\ref{Fig:monodromy triangle}.
	The quantum analog of Theorem~\ref{Th:transport triangle} reads \cite{Chekhov:2020darboux}:
	\begin{gather}
		R \widehat{T}_1 \widehat{T}_2=\widehat{T}_2 \widehat{T}_1R,
		\\
		\widehat{T}_{CA,BA,2}\widehat{T}_{BC,BA,1}=R \widehat{T}_{BC,BA,1}\widehat{T}_{CA,BA,2},
		\\
		R\widehat{T}_{BC,AC,2} \widehat{T}_{BC,BA,1}=\widehat{T}_{BC,BA,1}\widehat{T}_{BC,AC,2}.
	\end{gather}
	Here \(\widehat{T}\) in the first formula is any of the matrices \(\widehat{T}_{BC,BA}\), \(\widehat{T}_{CA,BA}\), \(\widehat{T}_{BC,AC}\) and \(R\) is a quantum \(R\)-matrix given by 
	\begin{equation}
		R=\sum\nolimits_{a} q^{1/2} E_{a,a}\otimes E_{a,a}+ \sum\nolimits_{a<b} \left( q^{-1/2} \big(E_{a,a}\otimes E_{b,b}+E_{b,b}\otimes E_{a,a}\big) +(q^{1/2}-q^{-3/2}) \big(E_{a,b}\otimes E_{b,a}\big) \right ).
	\end{equation}
	
	Using these formulas, one can also study quantum parallel transport matrices for the moduli space of framed local systems, discussed in Section~\ref{Sec:FG}.
\end{Example}

\begin{Example}
	In these notes, the main example of a cluster integrable system is an open relativistic Toda system.
	One can study it using the quantum parallel transport discussed in the previous example, see \cite{Schrader:2017continuous}.
	Quantum mutations also give a remarkable construction of another important ingredient of an integrable system: the Baxter operator \cite{Schrader:2018b}.
	
	In order to define it, we add a vertex to the Toda quiver in Fig.~\ref{Fig:Toda open}.
	The resulting quiver is depicted in Fig.~\ref{Fig:quiver Toda Baxter}.
	The variable corresponding to the additional node is denoted by \(Z\).
	We also specified in the figure the \emph{polarization}: the expression of the variables \(\widehat{x}_j\) in terms of quantum Darboux variables \(\widehat{\mathsf{x}}_1,\dots, \widehat{\mathsf{x}}_N, \widehat{\mathsf{p}}_1,\dots, \widehat{\mathsf{p}}_N\) and the Casimir (or spectral parameter) variable \(\mathsf{u}\).
	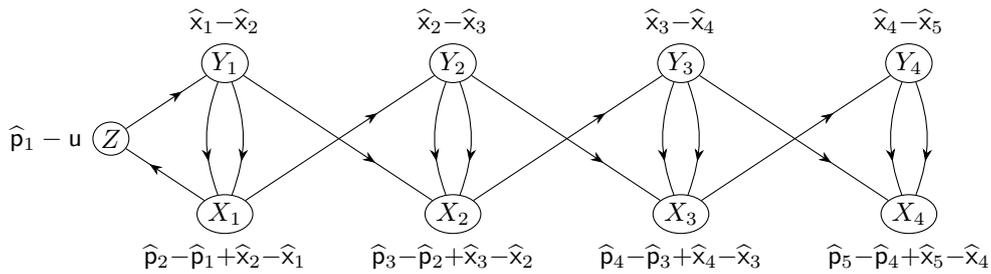
\begin{figure}[h]
		\centering 
		\begin{tikzpicture}[font=\small]
			\def\xs{3} 		
			\def\ys{2}	
			\def\NN{4} 
			\foreach \j in {1,...,\NN}
			{
				\pgfmathtruncatemacro{\jplusone}{\j+1}
				\node[styleNode,label=below:{$\widehat{\mathsf{p}}_{\jplusone}{-}\widehat{\mathsf{p}}_{\j}{+}\widehat{\mathsf{x}}_{\jplusone}{-}\widehat{\mathsf{x}}_{\j}$}] (X\j) at(\j*\xs,0){$X_\j$};
				\node[styleNode, label=above:{$\widehat{\mathsf{x}}_{\j}{-}\widehat{\mathsf{x}}_{\jplusone}$}] (Y\j) at(\j*\xs,\ys){$Y_\j$};
			}
			\node[styleNode,label=left:{$\widehat{\mathsf{p}}_1-\mathsf{u}$}] (Z) at (0.5*\xs,0.5*\ys) {$Z$};
			
			\draw[styleArrow](Z) to (Y1);
			\draw[styleArrow](X1) to (Z);
			
			\foreach \j in {2,...,\NN}
			{
				\pgfmathtruncatemacro{\jminusone}{\j-1}
				\draw[styleArrow](Y\j) to[bend left=20] (X\j);
				\draw[styleArrow](Y\j) to[bend right=20] (X\j);
				\draw[styleArrow](Y\jminusone) to (X\j);
				\draw[styleArrow](X\jminusone) to (Y\j);
			}
			\draw[styleArrow](Y1) to[bend left=20] (X1);
			\draw[styleArrow](Y1) to[bend right=20] (X1);
			
		\end{tikzpicture}
		\caption{\label{Fig:quiver Toda Baxter} The Toda quiver with additional vertex and polarization of cluster variables for \(N=5\)}
	\end{figure}
	
	The Baxter operator is defined to be a composition of mutations at all vertices of the quiver:
	\begin{equation}
		Q(\mathsf{u})={\mu}_{Y_{N-1}} {\mu}_{X_{N-1}} \cdot \ldots \cdot {\mu}_{Y_1} {\mu}_{X_1} {\mu}_{Z}.
	\end{equation}
	The corresponding sequence of monomial mutations \(\mu'_{Y_{N-1}}\mu'_{X_{N-1}} \cdot \ldots \cdot \mu'_{Y_1} \mu'_{X_1} \mu'_{Z}\) sends the quiver to the one depicted in Fig.~\ref{Fig:quiver Toda Baxter after}.
	It also has the form of a Toda quiver, but now with an additional vertex on the other side.
	We also give the variables after this monomial transformation.
	
	\begin{figure}[h]
		\centering 
		\begin{tikzpicture}[font=\small]
			\def\xs{3} 		
			\def\ys{2}	
			\def\NN{4} 
			\pgfmathtruncatemacro{\Nminusone}{\NN-1}
			\pgfmathtruncatemacro{\Nplusone}{\NN+1}
			\foreach \j in {1,...,\Nminusone}
			{
				\pgfmathtruncatemacro{\jplusone}{\j+1}
				\pgfmathtruncatemacro{\jplustwo}{\j+2}
				\node[styleNode, label=below:{$\widehat{\mathsf{x}}_{\j}{-}\widehat{\mathsf{x}}_{\jplusone}$}] (X\j) at(\j*\xs,0){$X_\j$};
				\node[styleNode, label=above:{$\widehat{\mathsf{p}}_{\jplustwo}{-}\widehat{\mathsf{p}}_{\jplusone}{+}\widehat{\mathsf{x}}_{\jplustwo}{-}\widehat{\mathsf{x}}_{\jplusone}$}] (Y\j) at(\j*\xs,\ys){$Y_\j$};
			}
			\node[styleNode,label=left:{$\widehat{\mathsf{p}}_2{-}\widehat{\mathsf{p}}_1{+}\widehat{\mathsf{x}}_2{-}\widehat{\mathsf{x}}_1$}] (Z) at (0.5*\xs,0.5*\ys) {$Z$};
			\node[styleNode, label=below:{$\widehat{\mathsf{x}}_{\NN}{-}\widehat{\mathsf{x}}_{\Nplusone}$}] (X\NN) at(\NN*\xs,0){$X_{\NN}$} ;
			\node[styleNode, label=above:{$\mathsf{u}{-}\widehat{\mathsf{p}}_{\Nplusone}$}] (Y\NN) at(\NN*\xs,\ys){$Y_{\NN}$};

			\draw[styleArrow](X1) to[bend left=20] (Z);
			\draw[styleArrow](X1) to[bend right=20] (Z);
			\draw[styleArrow](Y1) to (X1);
			\draw[styleArrow](Z) to[bend left=15] (1.25*\xs,0.25*\ys) to[bend right=15] (X2);

			\foreach \j in {2,...,\NN}
			{
				\pgfmathtruncatemacro{\jminusone}{\j-1}
				\draw[styleArrow](Y\j) to (X\j);
%				\draw[styleArrow](Y\j) to[bend right=20] (X\j);
%				\draw[styleArrow](Y\jminusone) to (X\j);
				\draw[styleArrow](X\j) to[bend left=20] (Y\jminusone);
				\draw[styleArrow](X\j) to[bend right=20] (Y\jminusone);				
			}
			\foreach \j in {2,...,\Nminusone}
			{
				\pgfmathtruncatemacro{\jminusone}{\j-1}
				\pgfmathtruncatemacro{\jplusone}{\j+1}
				\draw[styleArrow](Y\jminusone) to[bend left=20] (\j*\xs,0.5*\ys) to[bend right=20] (X\jplusone);
			}
			\draw[styleArrow](Y\Nminusone) to (Y\NN);
			
		\end{tikzpicture}
		\caption{\label{Fig:quiver Toda Baxter after} The quiver in Fig.~\ref{Fig:quiver Toda Baxter} with the additional vertex and polarization for \(N=5\) after the Baxter transformation}
	\end{figure}
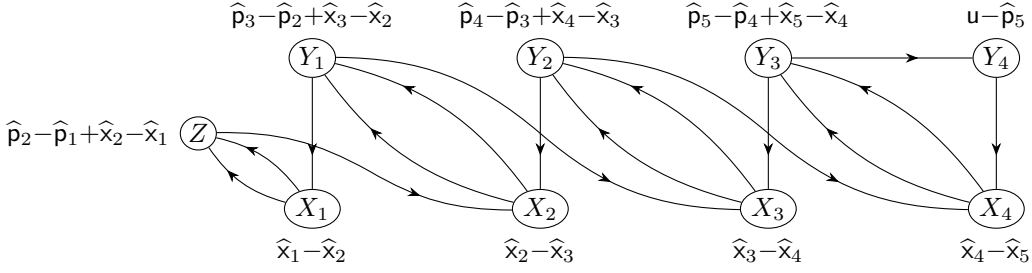
	
	Therefore, we can write 
	\begin{equation}\label{eq:Q in dilogs}
		Q(\mathsf{u})=\varphi(\mathsf{u}{-}\widehat{\mathsf{p}}_1)^{-1} \varphi(\mathsf{u}{-}\widehat{\mathsf{p}}_2{+}\widehat{\mathsf{x}}_1{-}\widehat{\mathsf{x}}_2)^{-1} \varphi(\mathsf{u}{-}\widehat{\mathsf{p}}_2)^{-1} \cdot \ldots \cdot \varphi(\mathsf{u}{-}\widehat{\mathsf{p}}_N{+}\widehat{\mathsf{x}}_{N{-}1}{-}\widehat{\mathsf{x}}_N)^{-1} \varphi(\mathsf{u}{-}\widehat{\mathsf{p}}_{N})^{-1}.
	\end{equation}	
	
	The key property of the Baxter operator proven in \cite{Schrader:2018b} is commutativity for different values of the spectral parameter: \([Q(\mathsf{u}), Q(\mathsf{v})]=0\).
	It follows essentially from the pentagon relation~\eqref{eq:pentagon}.
	Hence, the operators \(\{Q(\mathsf{u})|\mathsf{u}\in \mathbb{C}\}\) define a commutative subalgebra.
	Another way to define this algebra is to take the coefficients of the expansion of \(Q(\mathsf{u})\) in \(\re^{\mathsf{u}}\).
	In such a way, one can find Hamiltonians of the open quantum relativistic Toda system. 
	For example, using the expansion~\eqref{eq:Phi expansion}, one can see that the first term in the expansion of \(Q(\mathsf{u})\) is proportional to the Hamiltonian
	\begin{equation}
		H_1=\re^{-\widehat{\mathsf{p}}_1}+\re^ {\widehat{\mathsf{x}}_1{-}\widehat{\mathsf{x}}_2{-}\widehat{\mathsf{p}}_2}+\re^{-\widehat{\mathsf{p}}_2}+ \ldots + \re^{\widehat{\mathsf{x}}_{N{-}1}{-}\widehat{\mathsf{x}}_N{-}\widehat{\mathsf{p}}_N}+\re^{-\widehat{\mathsf{p}}_{N}}.
	\end{equation}
	This Hamiltonian is equivalent to the conventional Toda Hamiltonian (see, e.g.,
	\cite[Sec. 4.1]{Di:2018difference}).
	See \cite{Di:2024macdonald} for more results about the relation between cluster structures and relativistic Toda systems.
\end{Example}

%\begin{Remark}
%	Contain both \(b,b^{-1}\), defines unitary operator.
%\end{Remark}

\begin{Remark}
	For the quantization of Goncharov--Kenyon integrable systems (discussed in Section~\ref{Sec:GK}), see \cite{Goncharov:2013}.
\end{Remark}

\bibliographystyle{alpha}
\addcontentsline{toc}{section}{\refname}  
\bibliography{ClusterIS}

@incollection {Fock:2016,
	AUTHOR = {Fock, Vladimir and Marshakov, Andrei},
	TITLE = {{Loop groups, clusters, dimers and integrable systems}},
	BOOKTITLE = {{Geometry and quantization of moduli spaces}},
	SERIES = {Adv. Courses Math. CRM Barcelona},
	PAGES = {1--65},
	PUBLISHER = {Birkh\"{a}user/Springer, Cham},
	YEAR = {2016},
	MRCLASS = {37K30 (13F60 14H70 17B80 37K10)},
	MRNUMBER = {3675462},
	MRREVIEWER = {I. S. Krasil\cprime shchik},
	DOI = {10.1007/978-3-319-33578-0_1},
	note={[\doilink{10.1007/978-3-319-33578-0_1}; \href{https://arxiv.org/abs/1401.1606}{\texttt{arXiv:1401.1606}}]},
}

@article {Goncharov:2013,
	AUTHOR = {Goncharov, Alexander and Kenyon, Richard},
	TITLE = {{Dimers and cluster integrable systems}},
	JOURNAL = {Ann. Sci. \'{E}c. Norm. Sup\'{e}r. (4)},
	FJOURNAL = {Annales Scientifiques de l'\'{E}cole Normale Sup\'{e}rieure. Quatri\`eme
	S\'{e}rie},
	VOLUME = {46},
	YEAR = {2013},
	NUMBER = {5},
	PAGES = {747--813},
	ISSN = {0012-9593},
	MRCLASS = {37K30 (05C10 30F60 53D17 82B23)},
	MRNUMBER = {3185352},
	MRREVIEWER = {Guizhang Tu},
	DOI = {10.24033/asens.2201},
	URL = {https://doi.org/10.24033/asens.2201},
	NOTE={[\doilink{10.24033/asens.2201}; \href{https://arxiv.org/abs/1107.5588}{\texttt{arXiv:1107.5588}}]},
}

@article{George2022inverse,
	title={{The inverse spectral map for dimers}},
	author={George, Terrence and Goncharov, Alexander and Kenyon, Richard},
	journal={Mathematical Physics, Analysis and Geometry},
	volume={26},
	number={3},
	pages={24},
	year={2023},
	publisher={Springer},
	doi={10.1007/s11040-023-09466-5},
	note={[\doilink{10.1007/s11040-023-09466-5}; \href{https://arxiv.org/abs/2207.10146}{\texttt{arXiv:2207.10146}}]},
}

@article{Fock:2009cluster,
	title={{Cluster ensembles, quantization and the dilogarithm}},
	author={Fock, Vladimir and Goncharov, Alexander},
	journal={Annales scientifiques de l'{\'E}cole normale sup{\'e}rieure},
	volume={42},
	number={6},
	pages={865--930},
	year={2009},
	doi={10.24033/asens.2112},
	note={[\doilink{10.24033/asens.2112}; \href{https://arxiv.org/abs/math/0311245}{\texttt{arXiv:math/0311245}}]},
}

@incollection{Fock:2006cluster,
	title={{Cluster $\mathcal{X}$-varieties, amalgamation, and Poisson--Lie groups}},
	author={Fock, Vladimir and Goncharov, Alexander},
	booktitle={{Algebraic Geometry and Number Theory: In Honor of Vladimir Drinfeld's 50th Birthday}},
	series={Progress in Mathematics},
	volume={253},
	pages={27--68},
	year={2006},
	publisher={Birkh\"auser Boston},
	doi={10.1007/978-0-8176-4532-8_2},
	note={[\doilink{10.1007/978-0-8176-4532-8_2}; \href{https://arxiv.org/abs/math/0508408}{\texttt{arXiv:math/0508408}}]},
}

@article{Fock:2006moduli,
	title={{Moduli spaces of local systems and higher Teichm{\"u}ller theory}},
	author={Fock, Vladimir and Goncharov, Alexander},
	journal={Publications Math{\'e}matiques de l'IH{\'E}S},
	volume={103},
	pages={1--211},
	year={2006},
	doi={10.1007/s10240-006-0039-4},
	note={[\doilink{10.1007/s10240-006-0039-4}; \href{https://arxiv.org/abs/math/0311149}{\texttt{arXiv:math/0311149}}]},
}

@article{Schrader:2018b,
	title={{On $ b $-Whittaker functions}},
	author={Schrader, Gus and Shapiro, Alexander},
	year={2018},
	NOTE = {[{\href{https://arxiv.org/abs/1806.00747}   {\texttt{arXiv:1806.00747}}]}},
}

@article{Weinstein:1983,
	title={{The local structure of Poisson manifolds}},
	author={Weinstein, Alan},
	journal={Journal of Differential Geometry},
	volume={18},
	number={3},
	pages={523--557},
	year={1983},
	doi={10.4310/jdg/1214437787},
	note={[\doilink{10.4310/jdg/1214437787}]},
}

@book{Babelon:2003introduction,
  title={{Introduction to classical integrable systems}},
  author={Babelon, Olivier and Bernard, Denis and Talon, Michel},
  year={2003},
  publisher={Cambridge University Press},
  doi={10.1017/CBO9780511535024},
  note={[\doilink{10.1017/CBO9780511535024}]},
}

@article{Hoffmann:2000factorization,
  title={{Factorization Dynamics and Coxeter--Toda Lattices}},
  author={Hoffmann, Tim and Kellendonk, Johannes and Kutz, Nadja and Reshetikhin, Nicolai},
  journal={Communications in Mathematical Physics},
  volume={212},
  number={2},
  pages={297--321},
  year={2000},
  publisher={Springer},
	doi={10.1007/s002200000212},
	NOTE = {[\doilink{10.1007/s002200000212}; \href{https://arxiv.org/abs/solv-int/9906013}{\texttt{arXiv:solv-int/9906013}}]},
}

@article{Schrader:2019k,
  title={{$ K $-theoretic Coulomb branches of quiver gauge theories and cluster varieties}},
  author={Schrader, Gus and Shapiro, Alexander},
  year={2019},
	NOTE = {[{\href{https://arxiv.org/abs/1910.03186}{\texttt{arXiv:1910.03186}}]}},
}

@article{Fomin:1999double,
  title={{Double Bruhat cells and total positivity}},
  author={Fomin, Sergey and Zelevinsky, Andrei},
  journal={Journal of the American Mathematical Society},
  volume={12},
  number={2},
  pages={335--380},
  year={1999},
	doi={10.1090/S0894-0347-99-00295-7},
	NOTE = {[\doilink{10.1090/S0894-0347-99-00295-7}; \href{https://arxiv.org/abs/math/9802056}{\texttt{arXiv:math/9802056}}]},
}

@article{Fock:1997note,
    author = {Fock, Vladimir and Marshakov, Andrei},
    title = {{A note on quantum groups and relativistic Toda theory}},
    doi = "10.1016/S0920-5632(97)00328-9",
    journal = "Nucl. Phys. B Proc. Suppl.",
    volume = "56",
    pages = "208--214",
    year = "1997",
    note = {[\doilink{10.1016/S0920-5632(97)00328-9}]},
}

@article {Ruijsenaars:1990Toda,
    AUTHOR = {Ruijsenaars, Simon},
     TITLE = {{Relativistic {T}oda systems}},
   JOURNAL = {Comm. Math. Phys.},
  FJOURNAL = {Communications in Mathematical Physics},
    VOLUME = {133},
      YEAR = {1990},
    NUMBER = {2},
     PAGES = {217--247},
      ISSN = {0010-3616,1432-0916},
   MRCLASS = {58F07},
  MRNUMBER = {1090424},
MRREVIEWER = {Malcolm\ R.\ Adams},
       DOI = {10.1007/BF02097366},
       URL = {http://projecteuclid.org/euclid.cmp/1104201396},
      NOTE = {[\doilink{10.1007/BF02097366}]},
}

@article{Braverman:2019coulomb,
  title={{Coulomb branches of $3d$ $\mathcal{N}=4$ quiver gauge theories and slices in the affine Grassmannian}},
  author={Braverman, Alexander and Finkelberg, Michael and Nakajima, Hiraku},
  journal={Adv. Theor. Math. Phys.},
  volume={23},
  number={1},
  pages={75--166},
  year={2019},
	doi={10.4310/ATMP.2019.v23.n1.a3},
	NOTE = {[\doilink{10.4310/ATMP.2019.v23.n1.a3}; \href{https://arxiv.org/abs/1604.03625}{\texttt{arXiv:1604.03625}}]},
}

@incollection{Finkelberg:2019multiplicative,
  title={{Multiplicative slices, relativistic Toda and shifted quantum affine algebras}},
  author={Finkelberg, Michael and Tsymbaliuk, Alexander},
  booktitle={{Representations and Nilpotent Orbits of Lie Algebraic Systems}},
  series={Progress in Mathematics},
  volume={330},
  pages={133--304},
  year={2019},
  publisher={Birkh\"auser Cham},
  doi={10.1007/978-3-030-23531-4_6},
	NOTE = {[\doilink{10.1007/978-3-030-23531-4_6}; \href{https://arxiv.org/abs/1708.01795}{\texttt{arXiv:1708.01795}}]},
}

@article{Bershtein:2018cluster,
  title={{Cluster integrable systems, $q$-Painlev{\'e} equations and their quantization}},
  author={Bershtein, Mikhail and Gavrylenko, Pavlo and Marshakov, Andrei},
  journal={Journal of High Energy Physics},
  volume={2018},
  number={2},
  pages={077},
  year={2018},
  publisher={Springer},
	doi={10.1007/JHEP02(2018)077},
	NOTE = {[\doilink{10.1007/JHEP02(2018)077}; \href{https://arxiv.org/abs/1711.02063}{\texttt{arXiv:1711.02063}}]},
}

@book{Laurent:2012poisson,
	title={{Poisson structures}},
	author={Laurent-Gengoux, Camille and Pichereau, Anne and Vanhaecke, Pol},
	volume={347},
	series={Grundlehren der mathematischen Wissenschaften},
	year={2013},
	publisher={Springer},
	doi={10.1007/978-3-642-31090-4},
	note={[\doilink{10.1007/978-3-642-31090-4}]},
}

@book{Vaisman:2012lectures,
	title={{Lectures on the geometry of Poisson manifolds}},
	author={Vaisman, Izu},
	volume={118},
	series={Progress in Mathematics},
	year={1994},
	publisher={Birkh{\"a}user},
	doi={10.1007/978-3-0348-8495-2},
	note={[\doilink{10.1007/978-3-0348-8495-2}]},
}

@incollection{Kostant:2006gelfand,
	title={{Gelfand-Zeitlin theory from the perspective of classical mechanics. I}},
	author={Kostant, Bertram and Wallach, Nolan},
	booktitle={{Studies in Lie Theory}},
	series={Progress in Mathematics},
	volume={243},
	pages={319--364},
	year={2006},
	publisher={Birkh\"auser Boston},
	doi={10.1007/0-8176-4478-4_12},
	NOTE = {[\doilink{10.1007/0-8176-4478-4_12}; \href{https://arxiv.org/abs/math/0408342}{\texttt{arXiv:math/0408342}}]},
}

@book {Etingof:2002Lectures,
    AUTHOR = {Etingof, Pavel and Schiffmann, Olivier},
     TITLE = {{Lectures on quantum groups}},
    SERIES = {Lectures in Mathematical Physics},
   EDITION = {Second},
 PUBLISHER = {International Press, Somerville, MA},
      YEAR = {2002},
     PAGES = {xii+242},
      ISBN = {1-57146-094-2},
   MRCLASS = {17B37 (16W35 17-02 17B62 32G34 81R50)},
  MRNUMBER = {2284964},
}

@article{Semenov:1985dressing,
	title={{Dressing transformations and Poisson group actions}},
	author={Semenov-Tian-Shansky, Michael},
	journal={Publications of the Research Institute for Mathematical Sciences},
	volume={21},
	number={6},
	pages={1237--1260},
	year={1985},
	publisher={Research Institute for Mathematical Sciences},
	doi={10.2977/PRIMS/1195178514},
	note={[\doilink{10.2977/PRIMS/1195178514}]},
}

@article{Gross:2015birational,
	title={{Birational geometry of cluster algebras}},
	author={Gross, Mark and Hacking, Paul and Keel, Sean},
	journal={Algebraic Geometry},
	volume={2},
	number={2},
	pages={137--175},
	year={2015},
	publisher={Foundation Compositio Mathematica},
	doi={10.14231/AG-2015-007},
	NOTE = {[\doilink{10.14231/AG-2015-007}; \href{https://arxiv.org/abs/1309.2573}{\texttt{arXiv:1309.2573}}]},
}

@article{Berenstein:2005cluster,
	title={{Cluster algebras III: Upper bounds and double Bruhat cells}},
	author={Berenstein, Arkady and Fomin, Sergey and Zelevinsky, Andrei},
	journal={Duke Mathematical Journal},
	volume={126},
	number={1},
	pages={1--52},
	year={2005},
	publisher={Duke University Press},
	doi={10.1215/S0012-7094-04-12611-9},
	NOTE = {[\doilink{10.1215/S0012-7094-04-12611-9}; \href{https://arxiv.org/abs/math/0305434}{\texttt{arXiv:math/0305434}}]},
}

@article{Fomin:2017introduction,
	title={{Introduction to cluster algebras. Chapters 4--5}},
	author={Fomin, Sergey and Williams, Lauren and Zelevinsky, Andrei},
	year={2017},
	NOTE = {[{\href{https://arxiv.org/abs/1707.07190}{\texttt{arXiv:1707.07190}}]}},
}

@article{Fomin:2016introduction,
	title={{Introduction to cluster algebras. Chapters 1--3}},
	author={Fomin, Sergey and Williams, Lauren and Zelevinsky, Andrei},
	year={2016},
	NOTE = {[{\href{https://arxiv.org/abs/1608.05735}{\texttt{arXiv:1608.05735}}]}},
}

@article{Fomin:2020introduction,
	title={{Introduction to cluster algebras. Chapter 6}},
	author={Fomin, Sergey and Williams, Lauren and Zelevinsky, Andrei},
	year={2020},
	NOTE = {[{\href{https://arxiv.org/abs/2008.09189}{\texttt{arXiv:2008.09189}}]}},
}

@article{Fomin:2021introduction,
	title={{Introduction to cluster algebras. Chapter 7}},
	author={Fomin, Sergey and Williams, Lauren and Zelevinsky, Andrei},
	year={2021},
	NOTE = {[{\href{https://arxiv.org/abs/2106.02160}{\texttt{arXiv:2106.02160}}]}},
}

@article{Fomin:2002cluster,
	title={{Cluster algebras I: foundations}},
	author={Fomin, Sergey and Zelevinsky, Andrei},
	journal={Journal of the American mathematical society},
	volume={15},
	number={2},
	pages={497--529},
	year={2002},
	doi={10.1090/S0894-0347-01-00385-X},
	NOTE = {[\doilink{10.1090/S0894-0347-01-00385-X}; \href{https://arxiv.org/abs/math/0104151}{\texttt{arXiv:math/0104151}}]},
}

@Article{Fock:1999,
	Author = {Fock, Vladimir and Chekhov, Leonid},
	Title = {{A quantum {Teichm{\"u}ller} space}},
	FJournal = {Theoretical and Mathematical Physics},
	Journal = {Theor. Math. Phys.},
	Volume = {120},
	Number = {3},
	Pages = {1245--1259},
	Year = {1999},
	DOI = {10.1007/BF02557246},
	NOTE = {[\doilink{10.1007/BF02557246}; \href{https://arxiv.org/abs/math/9908165}{\texttt{arXiv:math/9908165}}]},
}

@article{Marshakov:2013,
	title={{Lie groups, cluster variables and integrable systems}},
	volume={67},
	ISSN={0393-0440},
	url={http://dx.doi.org/10.1016/j.geomphys.2012.12.003},
	DOI={10.1016/j.geomphys.2012.12.003},
	journal={Journal of Geometry and Physics},
	publisher={Elsevier BV},
	author={Marshakov, Andrei},
	year={2013},
	pages={16--36},
	note={[\doilink{10.1016/j.geomphys.2012.12.003}; \href{https://arxiv.org/abs/1207.1869}{\texttt{arXiv:1207.1869}}]},
}

@article{Postnikov2006total,
	title={{Total positivity, Grassmannians, and networks}},
	author={Postnikov, Alexander},
	note={[\href{https://arxiv.org/abs/math/0609764}{\texttt{arXiv:math/0609764}}]},
	year={2006},	 
}

@article{Goncharov:2019quantum,
	title={{Quantum geometry of moduli spaces of local systems and representation theory}},
	author={Goncharov, Alexander and Shen, Linhui},	
	note={[\href{https://arxiv.org/abs/1904.10491}{\texttt{arXiv:1904.10491}}]},
	year={2019},
}

@incollection{Goncharov:2017ideal,
	title={{Ideal webs, moduli spaces of local systems, and 3d Calabi--Yau categories}},
	author={Goncharov, Alexander},
	booktitle={{Algebra, Geometry, and Physics in the 21st Century: Kontsevich Festschrift}},
	series={Progress in Mathematics},
	volume={324},
	pages={31--97},
	year={2017},
	publisher={Birkh\"auser Cham},
	doi={10.1007/978-3-319-59939-7_2},
	note={[\doilink{10.1007/978-3-319-59939-7_2}; \href{https://arxiv.org/abs/1607.05228}{\texttt{arXiv:1607.05228}}]},
}

@article{Chekhov:2020darboux,
	title={{Log-canonical coordinates for symplectic groupoid and cluster algebras}},
	author={Chekhov, Leonid and Shapiro, Michael},	
	journal={International Mathematics Research Notices},
	volume={2023},
	number={11},
	pages={9565--9652},
	year={2023},
	doi={10.1093/imrn/rnac101},
	note={[\doilink{10.1093/imrn/rnac101}; \href{https://arxiv.org/abs/2003.07499}{\texttt{arXiv:2003.07499}}]},
}

@article{Fock:1998poisson,
	title={{Poisson structure on moduli of flat connections on Riemann surfaces and $ r $-matrix}},
	author={Fock, Vladimir and Rosly, Alexei},
	journal={American Mathematical Society Translations, Series 2},
	volume={191},
	pages={67--86},
	year={1999},
	note={[\href{https://arxiv.org/abs/math/9802054}{\texttt{arXiv:math/9802054}}]},
}

@article{Goldman:1986invariant,
	title={{Invariant functions on Lie groups and Hamiltonian flows of surface group representations}},
	author={Goldman, William},
	journal={Inventiones mathematicae},
	volume={85},
	number={2},
	pages={263--302},
	year={1986},
	publisher={Springer},
	doi={10.1007/BF01389091},
	note={[\doilink{10.1007/BF01389091}]},
}

@article{Gekhtman:2016integrable,
	title={{Integrable cluster dynamics of directed networks and pentagram maps}},
	author={Gekhtman, Michael and Shapiro, Michael and Tabachnikov, Serge and Vainshtein, Alek},
	journal={Advances in Mathematics},
	volume={300},
	pages={390--450},
	year={2016},
	publisher={Elsevier},	
	doi={10.1016/j.aim.2016.03.023},
	note={[\doilink{10.1016/j.aim.2016.03.023}; \href{https://arxiv.org/abs/1406.1883}{\texttt{arXiv:1406.1883}}]},
}

@article{Marshakov:2019cluster,
	title={{Cluster integrable systems and spin chains}},
	author={Marshakov, Andrei and Semenyakin, Mykola},
	journal={Journal of High Energy Physics},
	volume={2019},
	number={10},
	pages={100},
	year={2019},
	publisher={Springer},	
	doi={10.1007/JHEP10(2019)100},
	note={[\doilink{10.1007/JHEP10(2019)100}; \href{https://arxiv.org/abs/1905.09921}{\texttt{arXiv:1905.09921}}]},
}

@article{Bocklandt:2016dimer,
	title={{A dimer ABC}},
	author={Bocklandt, Raf},
	journal={Bulletin of the London Mathematical Society},
	volume={48},
	number={3},
	pages={387--451},
	year={2016},
	publisher={Oxford University Press},
	doi={10.1112/blms/bdv101},
	NOTE = {[\doilink{10.1112/blms/bdv101}; \href{https://arxiv.org/abs/1510.04242}{\texttt{arXiv:1510.04242}}]},
}

@article{Bershtein:2024cluster,
	title={{Cluster Reductions, Mutations, and $q$-Painlev\'e Equations}},
	author={Bershtein, Mikhail and Gavrylenko, Pavlo and Marshakov, Andrei and Semenyakin, Mykola},
	year={2024},
	NOTE = {[{\href{https://arxiv.org/abs/2411.00325}   {\texttt{arXiv:2411.00325}}]}},
}

@article{Gaiotto:2013wall,
	title={{Wall-crossing, Hitchin systems, and the WKB approximation}},
	author={Gaiotto, Davide and Moore, Gregory W and Neitzke, Andrew},
	journal={Advances in Mathematics},
	volume={234},
	pages={239--403},
	year={2013},
	publisher={Elsevier},
	doi={10.1016/j.aim.2012.09.027},
	NOTE = {[\doilink{10.1016/j.aim.2012.09.027}; \href{https://arxiv.org/abs/0907.3987}{\texttt{arXiv:0907.3987}}]},
}

@incollection {Ishii:2011note,
	AUTHOR = {Ishii, Akira and Ueda, Kazushi},
	TITLE = {{A note on consistency conditions on dimer models}},
	BOOKTITLE = {{Higher dimensional algebraic geometry}},
	SERIES = {RIMS K\^{o}ky\^{u}roku Bessatsu},
	VOLUME = {B24},
	PAGES = {143--164},
	PUBLISHER = {Res. Inst. Math. Sci. (RIMS), Kyoto},
	YEAR = {2011},
	NOTE={[\href{https://arxiv.org/abs/1012.5449}{\texttt{arXiv:1012.5449}}]},
}

@article {Broomhead:2010,
	AUTHOR = {Broomhead, Nathan},
	TITLE = {{Dimer models and {C}alabi-{Y}au algebras}},
	JOURNAL = {Mem. Amer. Math. Soc.},
	FJOURNAL = {Memoirs of the American Mathematical Society},
	VOLUME = {215},
	YEAR = {2012},
	NUMBER = {1011},
	PAGES = {viii+86},
	ISSN = {0065-9266,1947-6221},
	ISBN = {978-0-8218-5308-5},
	MRCLASS = {82B20 (14M25)},
	MRNUMBER = {2908565},
	DOI = {10.1090/S0065-9266-2011-00617-9},
	URL = {https://doi.org/10.1090/S0065-9266-2011-00617-9},
	note={ [\doilink{10.1090/S0065-9266-2011-00617-9}; \href{https://arxiv.org/abs/0901.4662}{\texttt{arXiv:0901.4662}}]},
}

@article{Gulotta:2008properly,
	title={{Properly ordered dimers, $R$-charges, and an efficient inverse algorithm}},
	author={Gulotta, Daniel R},
	journal={Journal of High Energy Physics},
	volume={2008},
	number={10},
	pages={014},
	year={2008},
	publisher={IOP Publishing},
	doi={10.1088/1126-6708/2008/10/014},
	note= "{[\doilink{10.1088/1126-6708/2008/10/014}; \href{https://arxiv.org/abs/0807.3012}{\texttt{arXiv:0807.3012}}]}",
}

@article {Ishii:2009:2,
	AUTHOR = {Ishii, Akira and Ueda, Kazushi},
	TITLE = {{Dimer models and the special {M}c{K}ay correspondence}},
	JOURNAL = {Geom. Topol.},
	FJOURNAL = {Geometry \& Topology},
	VOLUME = {19},
	YEAR = {2015},
	NUMBER = {6},
	PAGES = {3405--3466},
	ISSN = {1465-3060},
	MRCLASS = {14F05 (14D20 14E16 16G10 82B20)},
	MRNUMBER = {3447107},
	MRREVIEWER = {Howard Nuer},
	DOI = {10.2140/gt.2015.19.3405},
	URL = {https://doi.org/10.2140/gt.2015.19.3405},
	note={ [\doilink{10.2140/gt.2015.19.3405}; \href{https://arxiv.org/abs/0905.0059}{\texttt{arXiv:0905.0059}}]},
}

@article {Khovanskii:1978,
	AUTHOR = {Khovanski\u{\i}, Askold},
	TITLE = {{Newton polyhedra, and the genus of complete intersections}},
	JOURNAL = {Functional Analysis and Its Applications},
	FJOURNAL = {Akademiya Nauk SSSR. Funktsional\cprime ny\u{\i} Analiz i ego
	Prilozheniya},
	VOLUME = {12},
	YEAR = {1978},
	NUMBER = {1},
	PAGES = {38--46},
	ISSN = {0374-1990},
	MRCLASS = {14M10 (14B07)},
	MRNUMBER = {487230},
	MRREVIEWER = {K.\ Wolffhardt},
	DOI = {10.1007/BF01077562},
	NOTE = {[\doilink{10.1007/BF01077562}]},
}

@incollection{Gekhtman:2024integrable,
	title={{Integrable systems and cluster algebras}},
	author={Gekhtman, Michael and Izosimov, Anton},
	booktitle={{Encyclopedia of Mathematical Physics, Second Edition}},
	volume={1--5},
	pages={V3:294--V3:308},
	year={2024},
	publisher={Elsevier},
	doi={10.1016/B978-0-323-95703-8.00029-X},
	note={ [\doilink{10.1016/B978-0-323-95703-8.00029-X}; \href{https://arxiv.org/abs/2403.07287}{\texttt{arXiv:2403.07287}}]},
}

@article{Davison:2021strong,
	title={{Strong positivity for quantum theta bases of quantum cluster algebras}},
	author={Davison, Ben and Mandel, Travis},
	journal={Inventiones mathematicae},
	volume={226},
	pages={725--843},
	year={2021},
	publisher={Springer},
	doi={10.1007/s00222-021-01061-1},
	note= "{[\doilink{10.1007/s00222-021-01061-1}; \href{https://arxiv.org/abs/1910.12915}{\texttt{arXiv:1910.12915}}]}",
}

@article{George:2022discrete,
	title={{Discrete dynamics in cluster integrable systems from geometric $R$-matrix transformations}},
	author={George, Terrence and Ramassamy, Sanjay},
	JOURNAL = {Comb. Theory},
	FJOURNAL = {Combinatorial Theory},
	VOLUME = {3},
	YEAR = {2023},
	NUMBER = {2},
	PAGES = {Paper No. 12, 29},
	ISSN = {2766-1334},
	DOI = {10.5070/C63261990},
	note= "{[\doilink{10.5070/C63261990}; \href{https://arxiv.org/abs/2208.10306}{\texttt{arXiv:2208.10306}}]}",
}

@article{Gekhtman:2012Poisson,
	AUTHOR = {Gekhtman, Michael and Shapiro, Michael and Vainshtein, Alek},
	TITLE = {{Poisson geometry of directed networks in an annulus}},
	JOURNAL = {J. Eur. Math. Soc. (JEMS)},
	FJOURNAL = {Journal of the European Mathematical Society (JEMS)},
	VOLUME = {14},
	YEAR = {2012},
	NUMBER = {2},
	PAGES = {541--570},
	ISSN = {1435-9855,1435-9863},
	MRCLASS = {53D17 (14M15)},
	MRNUMBER = {2881305},
	MRREVIEWER = {Christian\ Ohn},
	DOI = {10.4171/JEMS/311},
	URL = {https://doi.org/10.4171/JEMS/311},
	note= "{[\doilink{10.4171/JEMS/311}; \href{https://arxiv.org/abs/0901.0020}{\texttt{arXiv:0901.0020}}]}",
}

@article{Izosimov:2022dimers,
	title={{Dimers, networks, and cluster integrable systems}},
	author={Izosimov, Anton},
	journal={Geometric and Functional Analysis},
	volume={32},
	number={4},
	pages={861--880},
	year={2022},
	publisher={Springer},
	doi={10.1007/s00039-022-00605-8},
	note= "{[\doilink{10.1007/s00039-022-00605-8}; \href{https://arxiv.org/abs/2108.04975}{\texttt{arXiv:2108.04975}}]}",
}

@article{Di:2024macdonald,
	title={{Macdonald Duality and the proof of the Quantum Q-system conjecture}},
	author={Di Francesco, Philippe and Kedem, Rinat},
	journal={Selecta Mathematica},
	volume={30},
	number={2},
	pages={23},
	year={2024},
	publisher={Springer},
	doi={10.1007/s00029-023-00909-z},
	note= "{[\doilink{10.1007/s00029-023-00909-z}; \href{https://arxiv.org/abs/2112.09798}{\texttt{arXiv:2112.09798}}]}",
}

@article{Di:2018difference,
	title={{Difference equations for graded characters from quantum cluster algebra}},
	author={Di Francesco, Philippe and Kedem, Rinat},
	journal={Transformation Groups},
	volume={23},
	pages={391--424},
	year={2018},
	publisher={Springer},
	doi={10.1007/s00031-018-9480-y},
	note= "{[\doilink{10.1007/s00031-018-9480-y}; \href{https://arxiv.org/abs/1505.01657}{\texttt{arXiv:1505.01657}}]}",
}

@article{Schrader:2017continuous,
	title={{Continuous tensor categories from quantum groups I: algebraic aspects}},
	author={Schrader, Gus and Shapiro, Alexander},
	year={2017},
	note= "{[\href{https://arxiv.org/abs/1708.08107}{\texttt{arXiv:1708.08107}}]}",
}

@article{Williams:2013cluster,
	title={{Cluster ensembles and Kac--Moody groups}},
	author={Williams, Harold},
	journal={Advances in Mathematics},
	volume={247},
	pages={1--40},
	year={2013},
	publisher={Elsevier},
	doi={10.1016/j.aim.2013.07.008},
	note= "{[\doilink{10.1016/j.aim.2013.07.008}; \href{https://arxiv.org/abs/1210.2533}{\texttt{arXiv:1210.2533}}]}",
}

@article{Williams:2013double,
	title={{Double Bruhat cells in Kac--Moody groups and integrable systems}},
	author={Williams, Harold},
	journal={Letters in Mathematical Physics},
	volume={103},
	pages={389--419},
	year={2013},
	publisher={Springer},
	doi={10.1007/s11005-012-0604-3},
	note= "{[\doilink{10.1007/s11005-012-0604-3}; \href{https://arxiv.org/abs/1204.0601}{\texttt{arXiv:1204.0601}}]}",
}

@article{Kogan:2002symplectic,
	title={{On symplectic leaves and integrable systems in standard complex semisimple Poisson-Lie groups}},
	author={Kogan, Mikhail and Zelevinsky, Andrei},
	journal={International Mathematics Research Notices},
	volume={2002},
	number={32},
	pages={1685--1702},
	year={2002},
	publisher={OUP},
	doi={10.1155/S1073792802203050},
	note= "{[\doilink{10.1155/S1073792802203050}; \href{https://arxiv.org/abs/math/0203069}{\texttt{arXiv:math/0203069}}]}",
}

\noindent 
\textsc{Scuola Internazionale Superiore di Studi Avanzati (SISSA), Trieste, Italy
\\  School of Mathematics, University of Edinburgh, Edinburgh, UK}

\emph{E-mail}:\,\,\textbf{mbersht@sissa.it}\\

\end{document}